\documentclass[a4paper]{article}
\pdfoutput=1
\usepackage{jheppub}

\usepackage[utf8]{inputenc}

\usepackage[table ]{ xcolor}

\usepackage{multirow}

\usepackage{amsmath}

\usepackage{tikz}
\usetikzlibrary{shapes,arrows}
\tikzstyle{startstop} = [rectangle,rounded corners, minimum width=3cm,minimum height=1cm,text centered, draw=black,fill=red!30]
\tikzstyle{io} = [trapezium, trapezium left angle = 70,trapezium right angle=110,minimum width=3cm,minimum height=1cm,text centered,draw=black,fill=blue!30]
\tikzstyle{process} = [rectangle,minimum width=3cm,minimum height=1cm,text centered,text width =3cm,draw=black,fill=orange!30]
\tikzstyle{decision} = [diamond,minimum width=3cm,minimum height=1cm,shape aspect=3,inner sep = 0.4pt,text centered,draw=black,fill=green!30]
\tikzstyle{arrow} = [thick,->,>=stealth]
\tikzstyle{shadow}=[preaction={fill=black,opacity=.5,transform canvas={xshift=0.5mm,yshift=-0.5mm},shading=radial,shading angle=20},fill=red]

\tikzstyle{ellipse}=[draw, rectangle, minimum width=2.8em, rounded corners=6pt,line width=0.5pt]
\tikzstyle{pxsbx}=[trapezium, trapezium left angle=75, trapezium right angle=105, minimum width=3em, text centered, draw = black, fill=white,line width=0.5pt] 
\tikzstyle{lingxing}=[draw,diamond,shape aspect=3,inner sep = 0.4pt,thick,font=\itshape,line width=0.5pt]

\usepackage{amssymb}
\usepackage{graphicx,color}

\usepackage{subfigure}
\usepackage{amsmath,amsthm}

\usepackage{mathrsfs}

\usepackage{bm}

\allowdisplaybreaks

\ifx\Ref\undefined
\newcommand{\Ref}[1]{(\ref{#1})}
\fi

\newcommand{\Z}{\mathbb{Z}}
\newcommand{\R}{\mathbb{R}}
\newcommand{\C}{\mathbb{C}}

\newcommand{\half}{\frac{1}{2}}

\newcommand{\ccirc}{\kern0.2ex\vcenter{\hbox{$\scriptstyle\circ$}}\kern0.2ex}

\newcommand{\Slc}{\mathrm{SL}(2,\mathbb{C})}

\newcommand{\Su}{\mathrm{SU}(2)}

\def\be{\begin{eqnarray}}
\def\ee{\end{eqnarray}}

\newcommand{\cc}{\mathcal C}

\newcommand{\ce}{\mathcal E}
\newcommand{\cf}{\mathcal F}
\newcommand{\cg}{\mathcal G}
\newcommand{\ch}{\mathcal H}

\newcommand{\cp}{\mathcal P}

\newcommand{\cs}{\mathcal S}

\newcommand{\cv}{\mathcal V}

\newcommand{\cx}{\mathcal X}
\newcommand{\cy}{\mathcal Y}

  \newcommand{\Fa}{\mathfrak{A}}
  \newcommand{\Fb}{\mathfrak{B}}
  \newcommand{\Fc}{\mathfrak{C}}
  \newcommand{\Fd}{\mathfrak{D}}

\renewcommand{\a}{\alpha}
\renewcommand{\b}{\beta}
\newcommand{\g}{\gamma}

\newcommand{\eps}{\varepsilon}

\newcommand{\sig}{\sigma}

\renewcommand{\L }{\Lambda}
\renewcommand{\o}{\omega}

\renewcommand{\t}{\tau}

\newcommand{\rmd}{\mathrm d}

\newcommand{\lt}{\left}
\newcommand{\rt}{\right}

\newcommand{\lag}{\left\langle}

\newcommand{\tr}{\mathrm{tr}}

\newcommand{\sgn}{\mathrm{sgn}}

\title{Manifestly Gauge-Invariant Cosmological Perturbation Theory from Full Loop Quantum Gravity}

\author[1,2]{Muxin Han}  

\affiliation[1]{Department of Physics, Florida Atlantic University, 777 Glades Road, Boca Raton, FL 33431-0991, USA}

\affiliation[2]{Institut f\"ur Quantengravitation, Universit\"at Erlangen-N\"urnberg, Staudtstr. 7/B2, 91058 Erlangen, Germany}

\author[3]{\ Haida Li} 

\affiliation[3]{Department of Physics, Beijing Normal University, Beijing 100875, China}

\author[2,4]{\ Hongguang Liu}

\affiliation[4]{Center for Quantum Computing, Pengcheng Laboratory, Shenzhen 518066, China}

\emailAdd{hanm(At)fau.edu}
\emailAdd{haidali(At)mail.bnu.edu.cn}
\emailAdd{liu.hongguang(At)cpt.univ-mrs.fr}

\abstract{We apply the full theory of Loop Quantum Gravity (LQG) to cosmology and present a top-down derivation of gauge-invariant cosmological perturbation theory from quantum gravity. The derivation employs the reduced phase space formulation of LQG and the new discrete path integral formulation defined in \cite{Han:2019vpw}. We demonstrate that in the semiclassical approximation and continuum limit, the result coincides with the existing formulation of gauge-invariant cosmological perturbation theory in e.g. \cite{Giesel:2007wk}. Time evolution of cosmological perturbations is computed numerically from the new cosmological perturbation theory of LQG, and various power spectrums are studied for scalar mode and tensor mode perturbations. Comparing these power spectrums with predictions from the classical theory demonstrate corrections in the ultra-long wavelength regime. These corrections are results from the lattice discretization in LQG. In addition, tensor mode perturbations at late time demonstrate the emergence of spin-2 gravitons as low energy excitations from LQG. The graviton has a modified dispersion relation and reduces to the standard graviton in the long wavelength limit. 
}


\begin{document}

\maketitle

\section{Introduction}

Loop Quantum Gravity (LQG) is a candidate for background independent and non-perturbative theory of quantum gravity \cite{book,review,review1,rovelli2014covariant}. Among successful sub-areas in LQG, applying LQG to cosmology is a fruitful direction in which LQG gives physical predictions and phenomenological impacts. Most studies of cosmology in LQG is based on Loop Quantum Cosmology (LQC): a LQG-like quantization of symmetry reduced model (quantization of homogeneous and isotropic degrees of freedom) \cite{Ashtekar:2006wn,Bojowald:2001xe,Agullo:2016tjh}. However, in this paper, we apply the full theory of LQG (quantizing all degrees of freedom) to cosmology and present a top-down derivation of cosmological perturbation theory from LQG.

A key tool in our work is the new path integral formulation of LQG proposed in \cite{Han:2019vpw}. This path integral is derived from the reduced phase space formulation of canonical LQG. The reduced phase space formulation couples gravity to matter fields such as dusts or scalar fields (clock fields), followed by a deparametrization procedure, in which gravity Dirac observables are parametrized by values of clock fields, and constraints are solved classically. The dynamics of Dirac observables is governed by the physical Hamiltonian ${\bf H}_0$ generating physical time evolution (the physical time is the value of a clock field) in the reduced phase space. Our work considers two popular scenarios of deparametrization: coupling gravity to Brown-Kucha\v{r} and Gaussian dusts \cite{Giesel:2007wi,Giesel:2007wk,Giesel:2007wn,Giesel:2012rb}. The path integral formulation is derived from discretizing the theory on a cubic lattice $\g$, followed by quantizing the reduced phase space and the Hamiltonian evolution generated by ${\bf H}_0$. We refer the readers to \cite{Han:2019vpw} for detailed derivation of the path integral formulation, and to \cite{Han2020} for the comparison with spin foam formulation.

The semiclassical approximation $\hbar\to0$ of LQG can be studied in this path integral formulation using the stationary phase analysis. It is shown in \cite{Han2020} that semiclassical equations of motion (EOMs) from the path integral consistently reproduces the classical reduced phase space EOMs of the gravity-dust system. These semiclassical EOMs take into account all degrees of freedom (DOFs) on $\g$, and govern the semiclassical dynamics of the full LQG. In addition, \cite{Han:2019vpw} shows that semiclassical EOMs contain the unique solution satisfying the homogeneous and isotropic symmetry. The solution reproduces the effective dynamics of $\mu_0$-scheme LQC, i.e. it recovers the Friedmann-Lema\^itre-Robertson-Walker (FLRW) cosmology at low energy density while replacing the big-bang singularity by a bounce at high energy density.

In this work, we study perturbations on the homogeneous and isotropic cosmology in this path integral formulation of full LQG. We focus on the cosmological perturbation theory at the semiclassical level. The dynamics of perturbations are studied by taking the above homogeneous and isotropic as the background and linearizing semiclassical EOMs of the full LQG. The resulting linearized EOMs are in terms of (perturbative) holonomies and fluxes on the cubic lattice $\g$. The initial condition of EOMs is imposed by the semiclassical initial state of the path integral, and uniquely determines a solution. In practice, we solve these linearized EOMs numerically and extract the physics of cosmological perturbations. The perturbation theory developed here is manifestly gauge invariant because it is derived from the reduced phase space formulation.

There are cosmological perturbation theories based on LQC instead of the full LQG, including the dressed metric approach \cite{Agullo:2016hap,Ashtekar:2016pqn,Ashtekar:2020gec}, deformed algebra approach \cite{Bojowald2008,Cailleteau2011,Mielczarek2011,Mielczarek2010}  and the hybrid model \cite{Gomar:2015oea,ElizagaNavascues:2016vqw}. In all those approaches, LQC quantum dynamics serves as the background for perturbations. However the dynamics of LQC is ambiguous by different treatments of Lorentzian terms in the Hamiltonian constraint. The ambiguity can have no nontrivial effects on predictions \cite{paramtalk,Li:2019qzr}. Our approach derives the cosmological perturbation theory from the full LQG Hamiltonian (proposed by Giesel and Thiemann \cite{Giesel:2007wn}) which specifies the Lorentzian term from the start. So ambiguities mentioned in \cite{paramtalk,Li:2019qzr} do not present in our approach.


As a consistency check, we take the continuum limit of linearized EOMs by refining the lattice $\g$, and find results agree with perturbative EOMs in \cite{Giesel:2007wk}, where the gauge-invariant cosmological perturbation theory is developed from classical gravity-dust theory on the continuum. Our result provides an example confirming the semiclassical consistency of the reduced phase space LQG. The cosmological perturbation theory from the reduced phase space formulation closely relates to the standard gauge-invariant treatment of cosmological perturbations \cite{Giesel:2007wk}.

Our top-down approach of the cosmological perturbation theory opens a new window for extracting physical predication from the full LQG and contacting with phenomenology. As the first step, we relate holonomy and flux perturbations to the standard decomposition into scalar, vector, and tensor modes, and numerically study their power spectrums determined by the semiclassical dynamics of LQG. Resulting power spectrums are compared with predictions from the classical theory on the continuum. This comparison demonstrates physical effects implied by the lattice discreteness and cosmic bounce in LQG.

Our analysis of power spectrums mainly focuses on scalar and tensor modes, since they have more phenomenological impact. Concretely, we study the power spectrum of the Bardeen potential $\Psi$ for the scalar mode perturbation (see Section \ref{Power Spectrum}), and the power spectrum of metric perturbations of the tensor mode (see Section \ref{Tensor Mode Perturbations}). Power spectrums are obtained by numerically evolving perturbations from certain initial conditions imposed at early time.

Firstly it is clear that predictions from LQG semiclassical EOMs are very different from the continuum classical theory in case that the wavelength is as short as the lattice spacing. However when we focus on wavelengths much longer than the lattice spacing, differences in power spectrums between LQG and the classical theory are much larger  in the ultra-long wavelength regime than they are in the regime where the wavelength is relatively short (but still much longer than the lattice spacing). Power spectrums from LQG coincide with the classical theory in the regime where the wavelength is relatively short. At late time, this difference of scalar mode power spectrums becomes smaller, while the difference of tensor mode power spectrums becomes larger. For the tensor mode, the long wavelength correction from LQG in the power spectrum has a similar reason as in the dressed metric approach \cite{Agullo:2016hap,Ashtekar:2020gec}, i.e. it is due to the LQG correction to the cosmological background. For the Bardeen potential $\Psi$, the difference of power spectrums is resulting from $\Psi\sim \text{wavelength}\times\text{perturbation}$ where corrections to perturbations from the lattice discreteness are amplified by ultra-long wavelengths. Differences in power spectrums between LQG and the classical theory vanish in the lattice continuum limit. Some more discussions about comparison are given in Sections \ref{Power Spectrum} and \ref{Tensor Mode Perturbations}.

At late time, tensor mode perturbations from LQG give a wave equation of spin-2 gravitons with a modified dispersion relation $\o(k)^2=k^2[1+O(k^2)]$ (see Section \ref{Tensor Mode Perturbations} for the expression). $\o(k)^2$ reduces to the usual dispersion relation of graviton in the long wavelength limit or small $k$. For larger $k$, gravitons travel in a speed less than the speed of light. Our result confirms that spin-2 gravitons are low energy excitations of LQG. It is in agreement with a recent result from the spin foam formulation \cite{Han:2018fmu}. The modified dispersion relation is in agreement with a recent result in \cite{Dapor:2020jvc} obtained from expanding the LQG Hamiltonian on the flat spacetime. The modified dispersion relation indicates an apparently spurious mode at large $k$ (at the wavelength comparable to the lattice scale). But in our opinion, the large $k$ is beyond the regime to valid our effective theory, so the dispersion relation should only be trusted in the long wavelength regime (see Section \ref{Tensor Mode Perturbations} for discussion).   

As another difference between LQG and the classical theory, the cosmological perturbation theory from LQG contain couplings among scalar, tensor, and vector modes, although these couplings are suppressed by the lattice continuum limit. For instance, the initial condition containing only scalar mode can excite tensor and vector modes in the time evolution at the discrete level. These tensor and vector modes have small amplitudes vanishing in the lattice continuum limit.

As a preliminary step toward making the full LQG theory contact with phenomenology, this work has following limitations: Firstly, our model focuses on pure gravity coupled to dusts, and does not take into account the radiative matter and inflation. However various matter couplings in the reduced phase space LQG have been worked out in \cite{Giesel:2007wn}. Deriving matter couplings in the path integral formulation is straight-forward. Generalizing the cosmological perturbation theory to including radiative matter and inflation is a work currently undergoing. Secondly, this work focuses on the semiclassical analysis, and does not take into account any $O(\ell_P^2)$ quantum correction (although effects from discreteness are discussed). By taking into account quantum corrections, the continuum limit at the quantum level is expected to be better understood.

Main computations in this work are carried out with Mathematica on High-Performance-Computing (HPC) servers. Some intermediate steps and results contain long formulae that cannot be shown in the paper. Mathematica codes and formulae can be downloaded from \cite{github}.

This paper is organized as follows: Section \ref{BK} reviews the reduced phase space formulation of LQG and the path integral formulation. Section \ref{Semiclassical Equations of Motion} discusses the semiclassical approximation of the path integral and semiclassical EOMs. Section \ref{Cosmological Background and Perturbations} discuss the cosmological solution, linearization of EOMs with cosmological perturbations, and lattice continuum limit. Section \ref{Power Spectrum} focuses on scalar mode perturbations, and discusses the initial condition and the power spectrum. Section \ref{Tensor Mode Perturbations} focuses on tensor mode perturbations, including discussions of the late time dispersion relation and the power spectrum.

\section{Reduced Phase Space Formulation of LQG}\label{BK}

\subsection{Classical Framework}\label{Classical Framework}

Reduced phase space formulations of LQG need to couple gravity to various matter fields at classical level. In this paper, we focus on two scenarios of matter field couplings: Brown-Kucha\v{r} (BK) dust and Gaussian dust \cite{Brown:1994py,Kuchar:1990vy,Giesel:2007wn,Giesel:2012rb}.

The action of BK dust model reads
\be
S_{BKD}[\rho,g_{\mu\nu},T,S^j,W_j]&=& -\frac{1}{2}\int\rmd^4x\ \sqrt{|\det(g)|}\ \rho\ [g^{\mu\nu}U_\mu U_\nu+1],\label{dustaction}\\
U_\mu&=&-\partial_\mu T+W_j\partial_\mu S^j,
\ee
where $T, S^{j=1,2,3}$ are scalars (dust coordinates of time and space) to parametrize physical fields, and $\rho,\ W_j$ are Lagrangian multipliers. $\rho$ is interpreted as the dust energy density. Coupling $S_{BKD}$ to gravity (or gravity coupled to some other matter fields) and carrying out Hamiltonian analysis \cite{Giesel:2012rb}, we obtain following constraints:
\be
\cc^{tot}&=&\cc+\frac{1}{2}\left[\frac{P^{2} / \rho}{\sqrt{\operatorname{det}(q)}}+\sqrt{\operatorname{det}(q)} \rho\left(q^{\a \b} U_{\a} U_{\b}+1\right)\right]=0,\label{C}\\
\cc^{tot}_\a&=&\cc_\a+PT_{,\a}-P_jS^j_{,\a}=0,\label{Ca}\\
\rho^2&=&\frac{P^2}{\det(q)}\lt(1+q^{\a\b}U_\a U_\b\rt)^{-1},\label{rhoP}\\
W_j&=&P_j/P,\label{WP}
\ee
where $\a,\b=1,2,3$ are spatial indices, $P,P_j$ are momenta conjugate to $T,S^j$, and $\cc,\cc_\a$ are Hamiltonian and diffeomorphism constraints of gravity (or gravity coupled to some other matter fields). Eq.\Ref{rhoP} is solved by
\be
\rho=\eps\frac{P}{\sqrt{\det(q)}}\lt(1+q^{\a\b}U_\a U_\b\rt)^{-1/2}, \quad \eps=\pm1.
\ee
The dust 4-velocity $U$ being timelike and future pointing fixes $\eps=1$ \cite{Giesel:2007wi}, so $\sgn(P)=\sgn(\rho)$. Inserting this solution to Eq.\Ref{C} and using Eq.\Ref{WP} lead to
\be
\cc=-P\sqrt{1+q^{\a\b}\cc_\a \cc_\b/P^2}.
\ee
Thus $-\sgn(\cc)=\sgn(P)=\sgn(\rho)$. For dust coupling to pure gravity, we must have $\cc<0$ and the physical dust $\rho,P>0$ to fulfill the energy condition \cite{Brown:1994py}. However, in presence of additional matter fields (e.g. scalars, fermions, gauge fields, etc), they can make $\cc>0$ and $\rho,P<0$ corresponding to the phantom dust \cite{Giesel:2007wn,Giesel:2007wi}. The case of phantom dust may not violate the usual energy condition due to presence of other matter fields. We solve $P,P_j$ from Eqs.\Ref{C} and \Ref{Ca}:
\be
&&P=\begin{cases} h &\  \text{physical dust}, \\
-h &\ \text{phantom dust},
\end{cases} \quad h=\sqrt{\cc^2-q^{\a\b}\cc_\a \cc_\b},\label{P=-h}\\
&&P_j=-S^\a_j\lt(\cc_\a-hT_{,\a}\rt),\label{Pj=cc}
\ee
are strongly Poisson commutative constraints. $S^\a_j$ is the inverse matrix of $\partial_\a S^j$ ($\a=1,2,3$). An intermediate step of the above derivation shows that $P^2 = \cc^2- q^{\a\b}\cc_\a \cc_\b>0$. It constrains the argument of the square root to be positive. Moreover the physical dust requires $\cc<0$ while the phantom dust requires $\cc>0$.

We use $A^a_\a(x),E^\a_a(x)$ as canonical variables of gravity. $A^a_\a(x)$ is the Ashtekar-Barbero connection and $E^\a_a(x)=\sqrt{\det q}\, e^\a_a(x)$ is the densitized triad. $a=1,2,3$ is the Lie algebra index of su(2). Gauge invariant Dirac observables are constructed relationally by parametrizing $(A,E)$ with values of dust fields $T(x)\equiv\t,S^j(x)\equiv\sig^j$, i.e. $A_j^a(\sig,\t)=A_j^a(x)|_{T(x)\equiv\t,\,S^j(x)\equiv\sig^j}$ and $E^j_a(\sig,\t)=E^j_a(x)|_{T(x)\equiv\t,\,S^j(x)\equiv\sig^j}$, where $\sig,\t$ are dust space and time coordinates, and $j=1,2,3$ is the dust coordinate index (e.g. $A_j=A_\a S^\a_j$).

$A_j^a(\sig,\t)$ and $E^j_a(\sig,\t)$ satisfy the standard Poisson bracket in the dust frame: 
\be
\{E^i_a(\sig,\t),A_j^b(\sig',\t)\}=\frac{1}{2}\kappa \b\ \delta^{i}_j\delta^b_a\delta^{3}(\sig,\sig'),\quad \kappa=16\pi G
\ee 
where $\b$ is the Barbero-Immirzi parameter. The phase space $\cp$ of $A_j^a(\sig,\t),E^j_a(\sig,\t)$ is free of Hamiltonian and diffeomorphism constraints. All SU(2) gauge invariant phase space functions are Dirac observables.

Physical time evolution in $\t$ is generated by the physical Hamiltonian ${\bf H}_0$ given by integrating $h$ on the constant $T=\t$ slice $\cs$. The constant $\t$ slice $\cs$ is coordinated by the value of dust scalars $S^j=\sig^j$ thus is called the dust space \cite{Giesel:2007wn,Giesel:2012rb}. By Eq.\Ref{P=-h}, ${\bf H}_0$ is negative (positive) for the physical (phantom) dust. We flip the direction of the time flow $\t\to -\t$ thus ${\bf H}_0 \to -{\bf H}_0$ for the physical dust. So we obtain positive Hamiltonians in both cases:
\be
{\bf H}_0=\int_\cs\rmd^3\sig\, \sqrt{\cc(\sig,\t)^2-\frac{1}{4}\sum_{a=1}^3\cc_a(\sig,\t)^2}.\label{bigH0}
\ee 
$\cc$ and $\cc_a=2e_a^\a \cc_\a$ are parametrized in the dust frame, and expressed in terms of $A_j^a(\sig,\t)$ and $E^j_a(\sig,\t)$:
\be
\cc&=&\frac{1}{\kappa}\lt[F^a_{jk}-\lt({\b^2+1}\rt)\eps_{ade}K^d_{j}K^e_{k}\rt]\eps_{abc}\frac{E^j_bE^k_c}{\sqrt{\det(q)}}+\frac{2\L}{\kappa}\sqrt{\det(q)}\\
\cc_a&=&\frac{4}{\kappa\b}F^b_{jk} \frac{E^j_aE^k_b}{\sqrt{\det(q)}},
\ee  
where $\L$ is the cosmological constant.

Coupling gravity to Gaussian dust model is similar, so we don't present the details here (while details can be found in \cite{Giesel:2012rb}). As a result the physical Hamiltonian has a simpler expression 
\be
\mathbf{H}_0=\int_\cs\rmd^3\sig\, \cc(\sig,\t). \label{gaussd}
\ee
The following Hamiltonian unifies both scenarios of the BK and Gaussian dusts:
\be
\mathbf{H}_0&=&\int_\cs\rmd^3\sig\, h(\sig,\t),\label{ham1}\\
 h(\sig,\t)&=&\sqrt{\cc(\sig,\t)^2-\frac{\a}{4}\sum_{a=1}^3\cc_a(\sig,\t)^2},\quad \begin{cases} 
\a =1& \text{BK dust},\\
\a =0& \text{Gaussian dust}.
\end{cases} \nonumber
\ee 
This physical Hamiltonian ${\bf H}_0$ is manifestly positive. However when $\cc<0$, Eq.\Ref{ham1} is different from Eq.\Ref{gaussd} by an overall minus sign, thus reverses the time flow $\t\to-\t$ for the Gaussian dust.

The physical Hamiltonian $\mathbf{H}_0$ generates the $\t$ evolution:
\be
\frac{\rmd f}{\rmd\t}=\lt\{ f, \mathbf{H}_0\rt\},
\ee
for all phase space function $f$. In particular, the Hamilton's equations are
\be
\frac{\rmd A^a_j(\sig,\t)}{\rmd\t}=-\frac{\kappa\b}{2}\frac{\delta\mathbf{H}_0}{\delta E^j_a(\sig,\t)},\quad \frac{\rmd E^j_a(\sig,\t)}{\rmd\t}=\frac{\kappa\b}{2}\frac{\delta\mathbf{H}_0}{\delta A^a_j(\sig,\t)}.\label{hamitoncon0}
\ee
Functional derivatives on the right-hand sides of Eq.\Ref{hamitoncon0} give
\be
\delta{\bf H}_0=\int_\cs\rmd^3\sig \lt(\frac{\cc}{h}\delta \cc -{\a}q^{ij}\frac{\cc_i}{h}\delta\cc_j+\frac{\a}{2h}q^{ij}q^{kl}\cc_j\cc_l\delta q_{ik}\rt),
\ee
where ${C}/{h}$ is negative (positive) for physical (phantom) dust. In this work we focus on the cosmological perturbation theory $q_{ij}=q^0_{ij}+h_{ij}$ ($q^0_{ij}$ is the homogeneous and isotropic cosmological background and $h_{ij}$ is the perturbation) and linearized EOMs. The last term gives $\frac{\a}{2h}q^{ij}q^{kl}\cc_j\cc_l= O(h_{ij}^2)$ since $\cc_j(q^0)=0$, thus does not affect linearized EOMs. Compare $\delta{\bf H}$ to the variation of Hamiltonian $H_{GR}$ of pure gravity in absence of dust motivates us to identify (dynamical) lapse function and shift vector
\be
N=\frac{\cc}{h}, \quad N_j=-{\a}\frac{\cc_j}{h}.\label{lapseshift0}
\ee
$N$ is negative (positive) for the physical (phantom) dust. Negative lapse indicates that $\t$ in Eq.\Ref{hamitoncon0} flows from future to past. Its origin is the flip $\t\to-\t$ before Eq.\Ref{bigH0}. In this paper we focus on gravity coupled to the physical dust. When we discuss the cosmological perturbation theory from the semiclassical limit of LQG, we are going to flip $\t\to-\t$ back such that $\t$ flows to the future again. In that case, the dynamical lapse function and shift vector Eq.\Ref{lapseshift0} have to change to
\be
N=-\frac{\cc}{h}, \quad N_j={\a}\frac{\cc_j}{h}.\label{lapseshift}
\ee
They can be obtained directly from the variation $\delta(-{\bf H}_0)$ ($-{\bf H}_0$ is the physical Hamiltonian of physical dust if we don't flip $\t\to-\t$ before Eq.\Ref{bigH0}.

In the gravity-dust models, we have resolved the Hamiltonian and diffeomorphism constraints classically, while the SU(2) Gauss constraint $\cg_a(\sig,\t)=D_j E^j_a(\sig,\t)=0$ still has to be imposed to the phase space. In addition, There are non-holonomic constraints: $\cc(\sig,\t)^2-\frac{\a}{4}\sum_{a=1}^3\cc_a(\sig,\t)^2\geq 0$ and $\cc<0$ for physical dust ($\cc>0$ for phantom dust).

These constraints are preserved by $\t$-evolution for gravity coupling to the BK dust. Indeed, firstly $\t$-evolution cannot break Gauss constraint since $\lt\{\cg_a(\sig,\t),\,{\bf H}_0\rt\}=0$. Secondly both $h(\sig,\t)$ and $\cc_j(\sig,\t)$ are conserved densities (thus $N_j$ is conserved) \cite{Giesel:2007wn}:
\be
\frac{\rmd h(\sig,\t)}{\rmd \t}=\lt\{h(\sig,\t),\,{\bf H}_0\rt\}=0,\quad \frac{\rmd \cc_j(\sig,\t)}{\rmd \t}=\lt\{\cc_j(\sig,\t),\,{\bf H}_0\rt\}=0\label{conserv0}
\ee
Therefore $\cc(\sig,\t)^2-\frac{1}{4}\sum_{a=1}^3\cc_a(\sig,\t)^2\geq 0$ is conserved. About $\cc<0$ ($\cc>0$), suppose $\cc<0$ ($\cc>0$) was violated in $\t$-evolution, there would exist a certain $\t_0$ that $\cc(\sig,\t_0)=0$, but then $\cc(\sig,\t)^2-\frac{1}{4}\sum_{a=1}^3\cc_a(\sig,\t)^2$ would becomes negative if $\cc_j(\sig,\t)\neq 0$, contradicting the conservation of $h(\sig,\t)$ and the other nonholonomic constraint. If the conserved $\cc_j(\sig,\t)=0$, $h(\sig,\t)^2=\cc(\sig,\t)^2$ is conserved and thus cannot evolve from nonzero to zero. For gravity coupled to the Gaussian dust, $\cc_j(\sig,\t)$ is conserved. $h(\sig,\t)$ and $\cc(\sig,\t)$ are conserved only when $\cc_j(\sig,\t)=0$. $\cc<0$ ($\cc>0$) may be violated in $\t$-evolution for coupling to the Gaussian dust if $\cc_j(\sig,\t)\neq 0$. 

In our following discussion, we focus on pure gravity coupled to dusts, thus we only work with physical dusts in order not to violate the energy condition.


\subsection{Quantization}

We construct a fixed finite cubic lattice $\g$ which partitions the dust space $\cs$. In this work, $\cs$ is compact and has no boundary. $E(\g)$ and $V(\g)$ denote sets of (oriented) edges and vertices in $\g$. By the dust coordinate on $\cs$, every edge has a constant coordinate length $\mu$. $\mu\to 0$ relates to the lattice continuum limit. Every vertex $v\in V(\g)$ is 6-valent, having 3 outgoing edges $e_I(v)$ ($I=1,2,3$) and 3 incoming edges $e_I(v-\mu \hat{I})$ where $\hat{I}$ is the coordinate basis vector along the $I$-th direction. It is sometimes convenient to orient all 6 edges to be outgoing from $v$, and denote them by $e_{v;I,s}$ ($s=\pm$):
\be
e_{v;I,+}=e_I(v),\quad e_{v;I,-}=e_I(v-\mu \hat{I})^{-1}.
\ee

Canonical variables $A^a_j(\sig,\t),E^j_a(\sig,\t)$ are regularized by holonomy $h(e)$ and gauge covariant flux $p^a(e)$ at every $e\in E(\g)$:
\be
h(e)&:=&\cp \exp \int_{e}A^a\t^a/2,\nonumber\\
p^a(e)&:=&-\frac{1}{2\b a^2}\tr\lt[\t^a\int_{S_e}\eps_{ijk}\rmd \sig^i\wedge\rmd \sig^j\ h\lt(\rho_e(\sig)\rt)\, E_b^k(\sig)\t^b\, h\lt(\rho_e(\sig)\rt)^{-1}\rt],\label{hpvari}
\ee 
where $\t^a=-i(\text{Pauli matrix})^a$. $S_e$ is a 2-face intersecting $e$ in the dual lattice $\g^*$. $\rho_e$ is a path starting at the source of $e$ and traveling along $e$ until $e\cap S_e$, then running in $S_e$ until $\vec{\sig}$. $a$ is a length unit for making $p^a(e)$ dimensionless. Because $p^a(e)$ is gauge covariant flux, we have
\be
p^{a}\left(e_{v ; I,-}\right)=\frac{1}{2} \operatorname{Tr}\left[\tau^{a} h\left(e_{v-\hat{I} ; I,+}\right)^{-1} p^{b}\left(e_{v-\hat{I} ; I,+}\right) \tau^{b} h\left(e_{v-\hat{I} ; I,+}\right)\right].
\ee
The Poisson algebra of $h(e)$ and $p^a(e)$ are called the holonomy-flux algebra:
\be
\left\{h(e), h\left(e^{\prime}\right)\right\} &=&0 ,\label{handh}\\
\left\{p^{a}(e), h\left(e^{\prime}\right)\right\} &=&\frac{\kappa}{a^{2}} \delta_{e, e^{\prime}} \frac{\tau^{a}}{2} h\left(e^{\prime}\right) ,\label{pandtheta}\\
\left\{p^{a}(e), p^{b}\left(e^{\prime}\right)\right\} &=&-\frac{\kappa}{a^{2}} \delta_{e, e^{\prime}} \varepsilon_{a b c} p^{c}\left(e^{\prime}\right),\label{pandp}
\ee
$h(e)$ and $p^a(e)$ are coordinates of the reduced phase space $\cp_\g$ for the theory discretized on $\g$.

In quantum theory, the Hilbert space $\ch_\g$ is spanned by gauge invariant (complex valued) functions of all $h(e)$'s, and is a proper subspace of $\ch_\g^0=\otimes_e L^2(\Su)$. $\hat{h}(e)$ becomes multiplication operators on functions in $\ch^0_\g$. $\hat{p}^a(e)=i t\,{R}_e^a/2$ where ${R}_e^a$ is the right invariant vector field on SU(2): $R^a f(h)=\frac{\rmd}{\rmd \eps}\big|_{\eps=0} f(e^{\eps\t^a}h)$. $t=\ell^2_p/a^2$ is a dimensionless semiclassicality parameter ($\ell^2_p=\hbar\kappa$). $\hat{h}(e),\hat{p}^a(e)$ satisfy the commutation relations:
\be
\lt[\hat{h}(e),\hat{h}(e')\rt] &=&0\nonumber\\
\lt[\hat{p}^a(e),\hat{h}(e')\rt] &=&i t \delta_{e,e'} \frac{\t^a}{2} {h}(e')\nonumber\\
\lt[\hat{p}^a(e),\hat{p}^b(e')\rt]&=&-it \delta_{e,e'} \eps_{abc} {p}^c(e'), \label{ph}
\ee
as quantization of the holonomy-flux algebra.

The physical Hamiltonian operators $\hat{\bf H}$ are given by \cite{Giesel:2007wn}:
\be
\hat{\mathbf{H}}&=&\sum_{v\in V(\g)}\hat{H}_v,\quad \hat{H}_v:=\lt[\hat{M}_-^\dagger(v) \hat{M}_-(v)\rt]^{1/4},\label{physHam}\\ 
\hat{M}_-(v)&=&\hat{C}_{v}^{\ \dagger}\hat{C}_{v}-\frac{\a}{4}\sum_{a=1}^3\hat{C}_{a,v}^{\ \dagger}\hat{C}_{a,v},\quad \a=\begin{cases}
1,&\text{BK dust,}\\
0,&\text{Gaussian dust.}
\end{cases}
\ee 
In our notation, ${\bf H}_0=\int_\cs\rmd^3\sig\, h$, $\cc$, and $\cc_{a}$ are the physical Hamiltonian, scalar constraint, and vector constraint on the continuum. ${\bf H}=\sum_v H_v$, $C_v$, and $C_{a,v}$ are their discretizations on $\g$. $\hat{\bf H}=\sum_v \hat{H}_v$, $\hat{C}_v$, and $\hat{C}_{a,v}$ are quantizations of ${\bf H}$, $C_v$, and $C_{a,v}$:
\be
\hat{C}_{0,v}&=&-\frac{2}{i\b\kappa\ell_p^2}\sum_{s_1,s_2,s_3=\pm1}s_1s_2s_3\ \eps^{I_1I_2I_3}\ \mathrm{Tr}\Bigg(\hat{h}(\a_{v;I_1s_1,I_2s_2}) \hat{h}(e_{v;I_3s_3})\Big[\hat{h}(e_{v;I_3s_3})^{-1},\hat{V}_v\Big] \Bigg)\label{Cd}\\
\hat{C}_{a,v}&=&-\frac{4}{i\b^2\kappa\ell_p^2}\sum_{s_1,s_2,s_3=\pm1}s_1s_2s_3\ \eps^{I_1I_2I_3}\ \mathrm{Tr}\Bigg(\t^a \hat{h}(\a_{v;I_1s_1,I_2s_2}) \hat{h}(e_{v;I_3s_3})\Big[\hat{h}(e_{v;I_3s_3})^{-1},\hat{V}_v\Big] \Bigg)\\
\hat{C}_v&=&\hat{C}_{0,v}+{(1+\b^2)}\hat{C}_{L,v}+\frac{2\L}{\kappa}\hat{ V}_v,\quad\quad \hat{K}=\frac{i}{\hbar\b^2}\lt[\sum_{v\in V(\g)}\hat{C}_{0,v},\sum_{v\in V(\g)}V_v\rt]\nonumber\\
\hat{C}_{L,v}&=&\frac{16}{\kappa\lt(i\b\ell_p^2\rt)^3}\sum_{s_1,s_2,s_3=\pm1}s_1s_2s_3\ \eps^{I_1I_2I_3}\label{HCO}\\
&&\mathrm{Tr}\Bigg( \hat{h}(e_{v;I_1s_1})\Big[\hat{h}(e_{v;I_1s_1})^{-1},\hat{K}\Big]\ \hat{h}(e_{v;I_2s_2})\Big[\hat{h}(e_{v;I_2s_2})^{-1},\hat{K}\Big]\ \hat{h}(e_{v;I_3s_3})\Big[\hat{h}(e_{v;I_3s_3})^{-1},\hat{V}_v\Big]\ \Bigg).\nonumber
\ee
where $\hat{V}_v$ is the volume operator at $v$:
\be
\hat{V}_v&=&\lt(\hat{Q}_v^2\rt)^{1/4},\\ 
\hat{Q}_v
&=&\b^3a^6\eps_{abc}\frac{\hat{p}^a({e_{v;1+}})-\hat{p}^a({e_{v;1-}})}{4}\frac{\hat{p}^b({e_{v;2+}})-\hat{p}^b({e_{v;2-}})}{4}\frac{\hat{p}^c({e_{v;3+}})-\hat{p}^c({e_{v;3-}})}{4}.\label{Qv}
\ee 

The Hamiltonian operator $\hat{\mathbf{H}}$ is positive semi-definite and self-adjoint because $\hat{M}_-^\dagger(v) \hat{M}_-(v)$ is manifestly positive semi-definite and Hermitian, therefore admits a canonical self-adjoint extension. 

Classical discrete $C_v$, and $C_{a,v}$ can be obtained from Eqs.\Ref{C} - \Ref{HCO} by mapping operators to their classical counterparts and $[\hat{f}_1,\hat{f}_2]\to i\hbar\{f_1,f_2\} $. Hence classical discrete physical Hamiltonian ${\bf H}$ is
\be
{\bf H}=\sum_{v\in V(\g)} H_v,\quad H_v=\sqrt{\lt|C_v^2-\frac{\a}{4}\sum_{a=1}^3 C_{a,v}^2\rt|}.\label{physHamcl}
\ee
The absolute value in the square-root results from that ${\bf H}$ is the classical limit of $\hat{\bf H}$, while $\hat{\bf H}$ is defined on the entire $\ch_\g$ disregarding nonholonomic constraints in particular $\cc^2-\frac{\a}{4}\sum_{a=1}^3\cc_a^2\geq 0$ for $\a=1$.

The transition amplitude $A_{[g],[g']}$ plays the central role in the quantum dynamics of reduced phase space LQG:
\be
A_{[g],[g']}=\langle \Psi^t_{[g]}|\,\exp\lt[-\frac{i}{\hbar}T \hat{\bf H}\rt]\,|\Psi^t_{[g']}\rangle.
\ee
We focus on the semiclassical initial and final states $\Psi^t_{[g']}, \Psi^t_{[g]}$ for the purpose of semiclassical analysis. $\Psi^t_{[g']}, \Psi^t_{[g]}$ are gauge invariant coherent states \cite{Thiemann:2000bw,Thiemann:2000ca}:
\be
\Psi^t_{[g]}(h)
&=&\int_{\mathrm{SU(2)}^{|V(\g)|}}\rmd h\prod_{e\in E(\g)}{\psi}^t_{h_{s(e)}^{-1}g(e)h_{t(e)}}\lt(h(e)\rt),\quad \rmd h=\prod_{v\in V(\g)}\rmd\mu_H(h_v).\label{gaugeinv}
\ee
The gauge invariant coherent state is labelled by the gauge equivalence class $[g]$ generated by $g(e)\sim g^h(e)= h_{s(e)}^{-1}g(e)h_{t(e)}$ at all $e$. $g(e)\in\Slc$. $\psi^{t}_{g(e)}\lt(h(e)\rt)$ is the complexifier coherent state on the edge $e$:
\be
\psi^{t}_{g(e)}\lt(h(e)\rt)
&=&\sum_{j_e\in\Z_+/2\cup\{0\}}(2j_e+1)\ e^{-tj_e(j_e+1)/2}\chi_{j_e}\lt(g(e)h(e)^{-1}\rt),\label{coherent}
\ee
where $g(e)$ is complex coordinate of $\cp_\g$ and relates to $h(e),p^a(e)$ by
\be
g(e)=e^{-ip_a(e)\t_a/2}h(e)=
e^{-ip^a(e)\t^a/2}e^{\theta^a(e)\t^a/2}, \quad p^a(e),\ \theta^a(e)\in\R^3.\label{gthetap}
\ee

Applying Eq.\Ref{gaugeinv} and discretizing time $T=N\Delta\t$ with large $N$ and infinitesimal $\Delta\t$, 
\be
A_{[g],[g']}&=&\int\rmd h\lag\psi^t_{g}\rt|\lt[e^{ -\frac{i}{\hbar}\Delta\t \hat{\mathbf{H}}}\rt]^N |{\psi}^t_{{g'}^{h}}\rangle,\\
&=&\int\rmd h\prod_{i=1}^{N+1}\mathrm{d}g_{i}\,\langle\psi^t_{g}|\tilde{\psi}^t_{g_{N+1}}\rangle\langle \tilde{\psi}^t_{g_{N+1}}\big|e^{ -\frac{i\Delta\t}{\hbar} \hat{\mathbf{H}}}\big|\tilde{\psi}^t_{g_{N}}\rangle
\langle \tilde{\psi}^t_{g_{N}}\big|e^{ -\frac{i\Delta\t}{\hbar}\hat{\mathbf{H}}}\big|\tilde{\psi}^t_{g_{N-1}}\rangle\cdots\nonumber\\
&&\quad \cdots\ 
\langle \tilde{\psi}^t_{g_2}\big|e^{ -\frac{i\Delta\t}{\hbar}\hat{\mathbf{H}}}\big|\tilde{\psi}^t_{g_1}\rangle\langle\tilde{\psi}^t_{g_1}|{\psi}^t_{g'{}^{h}}\rangle 
\end{eqnarray}
where we have inserted $N+1$ resolutions of identities with normalized coherent state $\tilde{\psi}^t_{g}=\otimes_e{\psi}^t_{g(e)}/||{\psi}^t_{g(e)}||$:
\be
\int\rmd g_i\ |\tilde{\psi}^{t}_{g_i}\rangle\langle\tilde{\psi}^{t}_{g_i}|=1_{\ch_\g^0},\quad \rmd g_i=\lt(\frac{c}{t^3}\rt)^{|E(\g)|}\prod_{e\in E(\g)}\rmd\mu_H(h_i(e))\,\rmd^3p_i(e),\quad i=1,\cdots,N-1.
\ee

The above expression of $A_{[g],[g']}$ leads to a path integral formula (see \cite{Han:2019vpw} for derivation):
\be
A_{[g],[g']}=\left\|\psi_{g}^{t}\right\|\left\|\psi_{g^{\prime}}^{t}\right\| \int \mathrm{d} h \prod_{i=1}^{N+1} \mathrm{d} g_{i}\, \nu[g]\, e^{S[g, h] / t}\label{Agg}
\ee 
where we find the ``effective action'' $S[g,h]$ given by
\be
S[g, h]&=&\sum_{i=0}^{N+1} K\left(g_{i+1}, g_{i}\right)-\frac{i \kappa}{a^{2}} \sum_{i=1}^{N} \Delta \tau\left[\frac{\langle\psi_{g_{i+1}}^{t}|\hat{\mathbf{H}}| \psi_{g_{i}}^{t}\rangle}{\langle\psi_{g_{i+1}}^{i} | \psi_{g_{i}}^{t}\rangle}+i \tilde{\varepsilon}_{i+1, i}\left(\frac{\Delta \tau}{\hbar}\right)\right],\label{Sgh}\\
K\left(g_{i+1}, g_{i}\right)&=&\sum_{e \in E(\gamma)}\left[z_{i+1, i}(e)^{2}-\frac{1}{2} p_{i+1}(e)^{2}-\frac{1}{2} p_{i}(e)^{2}\right]
\ee
with $g_{0} \equiv g^{\prime h},\ g_{N+2} \equiv g$. $\tilde{\varepsilon}_{i+1, i}\left(\frac{\Delta \tau}{\hbar}\right)\to 0$ as $\Delta\t\to0$ and is negligible. In the above, $z_{i+1,i}(e)$ and $x_{i+1,i}(e)$ are given by 
\be
z_{i+1,i}(e)&=& \mathrm{arccosh}\lt(x_{i+1,i}(e)\rt),\quad x_{i+1,i}(e)=\half\tr\lt[g_{i+1}(e)^\dagger g_{i}(e)\rt].
\ee

\section{Semiclassical Equations of Motion}\label{Semiclassical Equations of Motion}

In the semiclassical limit $t\to0$ (or $\ell_P\ll a$), the dominant contribution to $A_{[g],[g']}$ comes from the semiclassical trajectories satisfying the semiclassical equations of motion (EOMs). Semiclassical EOMs has been derived in \cite{Han:2019vpw} by the variational principle $\delta S[g,h]=0$ (stationary phase approximation):

\begin{itemize}

\item For $i=1,\cdots,N$, at every edge $e\in E(\g)$,
\be
\frac{1}{\Delta\t}\lt[\frac{z_{i+1,i}(e)\,\tr\lt[\t^a g_{i+1}(e)^\dagger g_i(e)\rt]}{\sqrt{x_{i+1,i}(e)-1}\sqrt{x_{i+1,i}(e)+1}}-\frac{p_i(e)\,\tr\lt[\t^a g_{i}(e)^\dagger g_i(e)\rt]}{\sinh(p_i(e))}\rt]\nonumber\\
=\frac{i\kappa }{a^2}\frac{\partial}{\partial \varepsilon_{i}^{a}(e)} \frac{\langle\psi_{g_{i+1}^{\varepsilon}}^{t}|\hat{\mathbf{H}}| \psi_{g_{i}^{\varepsilon}}^{t}\rangle}{\langle\psi_{g_{i+1}^{\varepsilon}}^{t} | \psi_{g_{i}^{\varepsilon}}^{t}\rangle}\Bigg|_{\vec{\eps}=0}\label{eoms1}
\ee
where $g^\eps(e)=g(e) e^{\eps^a(e)\t^a}$ ($\eps^a(e)\in\C$) is a holomorphic deformation.

\item For $i=2,\cdots,N+1$, at every edge $e\in E(\g)$,
\be
\frac{1}{\Delta\t}\lt[\frac{z_{i,i-1}(e)\,\tr\lt[\t^a g_{i}(e)^\dagger g_{i-1}(e)\rt]}{\sqrt{x_{i,i-1}(e)-1}\sqrt{x_{i,i-1}(e)+1}}-\frac{p_i(e)\,\tr\lt[\t^a g_{i}(e)^\dagger g_i(e)\rt]}{\sinh(p_i(e))}\rt]\nonumber\\
=-\frac{i\kappa }{a^2}\frac{\partial}{\partial \bar{\varepsilon}_{i}^{a}(e)} \frac{\langle\psi_{g_{i}^{\varepsilon}}^{t}|\hat{\mathbf{H}}| \psi_{g_{i-1}^{\varepsilon}}^{t}\rangle}{\langle\psi_{g_{i}^{\varepsilon}}^{t} | \psi_{g_{i-1}^{\varepsilon}}^{t}\rangle}\Bigg|_{\vec{\eps}=0}.\label{eoms2}
\ee

\item The closure condition at every vertex $v\in V(\g)$ for initial data:
\be
G_v^a\equiv-\sum_{e, s(e)=v}p_1^a(e)+\sum_{e, t(e)=v}\L^a_{\ b}\lt(\vec{\theta}_1(e)\rt)\,p_1^b(e)=0.\label{closure0}
\ee
where $\L^a_{\ b}(\vec{\theta})\in\mathrm{SO}(3)$ is given by $e^{\theta^c\t^c/2}\t^a e^{-\theta^c\t^c/2}=\L^a_{\ b}(\vec{\theta})\t^b$.

\end{itemize}
\noindent
The initial and final conditions are given by $g_{1}=g'^h$ and $g_{N+1}=g$. Eqs.\Ref{eoms1} and \Ref{eoms2} come from $\delta S/\delta g=0$ and $\delta S/\delta\bar{g}=0$, while Eq.\Ref{closure0} comes from $\delta S/\delta h=0$. These semiclassical EOMs govern the semiclassical dynamics of LQG in the reduced phase space formulation.

We can take $\Delta\t\to0$ in these semiclassial EOMs since $\Delta\t$ is arbitrarily small. Solutions of EOMs with $\Delta\t\to0$ are time-continuous approximation of solutions of Eqs.\Ref{eoms1} - \Ref{closure0}.

It is proven in \cite{Han:2019vpw} that Eqs.\Ref{eoms1} - \Ref{eoms2} implies $g_{i}\to g_{i+1}$ as $\Delta\t\to0$, i.e. $g_i=g(\t)$ is a continuous function of $\t$. Therefore, matrix elements $\langle\psi_{g_{i}^{\varepsilon}}^{t}|\hat{\mathbf{H}}| \psi_{g_{i-1}^{\varepsilon}}^{t}\rangle$ on right-hand sides of Eqs.\Ref{eoms1} - \Ref{eoms2} reduces to the expectation values $\langle\psi_{g^{\varepsilon}}^{t}|\hat{\mathbf{H}}| \psi_{g^{\varepsilon}}^{t}\rangle$ as $\Delta\t\to0$. Coherent state expectation values of $\hat{\bf H}$ have correct semiclassical limit\footnote{Firstly we apply the semiclassical perturbation theory of \cite{Giesel:2006um} to $\hat{O}\equiv\hat{{H}}_v^4$ (recall Eq.\Ref{physHam}) and all $\hat{O}^n$ ($n>1$): $\langle\tilde{\psi}_{g}^{t}|\hat{O}^n| \tilde{\psi}_{g}^{t}\rangle={O}[g]^n+O(t)$. By Theorem 3.6 of \cite{Thiemann:2000bx}, $\lim_{t\to0}\langle\tilde{\psi}_{g}^{t}|f(\hat{O})| \tilde{\psi}_{g}^{t}\rangle=f({O}[g])$ for any any Borel measurable function on $\R$ such that $\langle\tilde{\psi}_{g}^{t}|f(\hat{O})^\dagger f(\hat{O})| \tilde{\psi}_{g}^{t}\rangle<\infty$.}
\be
\lim_{t\to0}\langle\tilde{\psi}_{g}^{t}|\hat{\mathbf{H}}| \tilde{\psi}_{g}^{t}\rangle={\bf H}[g]
\ee
where ${\bf H}[g]$ is the classical discrete Hamiltonian \Ref{physHamcl} evaluated at $p^a(e),h(e)$ determined by $g(e)$ in Eq.\Ref{gthetap}. Note that the above semiclassical behavior of $\langle\tilde{\psi}_{g}^{t}|\hat{\mathbf{H}}| \tilde{\psi}_{g}^{t}\rangle$ relies on the following semiclassical expansion of volume operator $\hat{V}_v$ \cite{Giesel:2006um}:
\be
\hat{V}_v=\langle \hat{Q}_v\rangle^{2q}\lt[1+\sum_{n=1}^{2k+1}(-1)^{n+1}\frac{q(1-q)\cdots(n-1+q)}{n!}\lt(\frac{\hat{Q}_v^2}{\langle\hat{Q}_v\rangle^2}-1\rt)^n\rt],\quad q=1/4\label{expandvolume}
\ee
where $\langle \hat{Q}_v\rangle=\langle\psi^t_g|\hat{Q}_v|\psi^t_g\rangle$, and this expansion is valid when $\langle \hat{Q}_v\rangle\gg \ell_P^6$.

The time continuous limit of semiclassical EOMs is computed in \cite{Han2020} and expressed in terms of ${\bm p}(e)=(p^1(e),p^2(e),p^3(e))^T$ and ${\bm \theta}(e)=(\theta^1(e),\theta^2(e),\theta^3(e))^T$ and their time derivatives:
\be
\left( \begin{array}{l}  {\rmd {\bm p}}(e)/{\rmd \tau} \\ {\rmd \bm{\theta}}(e)/{\rmd \tau}  \end{array} \right)  = \frac{i\kappa}{a^2}\, {T}\lt({\bm p},{\bm \theta}\rt) ^{-1}\left( \begin{array}{l}  {\partial {\bf H} }/{\partial {\bm p} (e)} \\ {\partial {\bf H} }/{\partial \bm{\theta} (e)} \end{array} \right) .\label{eom0}
\ee
The matrix elements $T$ is lengthy, and are given explicitly in \cite{github0}. It is shown in \cite{Han2020} that Eq.\Ref{eom0} is equivalent to that for any phase space function $f$ on $\cp_\g$, its $\t$-evolution is given by the Hamiltonian flow generated by ${\bf H} $:
\be
\frac{\rmd f}{\rmd \tau}=\lt\{f, \ {\bf H}\rt\}.\label{hamilton}
\ee
The closure condition is preserved by $\t$-evolution by $\{G^a_v,\,{\bf H}\}=0$.


The lattice continuum limit of Eq.\Ref{eom0} is studied in \cite{Han2020}. We define $\mu$ to be the coordinate length of every lattice edge, the lattice continuum limit is formally given by $\mu\to0$ and $|V(\g)|\to\infty$ while keeping $\mu^3|V(\g)|$ fixed. More precisely, recall that Eq.\Ref{eom0} are derived with $t=\ell_P^2/a^2\to0$ and the assumption $\langle \hat{Q}_v\rangle\sim \mu^6\gg \ell_P^6$ (see Eq.\Ref{expandvolume}), the lattice continuum limit are taken in the regime 
\be
\ell_P\ll\mu\ll a,
\ee
where $a$ is a macroscopic unit, e.g. 1 mm. When keeping $a$ fixed, the lattice continuum limit sends $\mu\to0$ after the semiclassical limit $\ell_P\to0$ so $\ell_P\ll\mu$ is kept. In the lattice continuum limit, EOMs.\Ref{eom0} reduce to the EOMs \Ref{hamitoncon0} of the continuum theory, when suitable initial conditions are imposed (see \cite{Han2020} for details).   

\section{Cosmological Background and Perturbations} \label{Cosmological Background and Perturbations}

\subsection{Cosmological Background}

As in \cite{Han:2019vpw}, we apply the following (homogeneous and isotropic) cosmological ansatz to the semiclassical EOMs
\be
\theta^a(e_I(v))=\mu \b K_0\delta^a_I,\quad p^a(e_I(v))=\frac{2\mu^2}{\b a^2}P_0\delta^a_I
\ee
Here $K_0=K_0(\t)$ and $P_0=P_0(\t)$ are constant on $\g$ but evolve with the dust time $\t$. Inserting the ansatz, left hand sides of EOMs \Ref{eom0} contain (1) $\rmd p^a(e_I(v))/\rmd \t$ and $\rmd \theta^a(e_I(v))/\rmd \t$ with $a= I$, which are proportional to $\dot{P}_0=\rmd P_0/\rmd\t$ and $\dot{K}_0=\rmd K_0/\rmd\t$, and (2) $\rmd p^a(e_I(v))/\rmd \t$ and $\rmd \theta^a(e_I(v))/\rmd \t$ with $a\neq I$, which are zero.

\begin{itemize}

\item EOMs of case (1) reduce to 
\be
\frac{4 \beta ^2 \left[-2 \mu ^2 \sqrt{{P}_0} \dot{K_0}+\sin ^4(\beta  \mu  {K_0})+\Lambda  \mu ^2 {P_0}\right]-\sin ^2(2 \beta  \mu  {K_0})}{\sqrt{{P_0}}}&=&0,\label{cosm1}\\
\sqrt{{P_0}} \left[2 \beta ^2 \sin (2 \beta  \mu  {K_0})-\left(\beta ^2+1\right) \sin (4 \beta  \mu  {K_0})\right]+2 \beta  \mu  \dot{P_0}&=&0.\label{cosm2}
\ee
where an effective Hamiltonian of cosmology can be extracted
\be
H_{eff}(P_0,K_0)=\frac{\left(\beta ^2+1\right) \sqrt{{P_0}} \sin ^2(2 \beta  \mu  {K_0})}{4 \beta ^2 \mu ^2}-\frac{\sqrt{{P_0}} \sin ^2(\beta  \mu  {K_0})}{\mu ^2}-\frac{1}{3} \Lambda P_0^{3/2}.
\ee
Eqs.\Ref{cosm1} and \Ref{cosm2} can be written as Hamilton's equations
\be
\dot{P_0}=\frac{\partial H_{eff}}{\partial K_0},\quad \dot{K_0}=-\frac{\partial H_{eff}}{\partial P_0}.\label{cosmhameq}
\ee

\item EOMs of case (2) are satisfied automatically, thus do not impose any constraints \cite{Han:2019vpw}.

\item Closure condition \Ref{closure0} is satisfied automatically.

\end{itemize}

By Hamilton's equations \Ref{cosmhameq}, $H_{eff}=H_v/6$ is conserved in $\t$-evolution:
\be
\mu^3\lt[\frac{\left(\beta ^2+1\right) \sqrt{{P_0}} \sin ^2(2 \beta  \mu  {K_0})}{4 \beta ^2 \mu ^2}-\frac{\sqrt{{P_0}} \sin ^2(\beta  \mu  {K_0})}{\mu ^2}-\frac{1}{3} \Lambda P_0^{3/2}\rt]=\frac{\kappa\mathscr{E}}{6}=\frac{\kappa\rho V}{6}.\label{scrE}
\ee
where $\mathscr{E}>0$ is the dust energy per lattice site, and $\rho=\mathscr{E}/V$ is the dust energy density (recall Eq.\Ref{P=-h}). $V=\mu^3{P_0}^{3/2}$ is the volume per lattice site. Both $\rho$ and $V$ evolve in $\t$ while $\mathscr{E}$ is conserved. Note that because we use the dust to deparametrize gravity, the physical lapse was negative and $\t$ flowed backward (recall Eq.\Ref{lapseshift}). But in Eqs.\Ref{cosm1}, \Ref{cosm2}, and all following equations, we have flipped the time orientation $\t\to -\t$ to make the dust time flow forward.

\begin{figure}[h]
\begin{center}
   \includegraphics[width=1\linewidth]{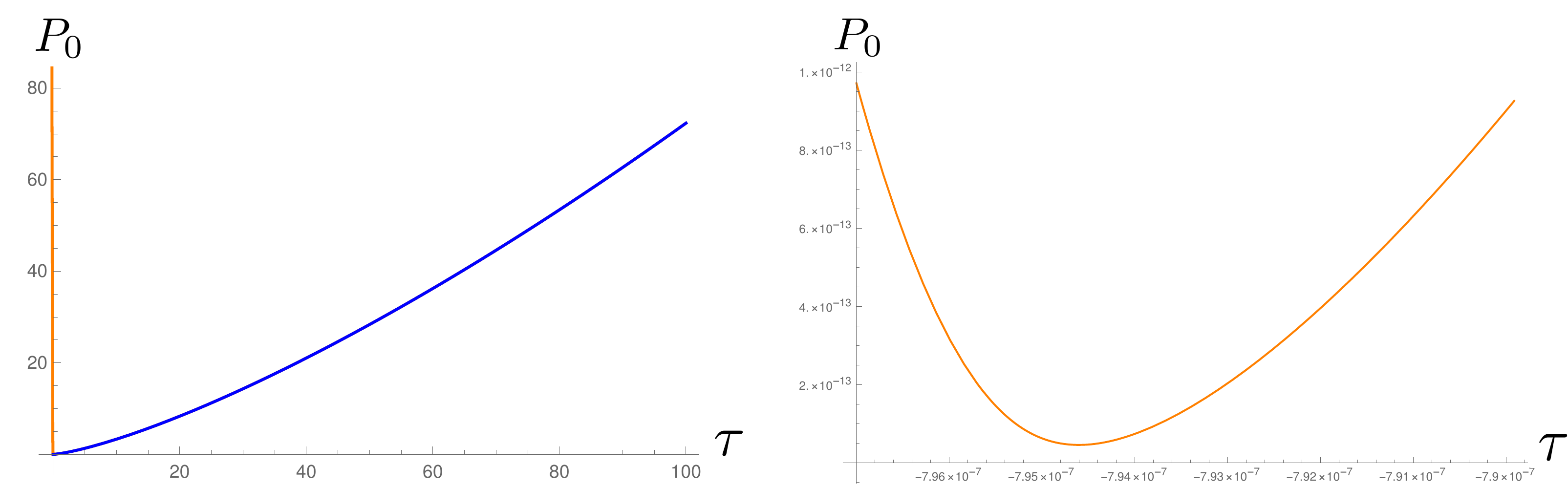}
   \caption{The left panel plots of $P_0(\t)$ solving Eqs.\Ref{cosm1} and \Ref{cosm2} (orange curve) and $P_0(\t)$ solving Friedmann equations (blue curve). Two solutions approximately coincide in $\t>0$ except for regime near the big-bang singularity. The solution of Eqs.\Ref{cosm1} and \Ref{cosm2} replaces the singularity by a bounce. The right panel zooms in the regime near where the bounce happens. The solutions use initial conditions $P_0(1)=0.153262$, $K_0(1)=0.260992$. Values of other parameters are $\L=10^{-5}$, $\b=1$, $\kappa=1$, and $\mu=10^{-3}$. The final time is $T=100$. }
   \label{back}
   \end{center}
 \end{figure}

The effective cosmological equations \Ref{cosm1} and \Ref{cosm2} reduce to classical Friedmann equations when $V$ is large (low density $\rho\ll1$). It may be seen by the following lattice continuum limit of $H_{eff}$ as $\mu\to0$, because the lattice spacing $\mu$ becomes negligible at large scale.
\be
\lim_{\mu\to0}H_{eff}=\sqrt{P_0}  K_0^2-\frac{1}{3} \Lambda {P_0}^{3/2}
\ee
reduces to $h/6=-\cc/6$ for cosmology, and Eqs.\Ref{cosmhameq} reduce to Friedmann equations. {FIG \ref{back}} compares solution $P_0(\t)$ of Eqs.\Ref{cosm1} and \Ref{cosm2} to solution $P_0(\t)$ of Friedmann equations.

Effective equations \Ref{cosm1} and \Ref{cosm2} with finite $\mu$ modify Friedmann equations at high density $\rho$ and lead to a unsymmetric bounce to replace the big bang singularity. {The critical volume and density is given by 
\be
V_c=\frac{8}{27}\beta^6(\beta ^2+1)^3 \kappa ^3 \mathscr{E}^3+O(\L),\quad \rho_c=\mathscr{E}/V_c.
\ee
} $\rho_c$ depending on the conserved quantity $\mathscr{E}$ indicates that the cosmological effective dynamics given by Eqs.\Ref{cosm1} and \Ref{cosm2} is an analog of the $\mu_0$-scheme LQC. The predicted effective dynamics is problematic near the singularity/bounce, because $V_c$ has to be of $O(\ell_P^3)$ in order to have Planckian critical density (for finite $\mathscr{E}$), but $V_c\sim \ell_P^3$ is inconsistent with Eq.\Ref{expandvolume} and invalidate the semiclassical approximation of ${\bf H}$. Otherwise if $V_c$ is much larger than $\ell_P^3$, the bounce can happen at a low critical density, and is not physically sound.

Therefore Eqs.\Ref{cosm1} and \Ref{cosm2} are only valid at the semiclassical regime where the density $\rho$ is low. Given our purpose of the semiclassical analysis, it is sufficient for us to only focus on solutions $(P_0(\t),K_0(\t))$ of Eqs.\Ref{cosm1} and \Ref{cosm2} in the semiclassical regime, and take them as backgrounds to study perturbations.

Cosmological effective dynamics with better behavior at the bounce is given by the $\bar{\mu}$-scheme LQC, where $\rho_c$ is a Planckian constant. Its relation to the full LQG theory is suggested recently in \cite{Han:2019feb}. However in this work, we focus on the cosmological perturbation theory based on solutions of Eqs.\Ref{cosm1} and \Ref{cosm2}, as analog of $\mu_0$-scheme.

\subsection{Cosmological Perturbations}

Given a cosmological background $P_0(\t),K_0(\t)$ satisfying Eqs.\Ref{cosm1} and \Ref{cosm2}, we perturb $p^a(e_I(v)),\theta^a(e_I(v))$ on this background:
\be
\theta^a(e_I(v))=\mu \lt[\b K_0\delta^a_I+\cx^a(e_I(v)) \rt],\quad p^a(e_I(v))=\frac{2\mu^2}{\b a^2}\lt[P_0\delta^a_I+\cy^a(e_I(v))\rt],\label{perturb}
\ee
where $\cx,\cy$ are perturbations. We introduce a vector $V^\rho(v)$ to contain both perturbations $\cx,\cy$ at $v$:
\be
V^\rho(v)=\lt(\cy^a(e_I(v)),\cx^a(e_I(v))\rt)^T,\quad \rho=1,\cdots,18.
\ee
The dictionary between $V^\rho(v)$ and $\cx^a(e_I(v)),\cy^a(e_I(v))$ is given below:
\be
V^1=\cy^1(e_1),\quad &V^2=\cy^2(e_2),&\quad V^3=\cy^3(e_3)\nonumber\\
V^4=\cy^2(e_1),\quad &V^5=\cy^3(e_1),&\quad V^6=\cy^3(e_2)\nonumber\\
V^7=\cy^1(e_2),\quad &V^8=\cy^1(e_3),&\quad V^9=\cy^2(e_3)\nonumber\\
V^{10}=\cx^1(e_1),\quad &V^{11}=\cx^2(e_2),&\quad V^{12}=\cx^3(e_3)\nonumber\\
V^{13}=\cx^2(e_1),\quad &V^{14}=\cx^3(e_1),&\quad V^{15}=\cx^3(e_2)\nonumber\\
V^{16}=\cx^1(e_2),\quad &V^{17}=\cx^1(e_3),&\quad V^{18}=\cx^2(e_3).
\ee

Thanks to the spatial homogeneity of $P_0(\t),K_0(\t)$, we make the following Fourier transformation on the cubic lattice $\g$
\be
V^\rho(\t,\vec{\sig})=\int_{-{\pi}/{\mu}}^{{\pi}/{\mu}}\frac{\rmd^3 k}{(2\pi)^3}\, e^{i \vec{k}\cdot \vec{\sig}}V^\rho(\t,\vec{k}),\quad \vec{\sig}\in (\mu\Z)^3,\label{fourier00}
\ee
where $\vec{\sig}$ are 3d coordinates at the vertex $v$.

Inserting perturbations Eq.\Ref{perturb} in semiclassical EOMs \Ref{eom0}, and applying Fourier transformation, we obtain the following linearized EOMs for each mode ${k}$:
\be
\frac{\rmd V^\rho\lt(\t,{k}\rt)}{\rmd\t}={\bf U}^\rho_{\ \nu}\lt(\mu,\t,{k}\rt)\,V^\nu\lt(\t,{k}\rt).\label{lineareom}
\ee
For simplicity we have assumed that 
\be
\vec{k}=(k,0,0) 
\ee 
has the only nonzero component $k^x=k$. Our discussion mainly focuses on the semiclassical regime where $\mu$ is negligible, this assumption doesn't lose generality in the continuum limit $\mu\to0$, because the background is $P_0(\t),K_0(\t)$ isotropic, the coordinate can always be chosen such that $\vec{k}=(k,0,0)$.

The computation of ${\bf U}^\rho_{\ \nu}(\mu,\t,{k})$ is carried out by expanding ${\bf H}$ up to quadratic order in perturbations followed by derivatives, and ${\bf H}$ contains $C_v$ with Lorentzian term shown in \Ref{HCO}. This computation is carried out on a HPC server and uses the parallel computing environment of Mathematica with 48 parallel kernels. The entire computation lasts for about 2 days. All Mathematica codes can be downloaded in \cite{github}. The explicit expression of $18\times 18$ matrix ${\bf U}^\rho_{\ \nu}(\mu,\t,{k})$ is too long to be shown in this paper but can be found in \cite{github}. Appendix \ref{H0} expands ${\bf U}^\rho_{\ \nu}(\mu,\t,{k})={\bf U}_0{}^\rho_{\ \nu}(\t,{k})+\mu\,{\bf U}_1{}^\rho_{\ \nu}(\t,{k})+O(\mu^2)$, and shows explicitly matrices ${\bf U}_0{}^\rho_{\ \nu}(\t,{k})$ and ${\bf U}_1{}^\rho_{\ \nu}(\t,{k})$.

The linearized closure condition Eq.\Ref{closure0} reads
{\be
0&=&P_0 \big[(V^{15}-V^{18}) \sin (\beta  \mu  K_0)-(V^{16}+V^{17}) (\cos (\beta  \mu  K_0)-1)\big]\nonumber\\
&&+\ \beta  K_0 \big[-i V^{1} \sin (k \mu )+V^{1} \cos (k \mu )-V^{6} \sin (\beta  \mu  K_0)+V^{9} \sin (\beta  \mu  K_0)\nonumber\\
&&+\ V^{7} \cos (\beta  \mu  K_0)+V^{8} \cos (\beta  \mu  K_0)-V^{1}-V^{7}-V^{8}\big],\nonumber\\
0&=&P_0 \big[\cos (k \mu ) (V^{14} \sin (\beta  \mu  K_0)+V^{13} \cos (\beta  \mu  K_0)-V^{13})-i \sin (k \mu ) (V^{14} \sin (\beta  \mu  K_0)\nonumber\\
&&+\ V^{13} \cos (\beta  \mu  K_0)-V^{13})-V^{17} \sin (\beta  \mu  K_0)+V^{18} \cos (\beta  \mu  K_0)-V^{18}\big]\nonumber\\
&&+\ \beta  K_0 \big[i V^{5} \sin (k \mu ) \sin (\beta  \mu  K_0)-\cos (k \mu ) (V^{5} \sin (\beta  \mu  K_0)+V^{4} \cos (\beta  \mu  K_0))\nonumber\\
&&+\ (-V^{9}+i V^{4} \sin (k \mu )) \cos (\beta  \mu  K_0)+V^{8} \sin (\beta  \mu  K_0)+V^{4}+V^{9}\big],\nonumber\\
0&=&P_0 \big[-\cos (k \mu ) (V^{13} \sin (\beta  \mu  K_0)-V^{14} (\cos (\beta  \mu  K_0)-1))+i \sin (k \mu ) (V^{13} \sin (\beta  \mu  K_0)\nonumber\\
&&-\ V^{14} \cos (\beta  \mu  K_0)+V^{14})+V^{16} \sin (\beta  \mu  K_0)+V^{15} \cos (\beta  \mu  K_0)-V^{15}\big]\nonumber\\
&&+\ \beta  K_0 \big[\cos (k \mu ) (V^{4} \sin (\beta  \mu  K_0)-V^{5} \cos (\beta  \mu  K_0))-i \sin (k \mu ) (V^{4} \sin (\beta  \mu  K_0)\nonumber\\
&&-\ V^{5} \cos (\beta  \mu  K_0))-V^{7} \sin (\beta  \mu  K_0)-V^{6} \cos (\beta  \mu  K_0)+V^{5}+V^{6}\big],\label{linearclosure}
\ee}
where $V^\rho=V^\rho(\t,k)$. Closure condition is preserved by $\t$-evolution, because of $\{G^a_v,\,{\bf H}\}=0$ and Eq.\Ref{hamilton}.

Eqs.\Ref{lineareom} and \Ref{linearclosure}, derived from the full LQG, govern the dynamics of cosmological perturbations. Given initial conditions of $V^{\rho=1,\cdots, 18}$ satisfying the closure condition \Ref{linearclosure}, the $\t$-evolution of $V^\rho$'s can be computed by numerically solving Eqs.\Ref{lineareom}. Some results of numerical solutions are discussed in Sections \ref{Power Spectrum} and \ref{Tensor Mode Perturbations}.

\subsection{Continuum Limit and Second Order Perturbative Equations}

Before we actually solve Eqs.\Ref{lineareom} and \Ref{linearclosure}, we would like to firstly derive their lattice continuum limits $\mu\to 0$ (keeping $k$ fixed), and compare with some existing results of the gauge invariant cosmological perturbation theory. 

First of all, the continuum limit of $C_v$, $C_{a,v}$, and $H_v$ reproduce $\cc$, $\cc_{a}$, and $h$:  
\be
C_v&=&\mu^3 \cc(v)+O(\mu^4),\label{Ccc}\\
C_{a,v}&=&\mu^3 \cc_{a}(v)+O(\mu^4),\label{Cjccj}\\
H_v&=&\mu^3 h(v)+O(\mu^4)=\mu^3 \sqrt{\lt|\cc(v)^2-\frac{\a}{4}\sum_{a=1}^3\cc_{a}(v)^2\rt|}+O(\mu^4)
\ee
The above relations not only can be checked perturbatively up to $O(V^2)$ but also can be derived even non-perturbatively as in \cite{Han2020}. Note that the absolute-value in $H_v$ can be remove here at the perturbative level. 

The lattice continuum limit $\mu\to0$ of linearized EOMs \Ref{lineareom} gives
\be
\frac{\rmd V^\rho(\t,{k})}{\rmd\t}+{\bf U}_0{}^\rho_{\ \nu}(\t,{k})\,V^\nu(\t,{k})=0,\quad {\bf U}_0{}^\rho_{\ \nu}(\t,{k})=\lim_{\mu\to0}{\bf U}{}^\rho_{\ \nu}\lt(\mu,\t,{k}\rt)\label{leomcon1}.
\ee
Matrix elements of ${\bf U}_0{}^\rho_{\ \nu}(\t,{k})$ are given explicitly in Appendix \ref{H0}. It is clear from Eq.\Ref{perturb} that in the continuum limit, $V^{\rho=1,\cdots,9}$ and $V^{\rho=10,\cdots,18}$ correspond to perturbations of $E^I_a$ and $A_I^a$ respectively.
\be
&&E^I_a(\t,\sig)=P_0(\t)\delta^I_a+\delta E^I_a(\t,\sig), \qquad\quad A_I^a(\t,\sig)=\b K_0(\t)\delta^I_a+\delta A_I^a(\t,\sig)\\
&&\delta E^I_a(\t,\vec{\sig})=\int_\infty^\infty\frac{\rmd^3 k}{(2\pi)^3}e^{i \vec{k}\cdot \vec{\sig}}\delta E^I_a(\t,\vec{k}), \quad\delta A_I^a(\t,\vec{\sig})=\int_\infty^\infty\frac{\rmd^3 k}{(2\pi)^3}e^{i \vec{k}\cdot \vec{\sig}}\delta A_I^a(\t,\vec{k})\\
&&V^{\rho}(\t,k)=\lt(\delta E^I_a(\t,{k}),\delta A_I^a(\t,{k})\rt)\lt[1+O(\mu k)\rt],\quad k\in\lt[-{\pi}/{\mu},{\pi}/{\mu}\rt]
\ee 
We ignore the difference between $V^{\rho}(\t,k)$ and $(\delta E^I_a(\t,{k}),\delta A_I^a(\t,{k}))$ in the context of lattice continuum limit $\mu\to0$ (fixing $k$).

Here we choose the dust coordinate adapted to the lattice $\g$ so that $I=1,2,3$ is the coordinate index, i.e. the tangent vector of $e_I$ is the $I$-th coordinate basis.  

The linearized closure condition \Ref{linearclosure} when $\mu\to0$ gives
\be
0&=&i k {V^{1}}+\beta  K_0 ( {V^{6}}- {V^{9}} )-{V^{15}}+{V^{18}},\\
0&=&i k {V^{4}}-\beta K_0 ( {V^{5}}- {V^{8}} )+{V^{14}}-{V^{17}},\\
0&=&i k {V^{5}}+\beta  K_0 ( {V^{4}}- {V^{7}} )-{V^{13}}+{V^{16}},
\label{leomcon2}
\ee
which coincide to the linearized Gauss constraint. 

We solve linear equations \Ref{leomcon1} with $\rho=1,\cdots,9$ (containing $\rmd \cy^a(e_I(v))/\rmd\t$) for $\cx^a(e_I(v))$ (perturbations of $\theta^a(e_I(v))$). Inserting solutions of $\cx^a(e_I(v))$ into Eqs \Ref{leomcon1} with $\rho=10,\cdots,18$ (containing $\rmd \cx^a(e_I(v))/\rmd\t$) we can obtain $\rmd \cx^a(e_I(v))/\rmd\t$ as functions of $\cy^a(e_I(v))$ and $\rmd \cy^a(e_I(v))/\rmd\t$. Then by taking time derivative to Eqs \Ref{leomcon1} with $\rho=1,\cdots,9$ and inserting solutions of $\cx^a(e_I(v))$ and $\rmd \cx^a(e_I(v))/\rmd\t$, we obtain $9$ linear second order differential equations of $\cy^a(e_I(v))=V^\rho(v),\ \rho=1,\cdots,9$ (perturbations of $p^a(e_I(v))$):
\be
\frac{\rmd^2 V^\rho(\t,{k})}{\rmd\t^2}+\Fa^\rho_{\ \nu}(\t,{k})\frac{\rmd V^\nu(\t,{k})}{\rmd\t}+\Fb^\rho_{\ \nu}(\t,{k})V^\nu(\t,{k})=0,\quad \rho,\nu=1,\cdots,9.\label{2ndeom}
\ee 
Inserting solutions of $\cx^a(e_I(v))$ into linearized closure condition \Ref{leomcon2} gives 3 first order differential equations of $\cy^a(e_I(v))=V^\rho(v),\ \rho=1,\cdots,9$
\be
G^a(\t,{k})=\Fc^a_{\ \nu}(\t,{k})\frac{\rmd V^\nu(\t,{k})}{\rmd\t}+\Fd^a_{\ \nu}(\t,{k})V^\nu(\t,{k})=0,\quad \nu=1,\cdots,9,\quad a=1,2,3.\label{2ndclosure}
\ee
\cite{github} contains explicit expressions of Eqs.\Ref{2ndeom} and \Ref{2ndclosure} and Mathematica codes for following derivations.

In order to relate to the standard language of cosmological perturbation theory, we construct spatial metric perturbations from the continuum limit of Eq.\Ref{perturb}
\be
q_{IJ}(\t,k)=P_0(\t)\delta_{IJ}+ \delta h_{IJ}(\t,k).
\ee
where $\delta h_{IJ}$ is linear to $V^{\rho=1,\cdots,9}$.
\be
\delta h_{IJ}=\left(
\begin{array}{ccc}
 -V^{1}+V^{2}+V^{3} & -V^{4}-V^{7} & -V^{5}-V^{8} \\
 -V^{4}-V^{7} & V^{1}-V^{2}+V^{3} & -V^{6}-V^{9} \\
 -V^{5}-V^{8} & -V^{6}-V^{9} & V^{1}+V^{2}-V^{3} \\
\end{array}
\right).\label{hVVVV}
\ee 
It is standard to decompose $\delta h_{IJ}$ into components corresponding to scalar, tensor, vector modes
\be
\delta h_{IJ}=P_0\lt(h^S_{IJ}+h^T_{IJ}+h^V_{IJ}\rt),
\ee 
each of which correspond to certain set of components of $V^\rho$ (see follows):

\begin{description}

\item[Scalar modes:] We impose the following ansatz
\be
V^\rho=0 \ \text{except for}\ \rho=1,2,3,6,9, \quad V^6 = - V^9, \quad V^2=V^3\equiv V^{1}-{ k^2 P_0}\ce.\label{scalaransatz}
\ee
$V^2-V^3$ and $V^6+V^9$ belongs to tensor modes (see below). The linearized closure condition Eq.\Ref{2ndclosure} gives only one nontrivial equation
\be
\frac{\rmd }{\rmd \t} \left( \frac{V^9(\tau,k)}{P_0(\t)} \right)  = \frac{4 i \alpha  \beta k \sqrt{P_0(\tau)} \dot{P}_0(\t)}{4 \Lambda P_0(\tau)^2-3  \dot{P}_0(\t)^2} \frac{\rmd\psi(\tau,k)}{\rmd \t}, \quad \psi(\tau,k) &=& \frac{V^{1}(\t,k)}{2 P_0(\t)} \label{closmode}
\ee

Metric perturbations in scalar modes read
\be
h^S_{IJ}(\t, k)=\left(
\begin{array}{ccc}
 2\psi(k)-2k^2\ce(\t, k) & 0 & 0 \\
 0 & 2\psi(k) & 0 \\
 0 & 0 & 2\psi( k) \\
\end{array}
\right)
\ee
$V^6,V^9$ doesn't appear in metric perturbations. Then Eq.\Ref{2ndeom} reduces to
\be
\frac{\rmd^2 \psi(\t,k)}{\rmd\t^2} &=&  -\frac{3 \dot{P}_0(\t) }{2 P_0(\t)} \frac{\rmd \psi(\t,k)}{\rmd\t} \, , \label{Smode0}\\
 \frac{\rmd^2 \ce(\t,k)}{\rmd\t^2} &=& \frac{\dot{P}_0(\t)}{P_0(\t) } \left(\frac{4 \alpha P_0(\t)  }{3\dot{P}_0(\t)^2-4 \Lambda  P_0(\t)^2} \frac{\rmd \psi(\t,k)}{\rmd\t} - \frac{3}{2} \frac{\rmd \ce(\t,k)}{\rmd\t} \right) + \frac{ \psi (\t,k) }{P_0(\t) } \label{Smode}
\ee
plus a few other equations indicating the conservation law of closure condition \Ref{closmode}.
This result holds for both BK and Gaussian dusts. 

\item[Tensor modes:] We impose the following ansatz
\be
V^\rho=0 \ \text{except for}\ \rho=2,3,6,9,\quad V^9=V^6\quad V^{3}=-V^2\ (\text{traceless}).\label{ansatzT}
\ee
Note that the mode $V^6-V^9$ has been considered above in scalar modes. The linearized closure condition Eq.\Ref{2ndclosure} is satisfied by the ansatz. 
Metric perturbations in tensor modes read
\be
h^T_{IJ}(\t, k)=\frac{1}{P_0(\t)}\left(
\begin{array}{ccc}
 0 & 0 & 0 \\
 0 & 2V^3(\t, k) & -2V^9(\t, k)+C(k) \\
 0 & -2V^9(\t, k)+C(k)P_0(\t) & -2V^3(\t, k)P_0(\t) \\
\end{array}
\right)
\ee
Eq.\Ref{2ndeom} reduces to
\be
k^2 h^T_{IJ}(\t,k)+\frac{3}{2} \dot{P}_0(\t) \frac{\rmd h^T_{IJ}(\t,k)}{\rmd\t}+P_0(\t) \frac{\rmd^2 h^T_{IJ}(\t,k)}{\rmd\t^2}=0.\label{Tmode}
\ee
This result holds for both BK and Gaussian dusts. 

\item[Vector modes:] We impose the following ansatz
\be
V^\rho=0 \ \text{except for}\ \rho=4,5,7,8.\label{vectoransatz}
\ee
Metric perturbations in vector modes read
\be
h^V_{IJ}(\t, k)=-\frac{1}{P_0(\t)}\left(
\begin{array}{ccc}
 0 &  V^4(\t, k)+V^7(\t, k) & V^5(\t, k)+V^8(\t, k) \\
V^4(\t, k)+V^7(\t, k) & 0 & 0 \\
V^5(\t, k)+V^8(\t, k) & 0 & 0 \\
\end{array}
\right),\label{vectorhV}
\ee
Firstly, we insert the ansatz \Ref{vectoransatz} and make the replacements $V^4\to -h^V_{12}-V^7$ and $V^5\to-h^V_{13}-V^8$ in both Eqs.\Ref{2ndeom} and \Ref{2ndclosure}. Secondly we solve the linearized closure condition \Ref{2ndclosure} for $\dot{V}^7,\dot{V}^8$. Thirdly, we insert solutions of $\dot{V}^7,\dot{V}^8$ in the resulting Eq.\Ref{2ndeom} from above replacements. As a result, we obtain in total 4 nontrivial equations, in which 2 equations can be expressed only in terms of $h_{IJ}$:
\be
\frac{\dot{P}_0(\t) \left[4\a {P}_0(\t) \left(k^2-3 \Lambda  {P}_0(\t)\right)+9 \dot{P}_0(\t)^2\right]}{4\a {P}_0(\t) \left(k^2-2 \Lambda  {P}_0(\t)\right)+6 \dot{P}_0(\t)^2}\frac{\rmd h^V_{IJ}(\t,k)}{\rmd \t} +{P}_0(\t) \frac{\rmd^2 h^V_{IJ}(\t,k)}{\rmd\t^2}=0,\label{Vmode}
\ee
where $\a=1,0$ corresponds to the BK or Gaussian dust. Other 2 equations with explicit ${V}^7,{V}^8$ are the conservation law of the closure condition. 

\end{description}

We count DOFs of $V^\rho$ (before imposing closure condition): Scalar modes have 3 DOFs ($\rho=1,2,6$), tensor modes have 2 DOFs ($\rho=3,9$), and vector modes have 4 DOFs ($\rho=4,5,7,8$). In total $2+3+4=9$ exhausts all DOFs of $V^{\rho=1,\cdots,9}$.

Scalar, tensor, and vector mode EOMs \Ref{Smode}, \Ref{Tmode}, and \Ref{Vmode} coincide with the ones derived in \cite{Giesel:2007wk}, where they are derived from classical gravity deparametrized by the BK dust and cosmological perturbations. Some details of comparing Eqs.\Ref{Smode}, \Ref{Tmode}, and \Ref{Vmode} to results in \cite{Giesel:2007wk} are presented in Section \ref{tina}. These results indicates that our cosmological perturbation theory derived from LQG has the correct semiclassical limit.

\subsection{Comparison with Results in \cite{Giesel:2007wk}}\label{tina}

This subsection focuses on the lattice continuum limit $\mu\to0$ (keeping $k$ fixed) of linearized semiclassical EOMs, and compares them to the results in \cite{Giesel:2007wk}.  

The metric perturbation $\delta h_{IJ}$ can be decomposed into scalar, tensor, and vector modes \cite{Giesel:2007wk}:
\be
  \delta h_{IJ} = P_0 (2 \psi \delta_{IJ}+2 \partial_{I} \partial_{J} \ce + 2 \partial_{(I} \cf_{J)} + h^T_{IJ})
\ee
where $\psi,\ce$ parametrize scalar modes, and $\cf,\ h^T$ parametrize vector and tensor modes. The above decomposition is in position space, while their Fourier transformations e.g. $\ce(\t,\vec{k})=\int_{-\infty}^{\infty}{\rmd^3 \sig}\,e^{-i \vec{k}\cdot \vec{\sig}}\ce(\t,\vec{\sig})$ are given by $\partial_I\to ik_I$ and 
\be
 \psi &=& \frac{1}{2 P_0} V^{1},\label{littlepsi}\\
 \ce&=&-\frac{-2 V^{1}+V^{2}+V^{3}}{2 k^2 P_0},\label{huaE}\\
 \cf&=&\left(0,\ \frac{i (V^{4}+V^{7})}{k P_0},\ \frac{i (V^{5}+V^{8})}{k P_0}\right)^T\label{huaF}\\
 h^T&=&\frac{1}{P_0}\left(
\begin{array}{ccc}
 0 & 0 & 0 \\
 0 & V^{1}-V^{2}+V^{3} & -V^{6}-V^{9} \\
0 & -V^{6}-V^{9} & V^{1}+V^{2}-V^{3} \\
\end{array}\label{hTVVV}
\right)
\ee
by comparing to Eq.\Ref{hVVVV}. Here we have assumed the only nonzero component of $\vec{k}$ is $k^x = k$. 

Following the standard cosmological perturbation theory, we define  
\be 
B&=&-\frac{-\beta  K_0  V^{1}+ P_0 ( V^{11}+V^{12} )}{\beta  P_0  \left(\Lambda  P_0-3 K_0^2\right)}\\
S&=&\left(0,\ -\frac{2i k (\beta K_0 V^{4}+  P_0 V^{16})}{\beta P_0 \left(3 K_0^2-\Lambda  P_0\right)},\ -\frac{2i k (\beta K_0 V^5 +  P_0 V^{17} )}{\beta P_0 \left(3 K_0^2-\Lambda  P_0\right)}\right)^T
\ee 
For gravity coupled to BK dust, the dynamical shift vector $N_I=\cc_I/h$ is conserved (see Eqs.\Ref{conserv0} and \Ref{lapseshift}). The background $\cc_I=0$ so $N_I=\delta N_I$. $ \delta N_{I}$ can be parametrized by $B$ and $S_I$:
\be
 \delta N_{I} = \sqrt{P_0}(ik_{I} B + S_{I})
\ee

We are going to express our linearized EOMs in terms of the conformal time $\eta$ by
\be 
\frac{\rmd f(\eta)}{\rmd \eta} \equiv f'(\eta) = \sqrt{P_0(\tau)} \frac{\rmd f(\tau)}{\rmd \tau}=\sqrt{P_0(\tau)} \dot{f}(\t).
\ee  

\begin{description} 

  \item[Scalar modes:] Eq.\Ref{scalaransatz} is equivalent to  
  \be
    \delta h_{IJ} = 2 P_0 (\psi \delta_{IJ} - k_{I} k_{J} \ce ), \quad S_J = 0
  \ee
  where $\psi$ and $\ce$ coincide to \Ref{scalaransatz} and \Ref{closmode} respectively. The ansatz implies 
  \be
    B = -\frac{8 {P_0}^{3/2} \dot{\psi} }{4 \Lambda P_0^2-3 \dot{P}_0^2} \label{Bpsi}
  \ee
  Using conformal time $\eta$ and changing variables, Eqs.\Ref{Smode0} and \Ref{Smode} can be rewrite as 
  \be
  2 \ch(\eta) \frac{\rmd \psi(\eta,k)}{\rmd\eta} +\frac{\rmd^2 \psi(\eta,k)}{\rmd\eta^2} &=& 0, \label{Smode1}\\
  \frac{\rmd^2 \ce(\eta,k)}{\rmd\eta^2} + 2 \ch(\eta) \frac{\rmd \ce(\eta,k)}{\rmd\eta}  -\alpha  \ch(\eta)  B(\eta,k ) -\psi (\eta,k ) &=& 0,\label{Smode2}
  \ee
Eqs.\Ref{Smode1} and \Ref{Smode2} at $\a=1$ recover scalar mode equations (3.38) in \cite{Giesel:2007wk} when the additional scalar field is absent. 

  \item[Tensor modes:] Eq.\Ref{ansatzT} is equivalent to $\delta h_{IJ} =P_0 h^T_{IJ},\ B = 0$, and $S_{J} = 0$.
  Eq.\Ref{Tmode} can be rewritten in terms of conformal time
  \be
  k^2 h^T_{IJ}(\eta,k)+ 2 \ch \frac{\rmd h^T_{IJ}(\eta,k)}{\rmd\eta}+ \frac{\rmd^2 h^T_{IJ}(\eta,k)}{\rmd\eta^2}=0.\label{k2hT}
  \ee
  where $\ch$ is the Hubble parameter in conformal time $\eta$:
  \be
  \ch = \frac{1}{\sqrt{P_0(\eta)}} \frac{\rmd \sqrt{P_0(\eta)}}{\rmd\eta}\, \label{hubblec}.
  \ee
  This equation is the Fourier transform of Eq.(3.31) in \cite{Giesel:2007wk}:
  \be
  -\nabla^2 h^T_{IJ}+ 2 \ch \frac{\rmd h^T_{IJ}}{\rmd\eta}+ \frac{\rmd^2 h^T_{IJ}}{\rmd\eta^2}=0.\label{nablahT}
  \ee

  \item[Vector modes:] Eq.\Ref{vectoransatz} is equivalent to 
  \be
    \delta h_{IJ} = 2 P_0 \partial_{(I} \cf_{J)}, \quad B = 0
  \ee
   After inserting solution of the linearized closure condition to $S_J$, we have 
{ \be
   S_1&=&0,\quad S_2=-\frac{4 i k P_0 \sqrt{P_0} }{2 k^2 P_0+3 \dot{P}_0^2-4 \Lambda  P_0^2} \frac{\rmd \partial_{(1} \cf_{2)}(\tau,k)}{\rmd \tau},\\
   S_3&=&-\frac{4 i k P_0 \sqrt{P_0} }{2 k^2 P_0+3 \dot{P}_0^2-4 \Lambda  P_0^2} \frac{\rmd \partial_{(1} \cf_{3)}(\tau,k)}{\rmd \tau}. 
   \ee
We check that Eq. \Ref{Vmode}, and can be rewrite as 
  \be
  2 \ch(\eta) \frac{\rmd \partial_{(I}\cf_{J)}(\eta,k)}{\rmd \eta} + \frac{\rmd \partial_{(I}\cf_{J)}(\eta,k)}{\rmd \eta^2} - \a\ch(\eta) {\partial_{(I} S_{J)}(\eta,k)} = 0 ,\label{FS=0}
  \ee
  which is the same as the vector mode equation (3.33) in \cite{Giesel:2007wk} when $\a=1$} Here e.g. $\partial_{(I}\cf_{J)}(\eta,k)=i k_{(I}\cf_{J)}(\eta,k)$. Furthermore, the conservation law $\frac{\rmd  \delta N_J(\tau)}{\rmd \tau} = \frac{\rmd (\sqrt{P_0} S_J)}{\rmd \tau} = 0$ reduces Eq.\Ref{FS=0} with $\a=1$ to
  \be
  2 \mathcal{H} \partial_{(I}\cv_{J)}+\partial_{(I}\cv_{J)}^{\prime}=0,
  \ee
where $\cv_I=S_I-\cf_I'$.

\end{description}

\section{Scalar Mode Perturbations}\label{Power Spectrum}

\subsection{Scalar Mode Perturbation Theory}\label{Scalar Mode Perturbation Theory}

In this subsection, we make some further analysis on scalar mode EOMs on the continuum. Entire Section \ref{Power Spectrum} specifically focus on gravity coupling to BK dust with $\a=1$. We define Bardeen potentials $\Phi$ and $\Psi$ which are used in the standard gauge-invariant cosmological perturbation theory, 
  \be
      \Phi = - (\ch(B - \ce') + (B -\ce')')= \ch \ce' + \ce '' , \qquad \Psi = \psi + \ch (B - \ce').\label{PhiPsi}
  \ee
  
Eqs.\Ref{Smode1} and \Ref{Smode2} can be expressed in terms of $\Phi$ and $\Psi$: 
  \be  
    2 \Phi \ch'+\ch \left( \Phi'+2  \Psi'\right)+\ch^2 \Phi + \Psi'' &=& 0, \label{Smode3}\\
    \Phi - \Psi &=& 0,\label{Smode4}
  \ee 
  where we have used {$  -\ch''+\ch \ch'+\ch^3 =0$ from background EOMs\footnote{Background EOMs $\dot{P}_0=2K_0\sqrt{P_0},\ 2\sqrt{P_0}\dot{K}_0=-K_0^2+\L P_0$ are given by continuum limits $\mu\to0$ of Eqs.\Ref{cosm1} and \Ref{cosm2}. Using conformal time, the 1st equation is written as $K_0=\ch$ while the 2nd equation is $2\ch'+\ch^2=\L P_0$, whose derivative gives $\ch''+\ch\ch'=\L P_0\ch$. Inserting $\L P_0=2\ch'+\ch^2$ in $\ch''+\ch\ch'=\L P_0\ch$ gives $\ch''-\ch\ch'-\ch^3=0$.} and $\ch B + B' =0$ from the conservation law $(\delta N_{I})' = 0$}. 
  
Moreover, recall that we have conserved quantities $h$ and $\cc_I$:
\be
h(k)&=&{\epsilon}_0+\delta\epsilon(k),\label{Econslin}\\
\delta N_1(k)&=&\delta\epsilon_1(k)/\epsilon_0,\quad \delta N_2(k)=\delta N_3(k)=0.\label{Nconslin}
\ee
${\epsilon}_0,\ \delta\epsilon(k)$, and $\delta\epsilon_1(k)$ are conserved. ${\epsilon}_0=\mathscr{E}/\mu^3$ is the coordinate energy density and $\delta\epsilon,\delta\epsilon_1$ are perturbations. $\delta N_2(k)=\delta N_3(k)=0$ because of $\vec{k}=(k,0,0)$. $h(k),\delta N_I(k)$ are Fourier transformations of $h(\sig),\delta N_I(\sig)$.  Conservation laws \Ref{Econslin} and \Ref{Nconslin} can be expressed in terms of $\Phi,\ \ce$, and $\psi$:
{\be
k^2 \Phi+3 \mathcal{H} \Phi^{\prime}+P_0{ \Lambda} \Phi
&=&\frac{\kappa}{4\sqrt{P_0}}\left[\delta \epsilon-{\epsilon}_0\left(5 \Phi-k^2 \ce\rt)\right],\label{Gpotential}\\
ik\, \psi^{\prime}&=&\frac{\kappa}{4 P_0} \delta \epsilon_{1},\label{psiepsilon}
\ee} 
where $k^2 \Phi$ and $ik\,\psi^{\prime}$ are Fourier transformations of $-\nabla^2\Phi$ and $\partial_I\psi^{\prime}$. In deriving above relations, we have used $\Psi=\Phi$, $\ch B + B' =0$, background EOMs $\dot{P}_0=2K_0\sqrt{P_0},\ 2\sqrt{P_0}\dot{K}_0=-K_0^2+\L P_0$ (continuum limits $\mu\to0$ of Eqs.\Ref{cosm1} and \Ref{cosm2}), and the background conservation law $3\sqrt{P_0} (2K_0^2-\Lambda  P_0)=\kappa \epsilon_0$ (continuum limit of Eq.\Ref{scrE}).

Background EOMs $\dot{P}_0=2K_0\sqrt{P_0},\ 2\sqrt{P_0}\dot{K}_0=-K_0^2+\L P_0$ can be solved analytically by
\be
{P_0}(\t)= \left(\frac{\kappa  \epsilon_0}{2 \Lambda }\right)^{\frac{2}{3}} \sinh ^{\frac{4}{3}}\left[\frac{\sqrt{3\Lambda }}{2} 
    (\t-\t_0)\right],\quad {K_0}(\t)=\left(\frac{\kappa  \epsilon_0}{2 \Lambda }\right)^{\frac{1}{3}}\frac{\sqrt{\Lambda } \cosh \left(\frac{\sqrt{3\Lambda }}{2}  (\t-\t_0)\right)}{\sqrt{3}\, {\sinh^{\frac{1}{3}} \left(\frac{\sqrt{3\Lambda }}{2}  (\t-\t_0)\right)}}
\ee
where the integration constant $\t_0$ is the dust time at big-bang. Prefactors of $P_0,K_0$ are determined by the background conservation law $3\sqrt{P_0} (2K_0^2-\Lambda  P_0)=\kappa \epsilon_0$. Applying the background solution $P_0(\t)$ to Eq.\Ref{Smode0}, we can solve Eq.\Ref{Smode0} for $\psi(\t,k)$
\be
\psi(\t,k)=C_2(k)-{ C_1(k) \frac{\sqrt{3\Lambda }}{2} \coth \left[\frac{\sqrt{3\Lambda }}{2}  (\t-\t_0)\right]},\label{psisol}
\ee
where $C_1(k),C_2(k)$ are arbitrary functions of $k$. Then the conservation law Eq.\Ref{psiepsilon} implies
\be
\delta \epsilon_1=  \frac{3 i k C_1(k)  \epsilon_0}{2}.\label{epsilonC1}
\ee

Furthermore, inserting the solution $\psi(\t ,k)$ into Eq.\Ref{Bpsi} and $\Phi=\Psi = \psi + \ch (B - \ce')$, we obtain $\ce'$ in terms of $\Phi$. Moreover we obtain $\Phi'$ in terms of $\Phi$ by $\Phi' = \psi' + [\ch (B - \ce')]'$ and $ \Phi-\ch \ce' =  \ce ''$. Inserting resulting $\Phi'$ in Eq.\Ref{Gpotential}, we solve $\ce$ in terms of $\Phi$. Resulting $\ce$ and $\dot{\ce}=\ce'/\sqrt{P_0}$ read
\be
\ce(\t,k)&=&\frac{3 C_2(k)}{k^2}-\frac{\delta \epsilon }{k^2 \epsilon_0}+\frac{2\times {2}^{\frac{2}{3}} \Phi (\t,k) \sinh ^{\frac{2}{3}}\left[\frac{\sqrt{3\Lambda }}{2}  (\t-\t_0)\right]}{\kappa ^{2/3} {\Lambda }^{1/3} \epsilon_0^{2/3}}\\
\frac{\rmd\ce(\t,k)}{\rmd\t}&=&\frac{2\times 2^{\frac{2}{3}} \sqrt{3\Lambda } \lt[C_2(k)-\Phi (\t,k)\rt] \sinh ^{\frac{2}{3}}\left[\frac{\sqrt{3\Lambda }}{2} (\t-\t_0)\right] \text{csch}\left[\sqrt{3\Lambda } (\t-\t_0)\right]}{(\kappa  \epsilon_0\sqrt{\L})^{2/3}}.
\ee

By above relations, a complete set of initial conditions is given by values of $\delta\epsilon_1,\ \delta\epsilon$ and the initial values $\Phi(\t_i,k)$ and $\psi(\t_i,k)$ ($\t_i$ is the initial time). In practically applying these initial conditions, $\delta\epsilon_1$ specifies $C_1(k)$ by Eq.\Ref{epsilonC1}, $\psi(\t_i,k)$ specifies $C_2(k)$, then $\delta\epsilon,\ \Phi(\t_i,k),\ C_2(k) $ determines $\ce(\t_i,k),\ \dot{\ce}(\t_i,k)$. After that, the solution $\psi(\t,k)$ is determined by $C_{1}(k),\ C_2(k)$ via Eq.\Ref{psisol}. The time evolution of $\ce(\t,k)$ is determined by Eq.\Ref{Smode} and initial values of $\ce(\t_i,k),\ \dot{\ce}(\t_i,k)$. 



\subsection{Initial Condition}

The time evolution of perturbations $V^\rho$ is determined by initial conditions. Our strategy for initial conditions is to firstly study initial conditions of the continuum theory discussed above, then translate these initial conditions to EOMs \Ref{lineareom} with finite $\mu$. In this section, we firstly focus on scalar mode perturbations.

Here is our choice of initial conditions for scalar modes: Firstly we require following properties of matter (the dust in our case) are not changed by perturbations:
\be
\delta\epsilon=\delta\epsilon_1=0,\label{initial1}
\ee
where $\delta\epsilon$ relates to the dust density, and $\delta\epsilon_1$ is the perturbation of $\cc_I$ and relates to the velocity of the dust (recall Eq.\Ref{Pj=cc}). Here $\delta\epsilon=0$ means that there is no additional matter energy\footnote{Here the notion of energy is fixed by our foliation with dust coordinates.} pumped into the background cosmological spacetime, and is an analog of the initial vaccum state of matter often used in cosmological perturbation theory. $ \delta\epsilon_1=0$ implies $C_1(k)=0$, then $\psi=C_2(k)$ is independent of $\t$.

Furthermore we assume that the Bardeen potential vanishes at initial time $\t_i$:
\be
\Psi(\t_i,k)=\Phi(\t_i,k)=0,
\ee
and the initial value of $\psi$ is a constant:
\be
\psi(\t_i,k)=C_2\label{psiC2}
\ee
Therefore $C_2(k)=C_2$ is a constant independent of $k$, and the solution $\psi(\t,k)=C_2$ is a constant at all time. 

The above specifies a complete set of initial conditions. They determine
\be
\ce(\t_i,k)&=&\frac{3 C_2}{k^2},\label{ceinitial1}\\
\frac{\rmd\ce(\t_i,k)}{\rmd\t}&=&\frac{2\times 2^{\frac{2}{3}} \sqrt{3\Lambda }\, C_2 \sinh ^{\frac{2}{3}}\left[\frac{\sqrt{3\Lambda }}{2} (\t_i-\t_0)\right] \text{csch}\left[\sqrt{3\Lambda } (\t_i-\t_0)\right]}{(\kappa  \epsilon_0\sqrt{\L})^{2/3}},\label{ceinitial2}
\ee
as the initial condition for Eq.\Ref{Smode}.  

We translate the above initial condition in the continuum to the initial condition for Eq.\Ref{lineareom} with finite $\mu$: Firstly we make following setup for initial values of $V^\rho,\dot{V}^\rho$ ($\rho=1,\cdots,9$) at the discrete level by relating to above initial values of $\psi,\dot{\psi},\ce,\dot{\ce}$:
\be
&&V^{4,7,5,8,6,9}(\t_i,k)=0,\\
&&V^2(\t_i,k)=V^3(\t_i,k)=2P_0(\t_i)\psi(\t_i,k)-k^2P_0(\t_i)\ce(\t_i,k),\\
&& V^1(\t_i,k)=2P_0(\t_i)\psi(\t_i,k)\\
&&\dot{V}^{4,7,5,8,6,9}(\t_i,k)=0,\\
&&\dot{V}^2(\t_i,k)=\dot{V}^3(\t_i,k)=\frac{\rmd}{\rmd\t}\lt[2P_0(\t)\psi(\t,k)-k^2P_0(\t)\ce(\t,k)\rt]_{\t=\t_i},\\
&& \dot{V}^1(\t_i,k)=\frac{\rmd}{\rmd\t}\lt[2P_0(\t)\psi(\t,k)\rt]_{\t=\t_i}
\ee
Next, we use components in Eq.\Ref{lineareom} with $\dot{V}^{\rho=1,\cdots,9}$ to solve $V^{\rho=10,\cdots,18}$ as a function of ${V}^{\rho=1,\cdots,9}$ and $\dot{V}^{\rho=1,\cdots,9}$. Initial values of $V^{\rho=10,\cdots,18}$ can be determined by using initial values of ${V}^{\rho=1,\cdots,9}$ and $\dot{V}^{\rho=1,\cdots,9}$. Initial values of $V^\rho$ solves linearized closure condition \Ref{linearclosure} approximately up to $O(\mu^4)$.

\subsection{Scalar Mode Power Spectrum}

\begin{figure}[h]
  \begin{minipage}[t]{0.58\textwidth}
    \centering\vspace{2.5em}
  \subfigure[\,]{ \includegraphics[width = 1\textwidth]{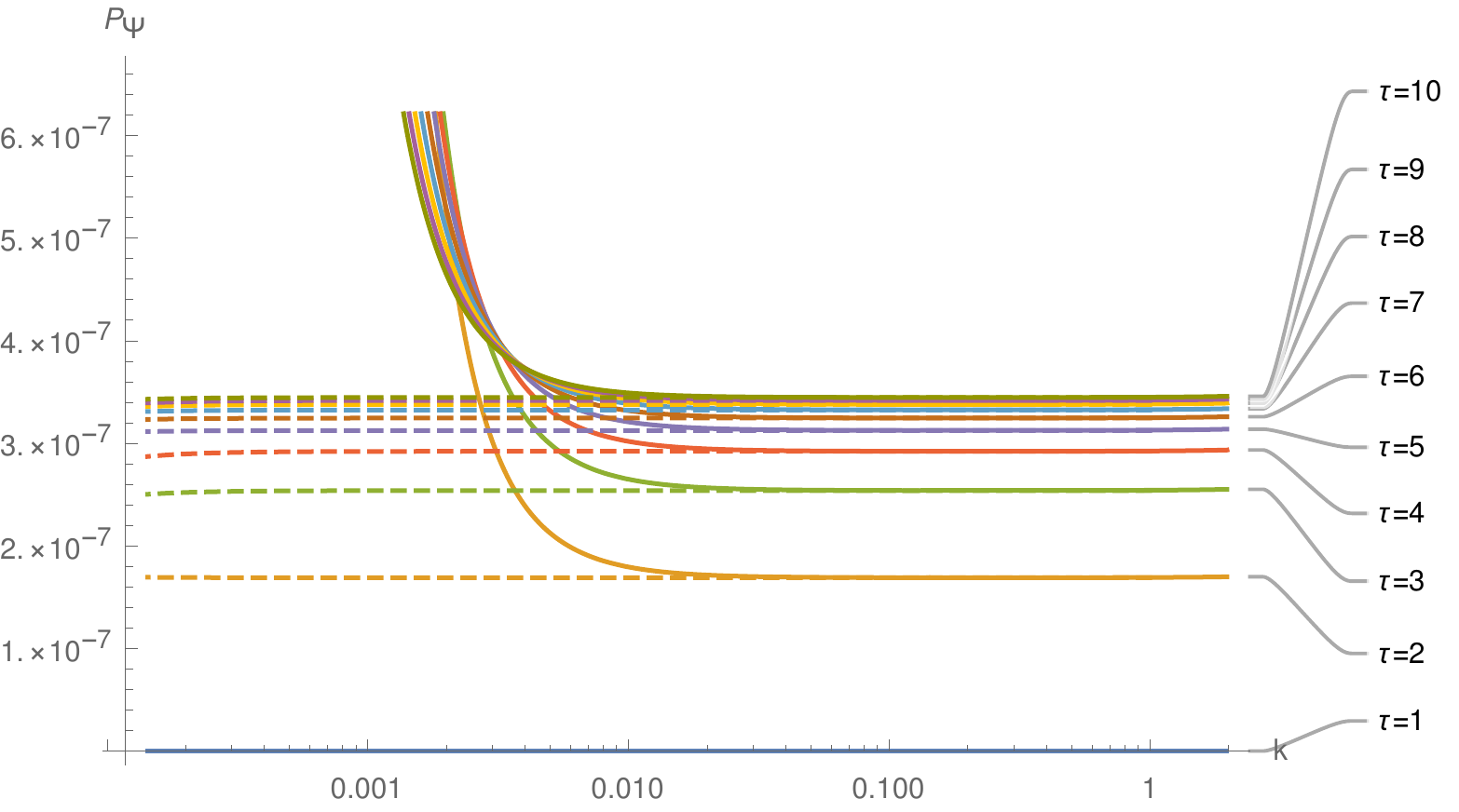} }
  \end{minipage}
  \begin{minipage}[t]{0.38\textwidth}
    \subfigure[\,]{ \includegraphics[width = 1\textwidth]{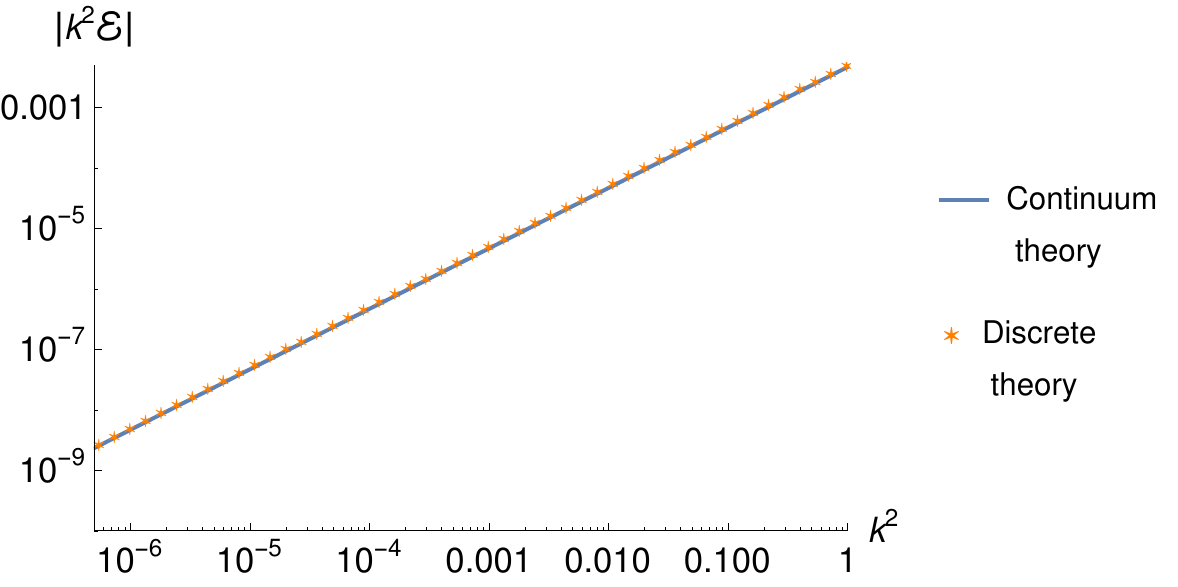} }
    \subfigure[\,]{ \includegraphics[width = 1\textwidth]{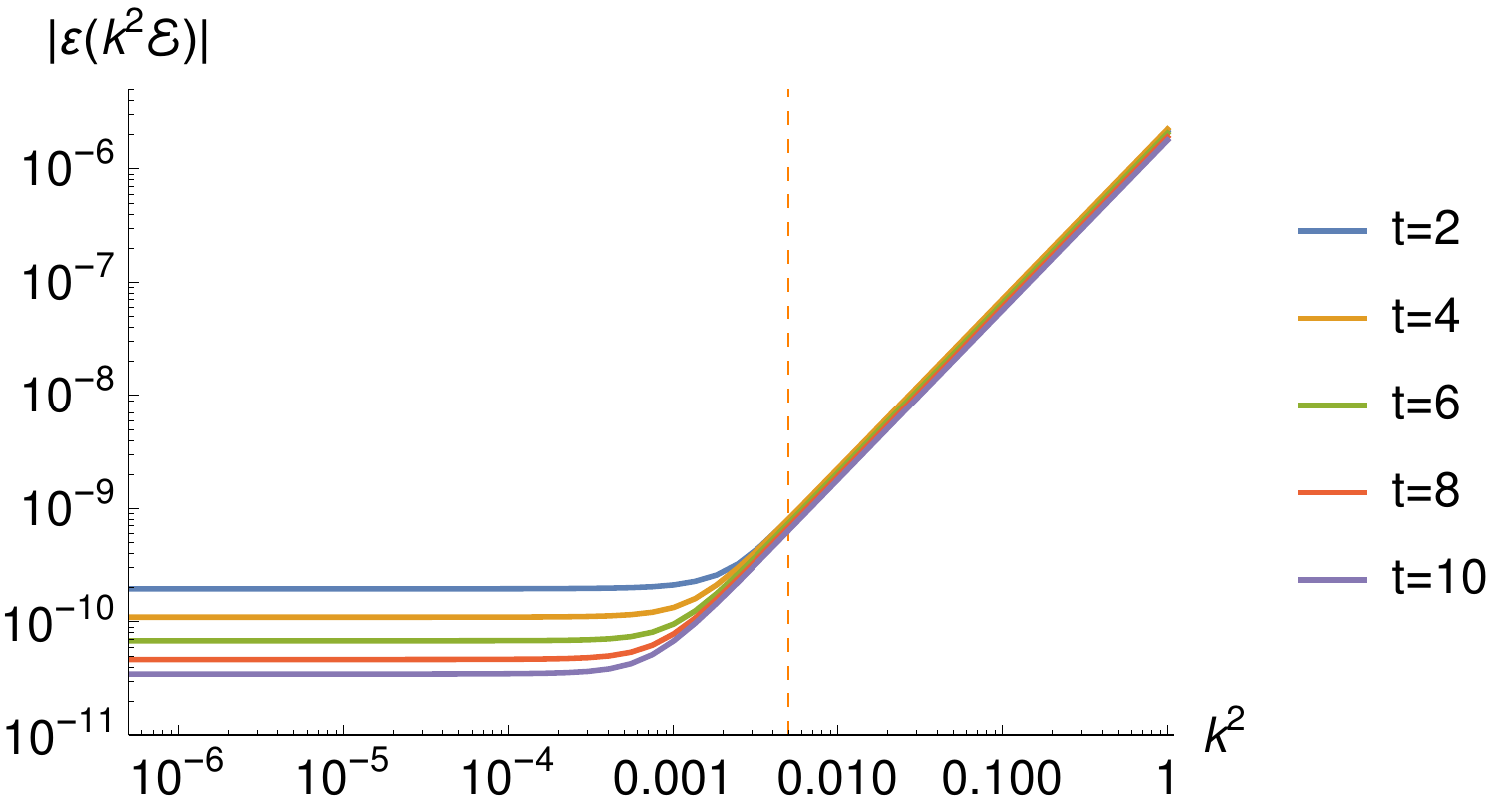} \label{scalarpowerc}}
    \end{minipage}
  \caption{(a): Comparing the scalar mode power spectrum $P_\Psi$ of Bardeen potential between the classical continuum theory and the discrete theory from LQG. Dashed lines are $P_\Psi$ from the classical continuum theory, while solid curves are from the discrete theory. Different colors illustrate $P_\Psi$ (as functions of $k$) at different time $\t$. (b): Plots of $|k^2 \ce| = | (h_{11}^S - h_{22}^S)/2 | $ vers $k^2$ at $\t=10$. In the continuum limit $\ce$ does not depend on $k$. (c): Plots of $ |\epsilon(k^2 \ce)|$ where $\epsilon(k^2 \ce) = k^2 (\ce - \ce |_{\mu \to 0})$ are differences between solutions of discrete EOMs and classical continuum theory. Orange dashed line separates approximately the $k$ dominant region and the background dominant region. Initial condition of those plots are imposed at $\t_i=1$. Initial values of $\psi,\dot{\psi},\ce,\dot{\ce}$ are given by Eqs.\Ref{psiC2}, \Ref{ceinitial1}, and \Ref{ceinitial2} with $C_2=0.001$, $\epsilon_0=0.16$, and $\t_0=0$. Values of other parameters used in numerical computations are $\L=10^{-5}$, $\a=1$, $\b=1$, $\kappa=1$, and $\mu=10^{-3}$.}
  \label{scalarpower}
\end{figure}

We evolve with Eq.\Ref{lineareom} from the initial condition of $V^{\rho=1,\cdots,18}$ using 4th-order implicit Runge-Kutta method. With the solution $V^\rho(\t,k)$, we obtain $\ce,\psi$ using Eqs.\Ref{littlepsi} and \Ref{huaE}, and Bardeen potential $\Psi$ using Eq.\Ref{PhiPsi}. FIG.\ref{scalarpower} demonstrates the power spectrum $P_\Psi=|\Psi(\t,k)|^2$ as a function of $k$ and how $P_\Psi$ evolves in time.

Note that in obtaining $\Psi$ at the discrete level, we apply Eqs.\Ref{littlepsi} and \Ref{huaE} to the discrete theory. Moreover we define a discrete version of shift vector $\delta N_I(v)=\half (C_{a,v}/H_v)\sqrt{P_0}$ linearized in perturbations $V^\rho$, followed by Fourier transform $\delta N_I(v)\to \delta N_I(k)$ as in \Ref{fourier00}. We define $B(\t,k):=\delta N_1(k)/(ik\sqrt{P_0})$ for the discrete theory. $\ch$ is given by Eq.\Ref{hubblec} with background $P_0$ from Eqs.\Ref{cosm1} and \Ref{cosm2}.



FIG.\ref{scalarpower} compares $P_\Psi$ from discrete EOMs \Ref{lineareom} (from LQG) and $P_\Psi$ from the continuum theory (in Section \ref{Scalar Mode Perturbation Theory}). We find that two $P_\Psi$'s coincide for relatively large $k$ while different for small $k$. The difference comes from $\ce\sim V^\rho/(k^2P_0)$ by Eq.\Ref{huaE}: although differences between the discrete and continuum $V^\rho$'s are small and of $O(\mu)$, the small $k^2P_0$ amplifies these differences in $\ce$. As shown in FIG.\ref{scalarpowerc}, the correction $|\epsilon(k^2 \ce)| =| k^2 \lt(\ce - \ce |_{\mu \to 0}\rt)|$  of $k^2 \ce \sim h_{11}^S - h_{22}^S$ is approximately time independent but depends on $k^2$ for relatively large $k$. However $|\epsilon(k^2 \ce)|$ becomes independent of $k$ for small $k$ where the $\mu$ corrections mainly come from the cosmological background, e.g. from terms of $O(\mu K_0)$ in semiclassical EOMs\footnote{If we expand EOMs \Ref{lineareom} in $\mu$, $O(\mu)$ terms are proportional either to $\mu k$ or to $\mu K_0$.}. This leads to the fact that, at late time when $K_0$ becomes smaller, $|\epsilon(k^2 \ce)|$ at small $k$ becomes smaller. 

Note that the ultra-large $k$ with $k\mu\sim1$ breaks the approximation to the continuum theory, and cause differences between the discrete and continuum $V^\rho$'s. Thus the discrete and continuum theory give different $P_\Psi$'s in the ultra-large $k$ regime, although this difference is not shown in FIG.\ref{scalarpower}. 

\begin{figure}[h]
  \begin{center}
  \includegraphics[width = 0.46\textwidth]{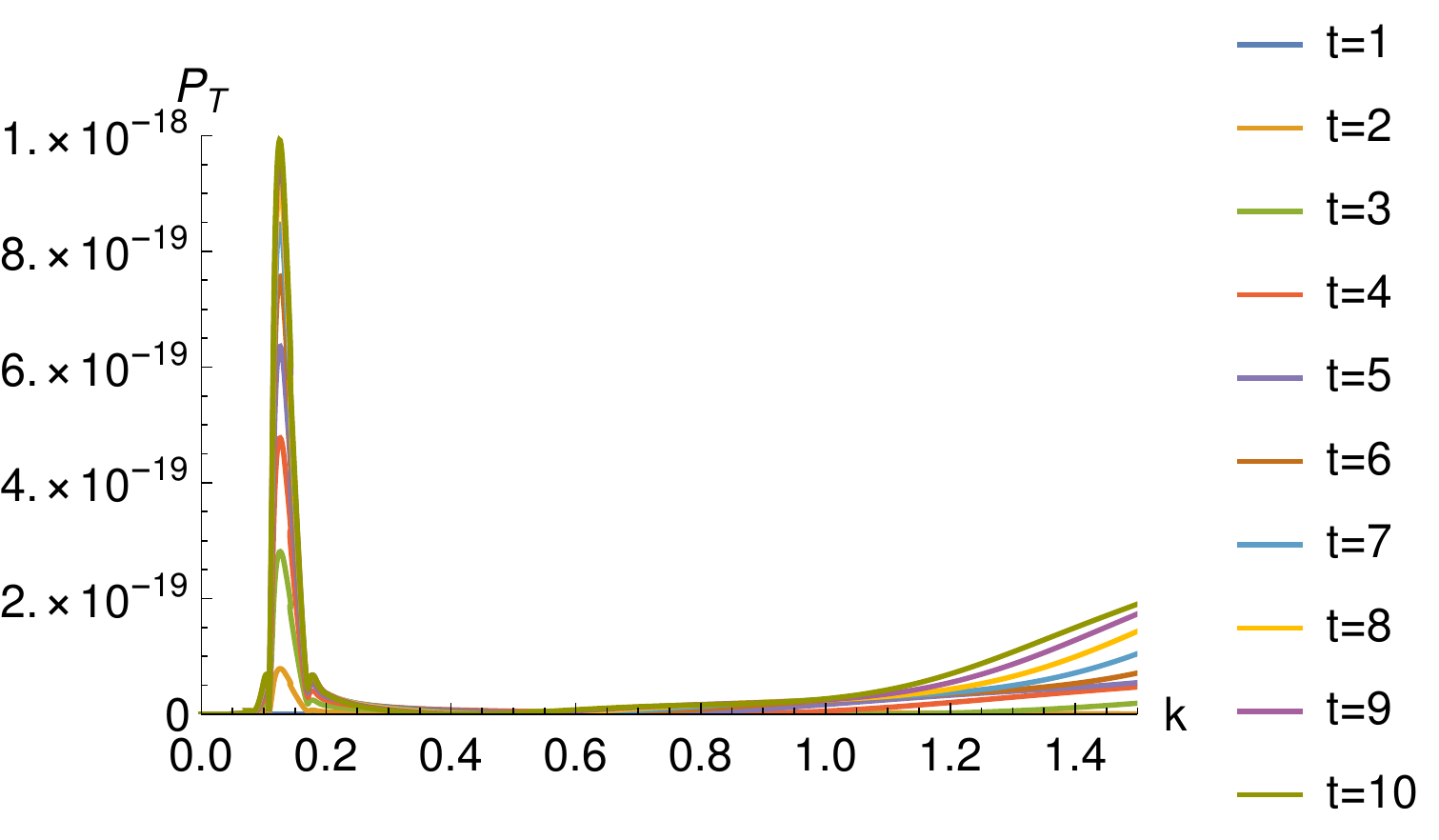}\quad
  \includegraphics[width = 0.46\textwidth]{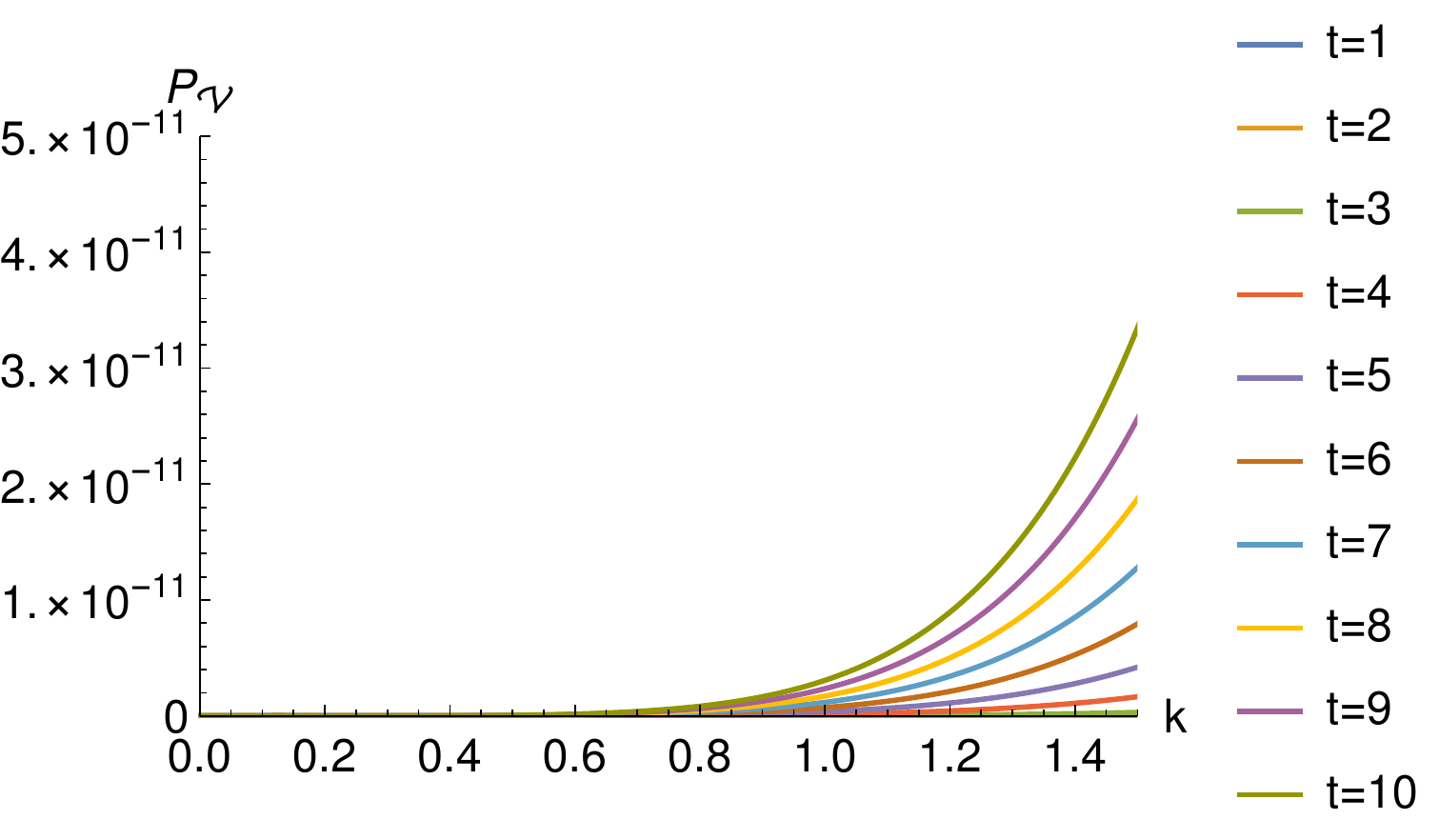}
  \end{center}
  \caption{The left panel plots tensor mode power spectrums $P_T=|h^T_{23}(\t,k)|^2$ as functions of $k$ at different $\t$, evolving from the scalar mode initial condition (the same as in FIG.\ref{scalarpower}). $P_T$ at different $\t$ are illustrated by different colors. The other $h^T_{IJ}$ component $|h^T_{22}(\t,k)|^2$ is even smaller than $|h^T_{23}(\t,k)|^2$, thus is not demonstrated. The right panel plots vector mode power spectrums $P_\cv=|\vec{\cv}(\t,k)|^2$ at different $\t$. }
  \label{phitvmode}
\end{figure}

\begin{figure}[h]
  \begin{center}
  \includegraphics[width = 0.5\textwidth]{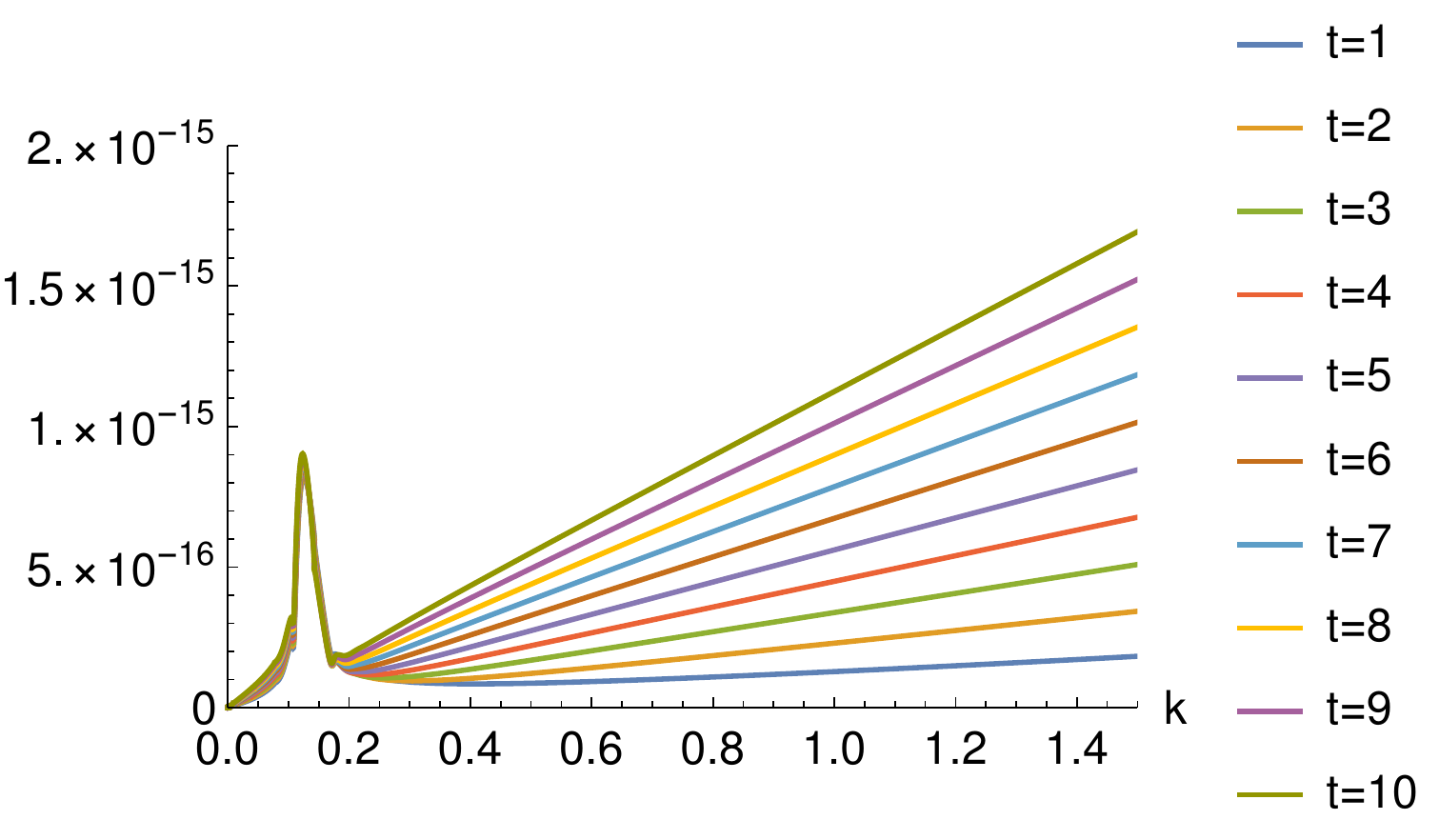}
  \end{center}
  \caption{This figure plots the error of closure condition $|G|^2=\sum_{i=1}^3G^2_i$ where $G_i$ are given by Eqs.\Ref{linearclosure}. $|G|^2$ at different $\t$ are illustrated by different colors. $|G|^2$ is much smaller than $\mu^4=10^{-12}$ in the plotted range of $k$.}
  \label{phiclo}
\end{figure}

Eq.\Ref{lineareom} with finite $\mu$ couples vector and tensor modes to scalar modes, while these couplings are turned off by the continuum limit $\mu\to0$. With finite $\mu$, the scalar model initial condition can excite tensor and vector modes in the time evolution. FIGs.\ref{phitvmode} plots power spectrums $P_T=|h^T_{23}(\t,k)|^2$ and $P_\cv=|\vec{\cv}(\t,k)|^2$ at different time $\t$ evolved from the scalar mode initial condition. Here $h^T_{IJ}$ are given by Eq.\Ref{hTVVV} with $V^\rho$ satisfying discrete EOMs. $\cv_I=S_I-\cf_I'$ where $\cf_I$ are given by Eq.\Ref{huaF} and $S=(0,\delta N_2,\delta N_3)/\sqrt{P_0}$ with $V^\rho$, $\delta N_I$, and $P_0$ satisfying EOMs with finite $\mu$. FIGs.\ref{phitvmode} demonstrates that the scalar mode initial condition excites both tensor and vector mode perturbations by EOMs with finite $\mu$. These tensor and vector modes are all small and of higher order in $k\mu$ (since $\mu$ is of length dimension, $\mu$-expansion is the same as $k\mu$-expansion for relatively large $k$), while they can smoothly grow as $k$ becoming large. FIG.\ref{phiclo} plots the error of linearized closure condition in the $\t$-evolution and finds that it is much smaller than $\mu^4$.

\section{Tensor Mode Perturbations}\label{Tensor Mode Perturbations}

\subsection{Modified Graviton Dispersion Relation}\label{Modified Graviton Dispersion Relation}

We consider Eq.\Ref{lineareom} in the late-time limit $K_0={P_0'}/{(2P_0)}\to0$ and absent of cosmological constant $\L=0$, and we insert the tensor mode ansatz: $V^{1}=0$, $V^2=-V^3$, and $V^6=V^9$ which turn off scalar modes at late time. 
The closure condition Eq.\Ref{linearclosure} and the compatibility of Eq.\Ref{lineareom} at late time leads to $V^{4,5,10,13,14,16,17}= 0,\ V^{12}=-V^{11},\ V^{15}= V^{18}$. Eq.\Ref{lineareom} at late time gives the following wave equation for the tensor modes metric $h^T_{IJ}$ (valid for both $\a=0,1$):
\be
\o (k)^2 h^T_{IJ}(\eta,k)+\frac{\rmd^2 h^T_{IJ}(\eta,k)}{\rmd \eta^2}=0,\qquad \omega(k)^2=\frac{\sin ^2(k \mu ) }{\mu ^2}\left[\left(\beta ^2+1\right) \cos (k \mu )-\beta ^2\right].\label{graviton}
\ee
The tensor mode metric perturbation relates to $V^{2,3,6,9}$ by
\be
h^T=\frac{1}{P_0}\left(
\begin{array}{ccc}
 0 & 0 & 0 \\
 0 & V^{2}+V^{3} & -V^{6}-V^{9} \\
0 & -V^{6}-V^{9} & V^{2}-V^{3} \\
\end{array}
\right)
\ee
Solutions of Eq.\Ref{graviton} are spin-2 gravitons with a modified dispersion relation $\o(k)^2$. We expand the $\o(k)^2$ in terms of $\mu$
\be
\o(k)^2=k^2\left[1-\frac{1}{6} \mu ^2k^2 \left(3 \beta ^2 +5 \right)+O\left(\mu ^3k^3\right)\right]
\ee
Gravitons travel in the speed of light in the continuum limit $\mu\to 0$ or the long wavelength limit $k\ll \mu^{-1}$, while less than speed of light for finite $\mu$. The finite $\mu$ generates a higher derivative term $O(k^4)$ in the wave equation of $h^T_{IJ}$
\be
\frac{\rmd^2 h^T_{IJ}(\eta,k)}{\rmd \eta^2}+k^2 h^T_{IJ}(\eta,k)-\frac{1}{6} \mu ^2k^4 \left(3 \beta ^2 +5 \right) h^T_{IJ}(\eta,k)+O\left(\mu ^3k^3\right)=0.
\ee

The result \Ref{graviton}, derived from top to down in the full theory of LQG, proves that LQG can give spin-2 gravitons as low energy excitations. The modified dispersion relation Eq.\Ref{graviton} is the same as the one in \cite{Dapor:2020jvc} obtained by expanding the LQG Hamiltonian on the flat spacetime. 

\begin{figure}[h]
  \begin{center}
  \includegraphics[width = 0.5\textwidth]{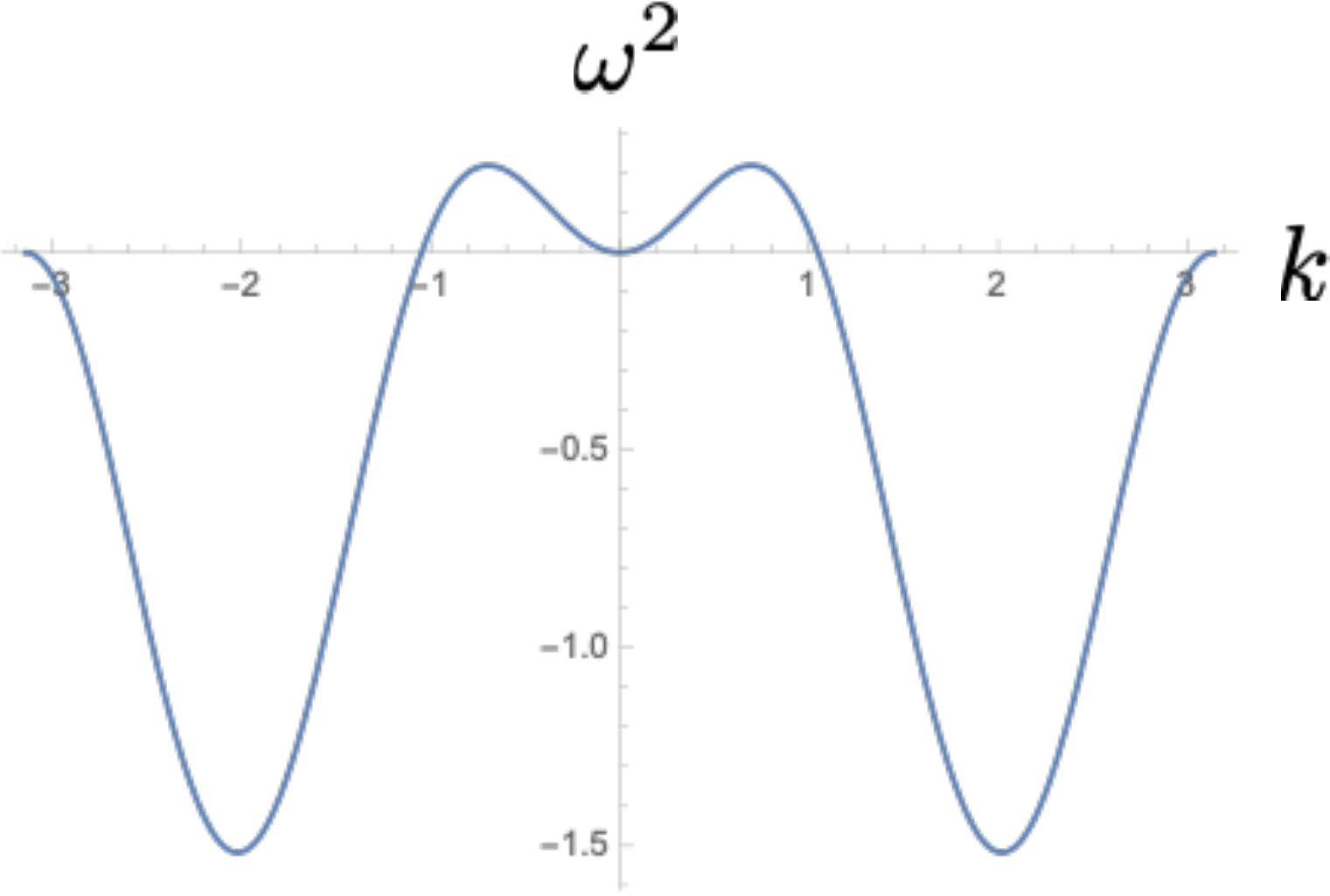}
  \end{center}
  \caption{Plot of the dispersion relation \Ref{graviton}. $\o^2=0$ corresponds to 2 non-negative $k$'s.}
  \label{doubledisper}
\end{figure}

In our opinion, the dispersion relation \Ref{graviton} is only valid in the long-wavelength regime $k\ll\mu^{-1}$. If we admit $k\sim\mu^{-1}$, there exists $k$ resulting $\o^2<0$ and corresponding to non-propagating modes. In addition, the dispersion relation \Ref{graviton} indicates that there are 2 non-negative $k$'s giving the same $\o^2\geq0$ (see Figure \ref{doubledisper}). For example, at low energy $\o^2= 0$ corresponds to
\be
k=0,\quad k=\mu^{-1}\arccos\lt(\frac{\b^2}{\b^2+1}\rt)
\ee 
where $k=0$ corresponds to the graviton, but the second mode is a spurious low energy excitation. $\o^2(k)$ expanded at this spurious mode gives
\be
\o^2(k)\mu^2&=&-\left(\beta ^2+1\right) \left(1-\frac{\beta ^4}{\left(\beta ^2+1\right)^2}\right)^{3/2} \mu\,  p-\frac{5}{2} \beta ^2 \left(1-\frac{\beta ^4}{\left(\beta ^2+1\right)^2}\right) \mu ^2\, p^2+O\left(\mu^3p^3\right)\\
p&=&k-{\mu^{-1} }\cos ^{-1}\left(\frac{\beta ^2}{\beta ^2+1}\right)
\ee
which has no analog in continuum field theories. The existence of this spurious mode should be due to the regime that $\ell_P\ll\mu$ on which our discussion have focused. Beyond this regime, e.g. in $\mu\sim\ell_P$ which may be more physically sensible, the dispersion relation \Ref{graviton} should be modified in large $k$ by $O(\ell_P^2)$ corrections so that the spurious mode may be removed/changed. Taking the continuum limit $\mu\to0$ before $\ell_P\to0$ might be physically relevant since it removes the lattice-dependence at the quantum level, but it is beyond the scope of this paper. However, the perturbation theory derived here from the semiclassical approximation $\ell_P\to0$ (while keeping $\mu$ finite) should be only viewed as an effective theory which only valid in the long wavelength regime $k\ll\mu^{-1}$, while behavior at $k\sim\mu^{-1}$ should not be trusted before $O(\hbar)$-corrections are implemented.

Eq.\Ref{lineareom} with finite $\mu$ contains another 2 nontrivial equations showing couplings between $V^{\rho=2,6}$ (tensor modes) and $V^{\rho=7,8}$ (vector modes). Defining $u^\rho\equiv V^\rho/P_0$, these equations (at late time) are shown below by expanding in $\mu$ 
\be
0&=&{u'^7}(\eta )+\frac{1}{2}k\mu  \left[-i  u'^2(\eta )+\beta  k  u^6(\eta )\right] \nonumber \\
& &\qquad+ \frac{2 i \alpha \beta  k}{3 K_0^2 \sqrt{P_0}} \left[2 u^8(\eta )+\mu\lt( i k   u^2(\eta )+\beta  u'^6(\eta )\rt)\right] +O\left(\mu ^2\right),\label{coupleVT1}\\
0&=&{u'^8}(\eta )-\frac{1}{2}k\mu  \left[-i  u'^2(\eta )+\beta  k  u^6(\eta )\right]\nonumber \\
& &\qquad+ \frac{2 i \alpha \beta  k}{3 K_0^2 \sqrt{P_0}} \left[ - 2 u^7(\eta ) + \mu \lt( i k   u^2(\eta )+\beta  u'^6(\eta )\rt)\right] +O\left(\mu ^2\right), \label{coupleVT2}
\ee
while equations before the $\mu$-expansion is too long to be shown here and they can be downloaded in \cite{github}. Couplings between $V^{\rho=2,6}$ and $V^{\rho=7,8}$ disappear in Eqs.\Ref{coupleVT1} and \Ref{coupleVT2} when $\mu\to0$.

\subsection{Tensor Mode Power Spectrum}

We set $\L=0$ in the discussion of tensor mode. The background EOMs \Ref{cosm1} and \Ref{cosm2} with $\L=0$ and $\mu\to0$ can be solved analytically with $\ch =4/\eta$. Then the tensor mode EOM \Ref{k2hT} at $\mu\to0$ can be written as a differential equation in terms of $x=k\eta$: 
\be
h^T_{IJ}+  \frac{8}{x } \frac{\rmd h^T_{IJ}}{\rmd x}+ \frac{\rmd^2 h^T_{IJ}}{\rmd x^2}=0,\quad x=k\eta
\ee
Therefore solutions at the continuum limit are functions of $k\eta$: $h^T_{IJ}=h^T_{IJ}(k\eta)$.

\begin{figure}[h]
  \begin{center}
  \includegraphics[width = 0.48\textwidth]{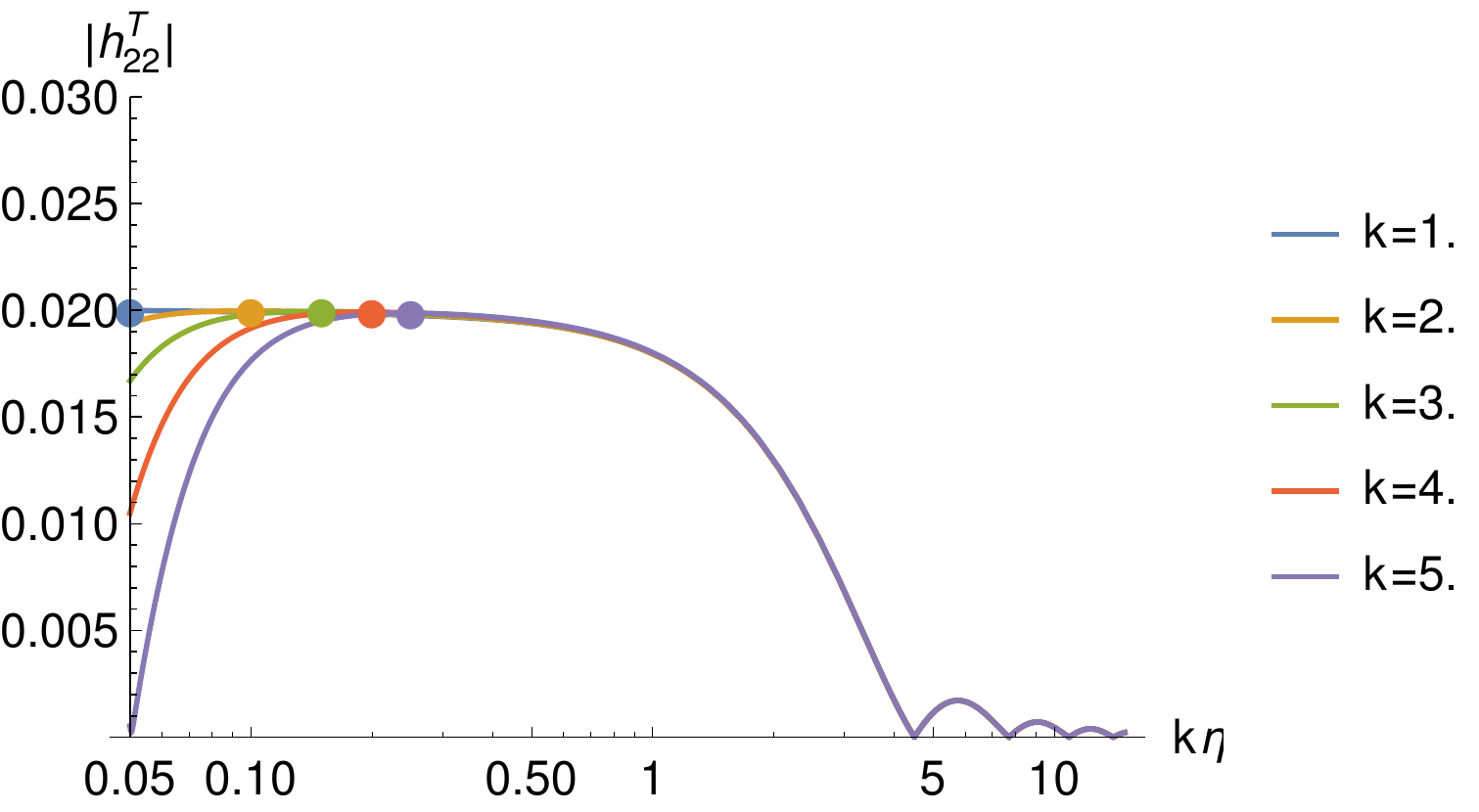}\quad
  \includegraphics[width = 0.48\textwidth]{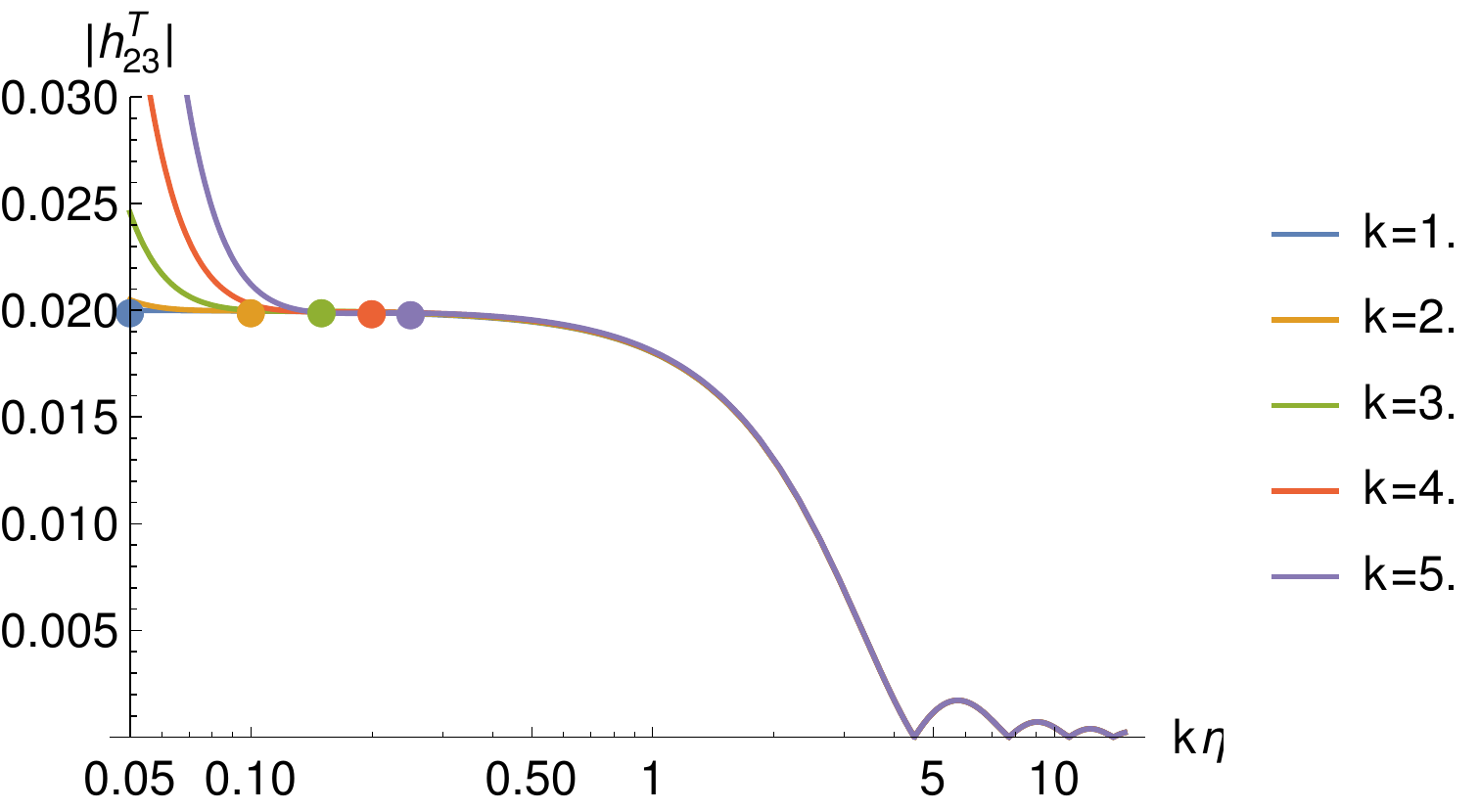}
  \end{center}
  \caption{Plots of $|h^T_{22}|$ and $|h^T_{23}|$ as functions of $k\eta$ at different $k$. Colored dots illustrate their initial values. The initial condition is imposed at $\eta_i=0.05$. Initial values are $u^{1,4,5,7,8}=0, u^3=-u^2=u^6=u^9=0.00999754$ and $u'^3=-u'^2=u'^6=u'^9=-0.000099816$. Values of parameters are $\L=0$, $\a=1$, $\b=1$, $\kappa=1$, and $\mu=10^{-3}$. }
  \label{tmodeini2}
\end{figure}

\begin{figure}[h]
  \begin{center}
  \includegraphics[width = 0.46\textwidth]{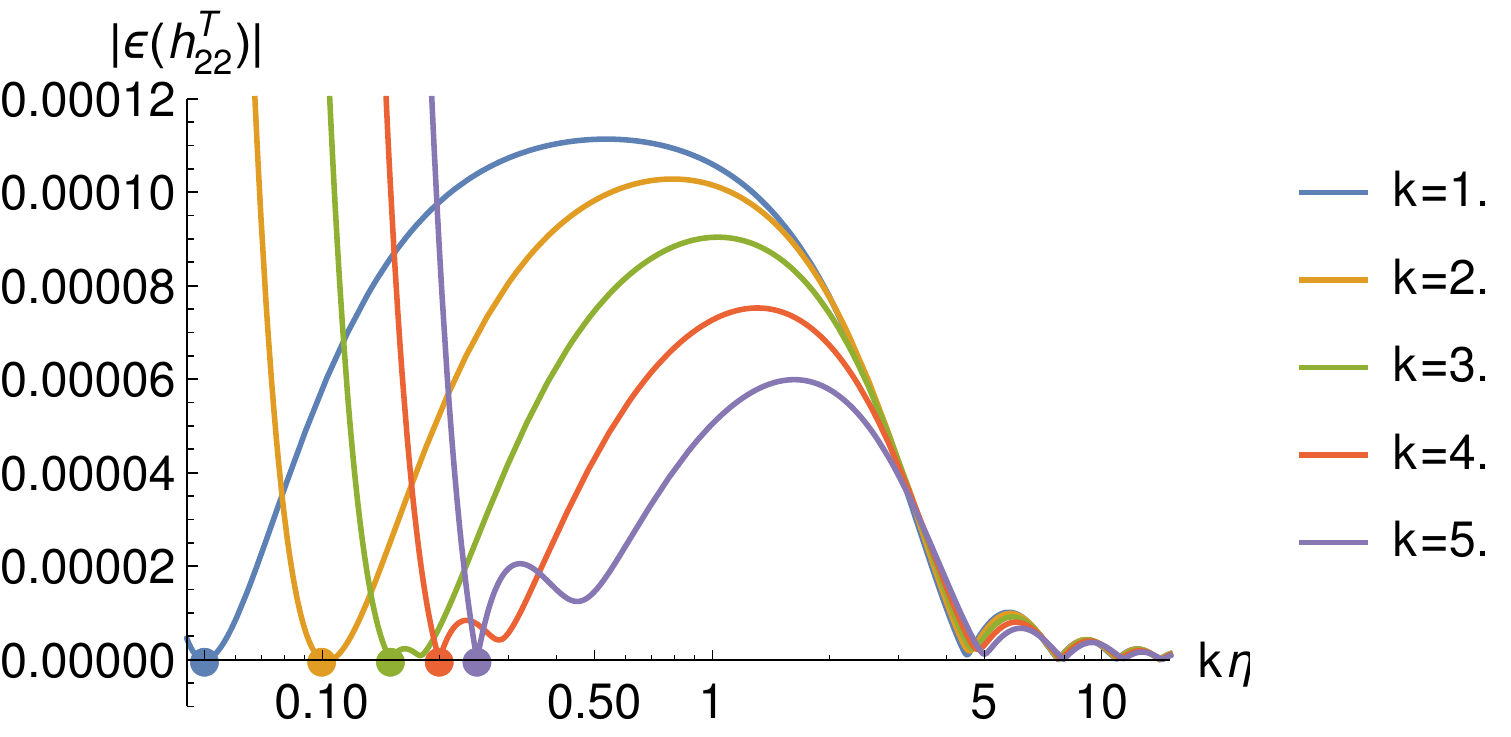}\quad
 \includegraphics[width = 0.46\textwidth]{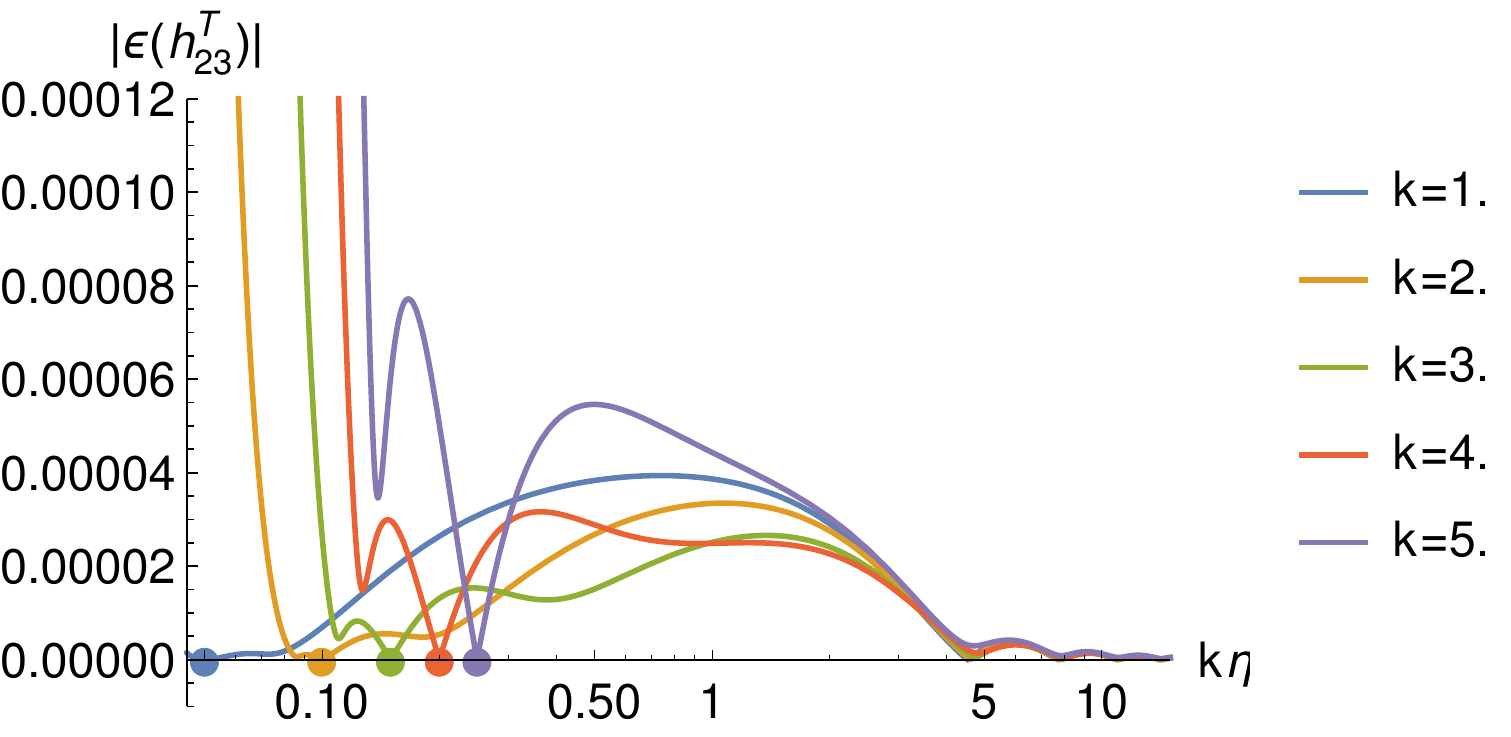}
  \end{center}
  \caption{Plots of $|\epsilon(h^T_{22})|$ and $|\epsilon(h^T_{23})|$ where $\epsilon(h^T_{IJ})=h^T_{IJ}-h^T_{IJ}|_{\mu\to0}$ are differences between solutions of discrete and continuum EOMs. Colored dots illustrate initial values. }
  \label{tmodediffh2}
\end{figure}

Semiclassical EOMs with finite $\mu$ can be solved numerically for both the cosmological background and tensor mode perturbations. Both initial conditions of the background $P_0,K_0$ and tensor mode perturbations are imposed at the conformal time $\eta_i$. The tensor mode initial condition is given by {$u^{1,4,5,7,8}=0, u'^{1,4,5,7,8}=0, u^3=-u^2=u^6=u^9\neq0$ and $u'^3=-u'^2=u'^6=u'^9\neq0$}. FIG.\ref{tmodeini2} plots time evolutions of tensor mode perturbations $h_{IJ}^T$ as functions of $k\eta$ (at different $k$), where we find approximately $h_{IJ}^T=h^T_{IJ}(k\eta)$ (depending on $k$ only through $k\eta$) at late time, and $h_{IJ}^T=h^T_{IJ}(k, k\eta)$ at early time (especially when we evolve from $\eta$ toward the bounce). FIG.\ref{tmodediffh2} plots the difference $\epsilon(h^T_{IJ})=h^T_{IJ}-h^T_{IJ}|_{\mu\to0}$ between solutions of discrete and continuum EOMs, and shows that  $|\epsilon(h^T_{IJ})|$ is small and less than $O(\mu)$. When we evolve from $\eta$ toward the bounce (with large curvature), $|\epsilon(h^T_{IJ})|$ becomes larger, and suggests that the continuum theory approximates well to the discrete theory only when the curvature is small.

\begin{figure}[h]
  \begin{center}
  \includegraphics[width = 0.46\textwidth]{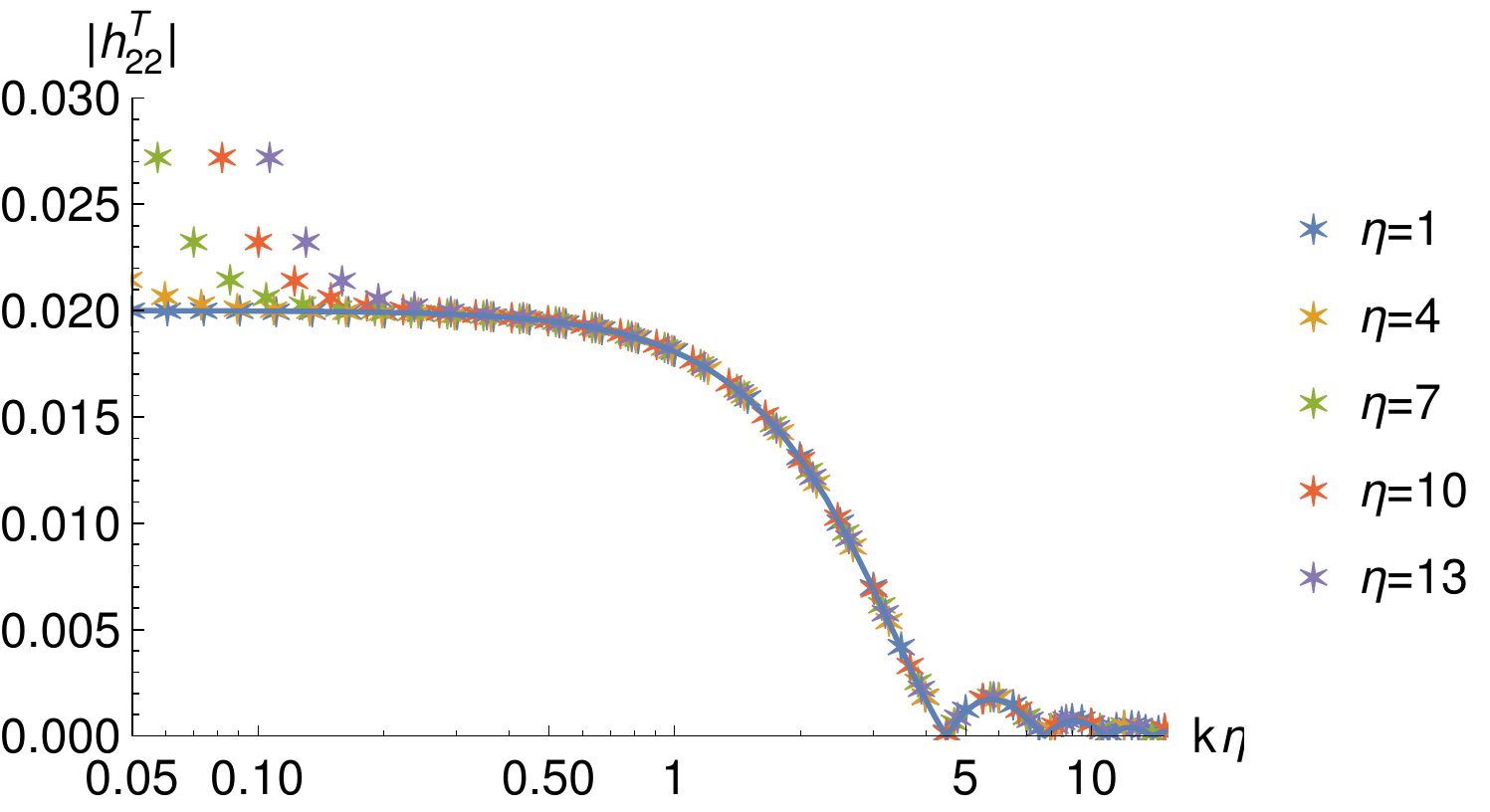}\quad
   \includegraphics[width = 0.46\textwidth]{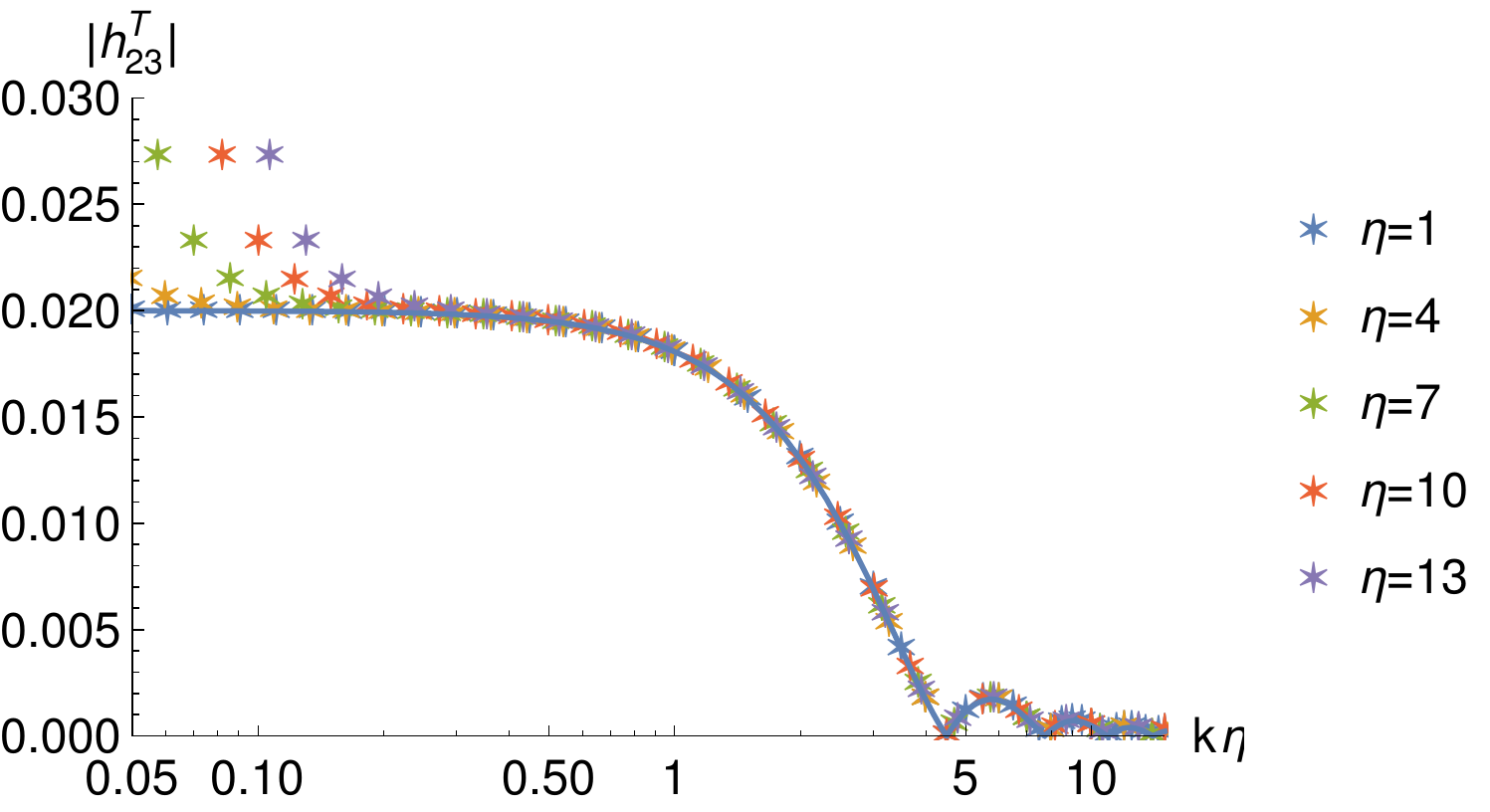}
  \end{center}
  \caption{Colored stars illustrate power spectrums $|h^T_{22}|$ and $|h^T_{23}|$, resulting from EOMs with $\mu=10^{-3}$, as functions of $k\eta$ at different conformal times $\eta$. Different colors label different $\eta$. The blue curve is the power spectrum from the continuum theory. The initial condition is the same as in FIG.\ref{tmodeini2}.}
  \label{powertt}
\end{figure}

FIG.\ref{powertt} plots power spectrums $|h^T_{22}(\eta, k\eta)|$ and $|h^T_{23}(\eta, k\eta)|$ as functions of $k\eta$ at different conformal time $\eta$. When $k$ are relatively large (but still much smaller than $\mu^{-1}$), power spectrums with finite $\mu$ approximately coincide with results from the continuum EOM \Ref{k2hT}, but depart from the continuum results for small $k$, similar to the scalar mode power spectrum FIG.\ref{scalarpower}. To understand this departure, we recall that Eq.\Ref{graviton} is an approximation of tensor mode EOMs at the late time, so at earlier time we have
\be
\o (k)^2 h^T_{IJ}(\eta,k)+\frac{\rmd^2 h^T_{IJ}(\eta,k)}{\rmd \eta^2}+O(K_0)=0.
\ee 
$O(K_0)$ collects terms vanishing as $K_0\to0$ while non-vanishing at earlier time. The small $k$ suppresses the first term and make the term with background $K_0$ stand out, while the background $K_0$ is different between the finite $\mu$ and $\mu\to0$. $\mu\to0$ removes the difference between discrete and continuum theory.   

\begin{figure}[h]
  \begin{center}
  \includegraphics[width = 0.46\textwidth]{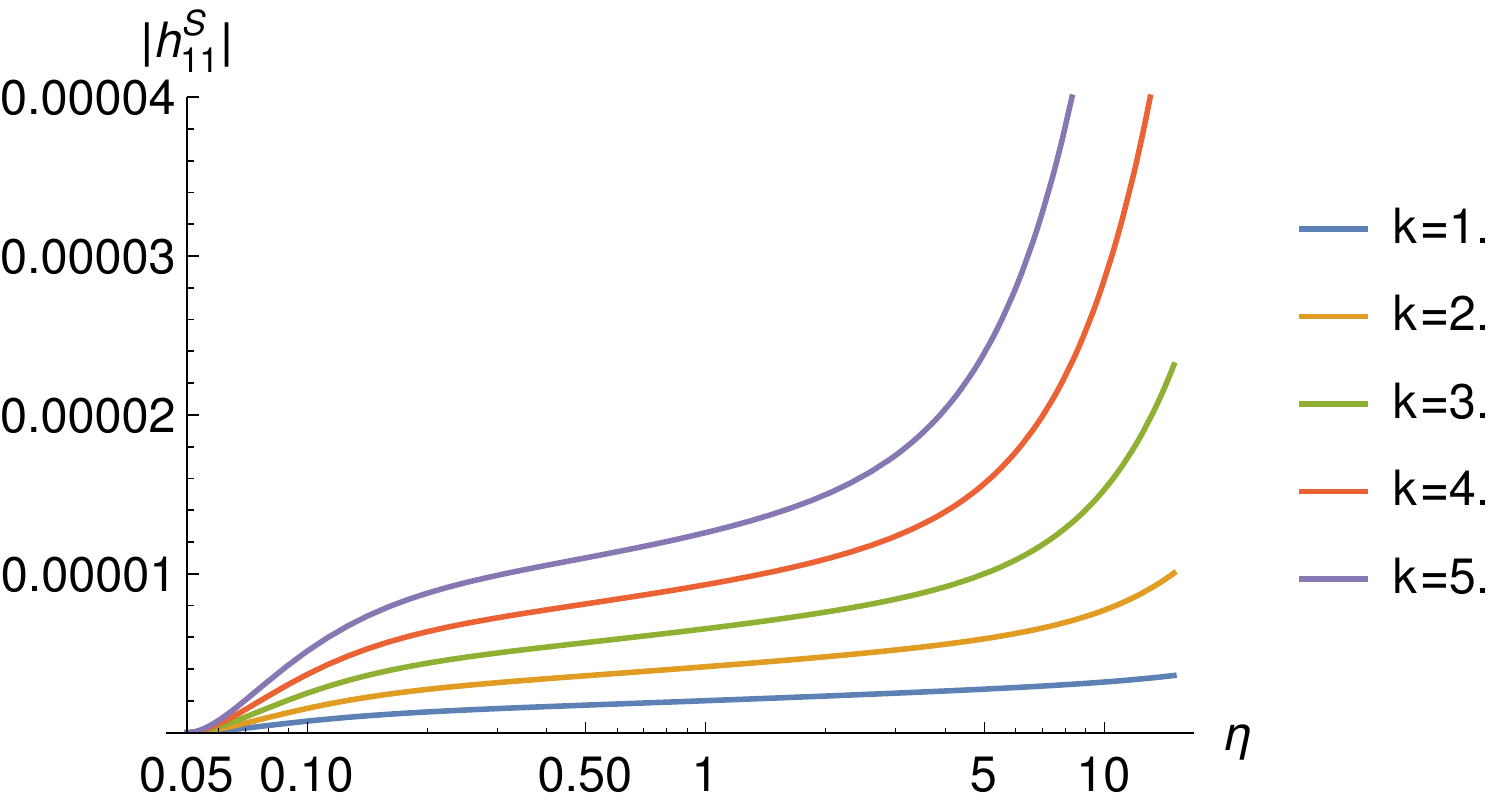}\quad
    \includegraphics[width = 0.46\textwidth]{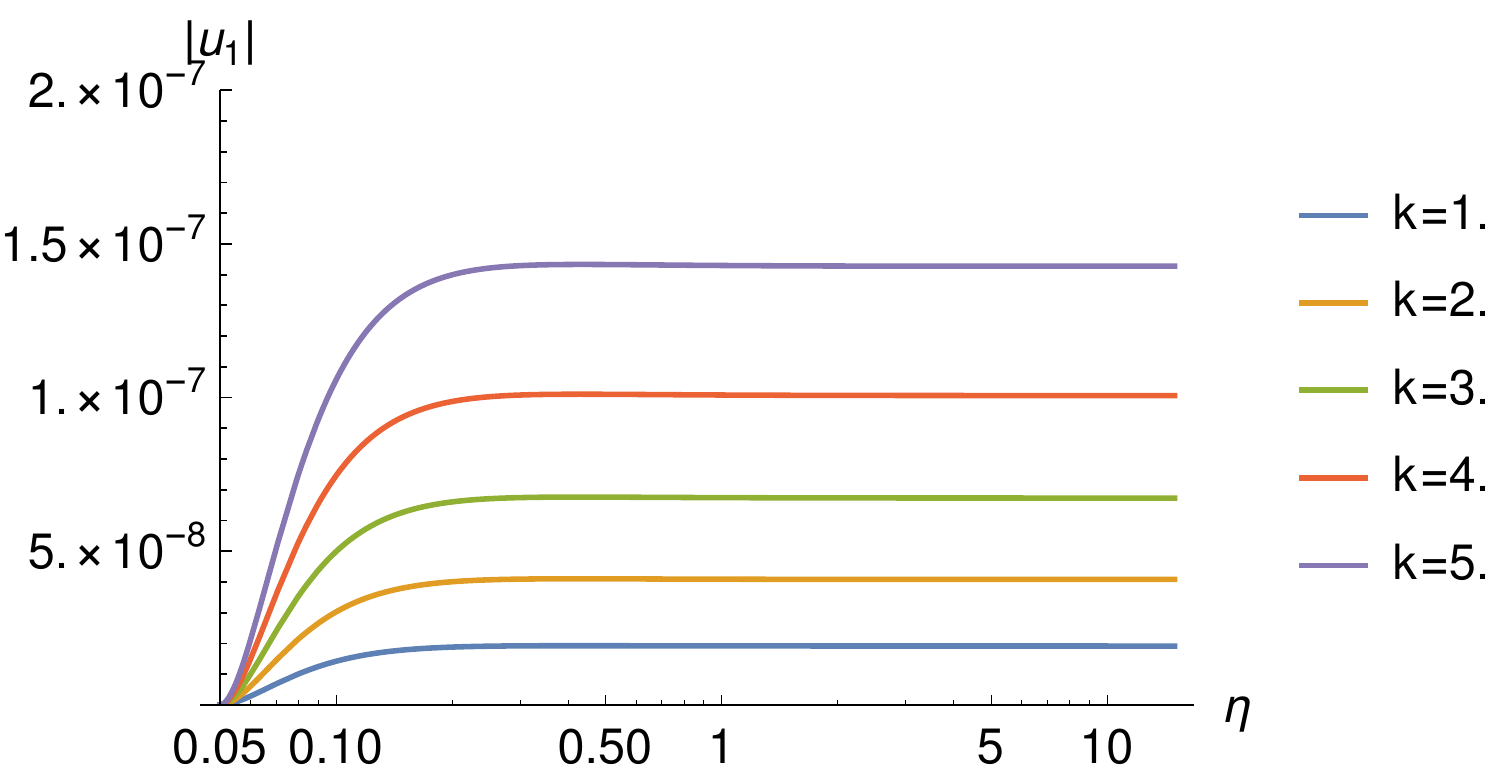}
  \end{center}
  \caption{Time evolution of the scalar modes $h^S_{11}=(-V^1+V^2+V^3)/P_0$ and $u^1=V^1/P_0$ excited by the tensor mode at different $k$, with the same initial condition as in FIG.\ref{tmodeini2}.}
  \label{tmodesini2}
\end{figure}

\begin{figure}[h]
  \begin{center}
  \includegraphics[width = 0.5\textwidth]{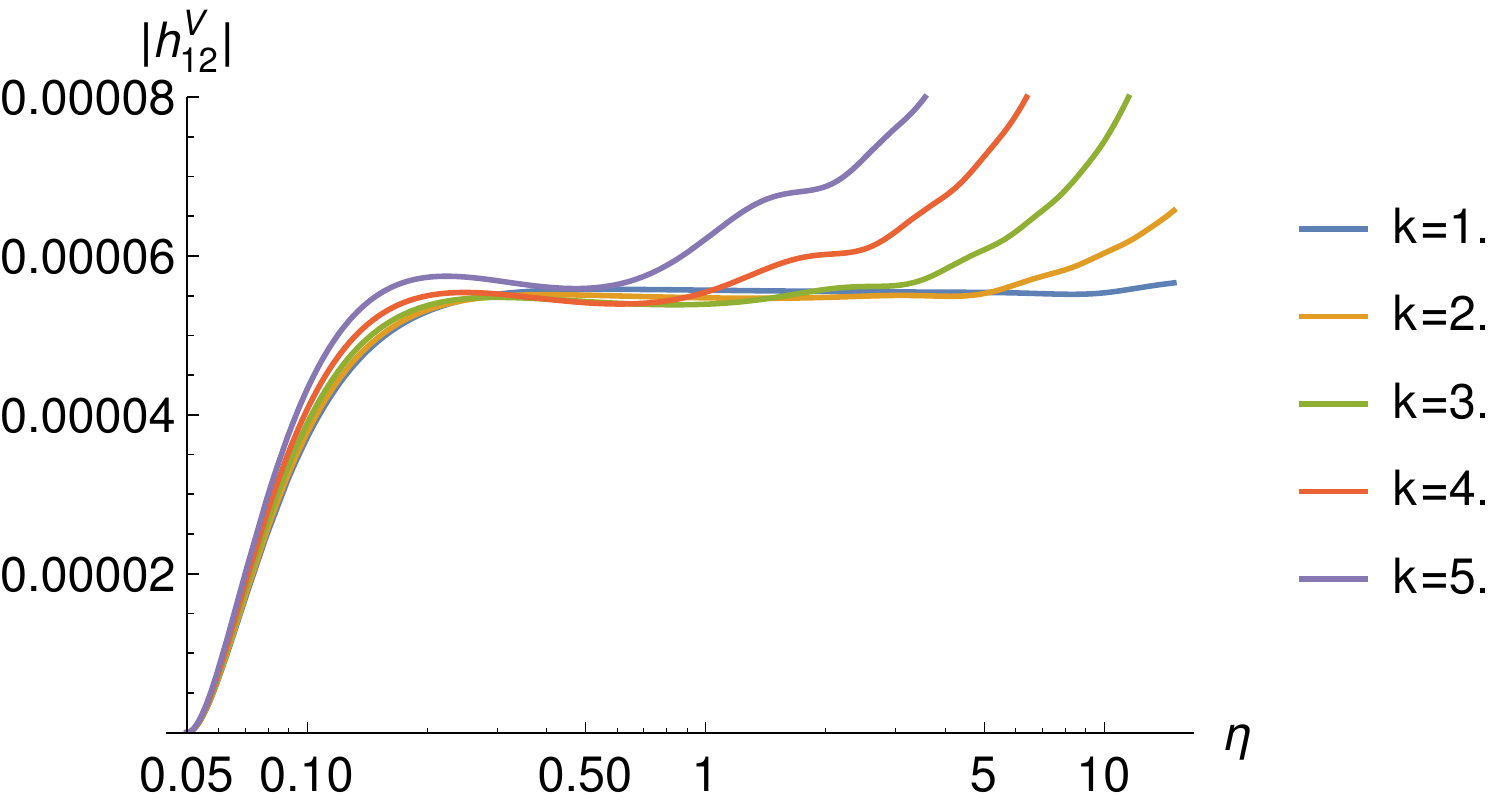}
  \end{center}
  \caption{Time evolution of the vector mode $h^V_{12}$ excited by the tensor mode at different $k$, with the same initial condition as in FIG.\ref{tmodeini2}.}
  \label{tmodevini2}
\end{figure}

Semiclassical EOMs couples tensor modes to scalar and vector modes when $\mu$ is finite. FIGs.\ref{tmodesini2} and \ref{tmodevini2} plot scalar mode perturbations $h^S_{11}=(-V^1+V^2+V^3)/P_0,\ u^1=V^1/P_0$ and vector mode perturbations $h^V_{12}$ (see Eq.\Ref{vectorhV}) excited by the tensor mode initial condition. Their amplitudes $|h^S_{11}|$, $|u_1|$, and $|h^V_{12}|$ are all less than $O(\mu)$, and suppressed by the lattice continuum limit $\mu\to0$. On the other hand, fixing the value of $\mu$, small effects from $\mu$ can accumulate and increase $|h^S_{11}|$, $|u_1|$, and $|h^V_{12}|$ when the evolution time is long. 

We note a different between the analysis here and in subsection \ref{Modified Graviton Dispersion Relation}: Here the tensor-mode initial condition is at early time, and there are scalar mode perturbations excited at late time, while in the discussion in subsection \ref{Modified Graviton Dispersion Relation}, we turn off scalar modes at late time. 

\section{Conclusion and Outlook}

In this work we derive the cosmological perturbation theory from the path integral formulation of the full LQG and the semiclassical approximation. In the lattice continuum limit, the result is consistent with the classical gravity-dust theory. Numerical studies of discrete semiclassical EOMs indicate some interesting corrections to power spectrums especially in the regime where wavelengths are very long. Our result provides a new routine of extracting physical predictions in cosmology from the full theory of LQG.

Our approach is a preliminary step toward relating LQG to observations, and at present has a few open issues which should be addressed in the future. These issues are summarized below:
\begin{enumerate} 

\item This work focuses on pure gravity coupling to dusts, while neglecting radiative matter. This work also doesn't take into account the inflation. We have to generalize our work to include these perspectives in order to make contact with observations of Cosmic Microwave Background (CMB). Fortunately, it is straight-forward to generalize the reduced phase space LQG to standard-model matter couplings \cite{Giesel:2007wn}. Deriving matter couplings in the path integral is a work currently undergoing. Therefore in the near future, we should be able to include the radiative matter and inflation in our analysis. The result should be compared with the recent work \cite{Giesel:2020bht}, where the inflationary cosmological perturbation theory is studied in the classical theory of gravity and matter coupling to dust. 

\item The initial state plays a crucial role in the cosmological perturbation theory. In above discussions, initial conditions of perturbations are translated from corresponding initial conditions in the classical continuum theory. We have neglect impacts on the initial condition of $O(\mu)$ from the discreteness and of $O(\ell_P^2)$ from quantum effects, while both of them are nontrivial at early time in cosmology. Therefore choices of initial states for cosmology, including their semiclassical and quantum properties, should be an important aspect to be understood in the future.

\end{enumerate}


\section*{Acknowledgements}

This work receives support from the National Science Foundation through grant PHY-1912278. Computations in this work is mainly carried out on the HPC server at Fudan University in China and the KoKo HPC server at Florida Atlantic University. The authors acknowledge Ling-Yan Hung for sharing the computational resource at Fudan University. 
	

\appendix

\section{${\bf U}_0{}^\rho_{\ \nu}(\t,{k})$ and ${\bf U}_1{}^\rho_{\ \nu}(\t,{k})$}\label{H0}

We expand the $18\times 18$ matrix ${\bf U}^\rho_{\ \nu}(\mu,\t,{k})$ as a power series in $\mu$
\be
{\bf U}^\rho_{\ \nu}(\mu,\t,{k})={\bf U}_0{}^\rho_{\ \nu}(\t,{k})+\mu\,{\bf U}_1{}^\rho_{\ \nu}(\t,{k})+O(\mu^2)
\ee
All nonzero matrix elements in ${\bf U}_0{}^\rho_{\ \nu}(\t,{k})$ are given by
{\tiny
\begin{align*}
({\bf U}_0)_{1,1}=\frac{2 K_0}{\sqrt{P_0}} & \qquad ({\bf U}_0)_{1,11}=\frac{2 \sqrt{P_0}}{\beta } \\
 ({\bf U}_0)_{1,12}=\frac{2 \sqrt{P_0}}{\beta } & \qquad ({\bf U}_0)_{2,2}=\frac{2 K_0}{\sqrt{P_0}} \\
 ({\bf U}_0)_{2,6}=\frac{2 i k \left(1+\frac{\alpha  \beta ^2 K_0^2}{3 K_0^2-\Lambda  P_0}\right)}{\beta  \sqrt{P_0}}
& \qquad ({\bf U}_0)_{2,9}=\frac{2 i k \alpha  \beta  K_0^2}{\sqrt{P_0} \left(-3 K_0^2+\Lambda  P_0\right)} \\
 ({\bf U}_0)_{2,10}=\frac{2 \sqrt{P_0}}{\beta } & \qquad ({\bf U}_0)_{2,11}=-\frac{2 k^2 \alpha  \sqrt{P_0}}{3 \beta  K_0^2-\beta  \Lambda  P_0}
\\
 ({\bf U}_0)_{2,12}=\frac{2 \sqrt{P_0} \left(1+\frac{k^2 \alpha }{-3 K_0^2+\Lambda  P_0}\right)}{\beta } & \qquad ({\bf U}_0)_{2,15}=-\frac{2 i
k \alpha  K_0 \sqrt{P_0}}{3 K_0^2-\Lambda  P_0} \\
 ({\bf U}_0)_{2,18}=\frac{2 i k \alpha  K_0 \sqrt{P_0}}{3 K_0^2-\Lambda  P_0} & \qquad ({\bf U}_0)_{3,3}=\frac{2 K_0}{\sqrt{P_0}}
\\
 ({\bf U}_0)_{3,6}=-\frac{2 i k \alpha  \beta  K_0^2}{\sqrt{P_0} \left(-3 K_0^2+\Lambda  P_0\right)} & \qquad ({\bf U}_0)_{3,9}=-\frac{2
i k \left(1+\frac{\alpha  \beta ^2 K_0^2}{3 K_0^2-\Lambda  P_0}\right)}{\beta  \sqrt{P_0}} \\
 ({\bf U}_0)_{3,10}=\frac{2 \sqrt{P_0}}{\beta } & \qquad ({\bf U}_0)_{3,11}=\frac{2 \sqrt{P_0} \left(1+\frac{k^2 \alpha }{-3 K_0^2+\Lambda  P_0}\right)}{\beta
} \\
 ({\bf U}_0)_{3,12}=-\frac{2 k^2 \alpha  \sqrt{P_0}}{3 \beta  K_0^2-\beta  \Lambda  P_0} & \qquad ({\bf U}_0)_{3,15}=-\frac{2 i k \alpha  K_0
\sqrt{P_0}}{3 K_0^2-\Lambda  P_0} \\
 ({\bf U}_0)_{3,18}=\frac{2 i k \alpha  K_0 \sqrt{P_0}}{3 K_0^2-\Lambda  P_0} & \qquad ({\bf U}_0)_{4,4}=\frac{2 K_0 \left(1+\frac{\alpha
 \beta ^2 K_0^2}{3 K_0^2-\Lambda  P_0}\right)}{\sqrt{P_0}} \\
 ({\bf U}_0)_{4,7}=\frac{2 \alpha  \beta ^2 K_0^3}{\sqrt{P_0} \left(-3 K_0^2+\Lambda  P_0\right)} & \qquad ({\bf U}_0)_{4,13}=-\frac{2
\alpha  \beta  K_0^2 \sqrt{P_0}}{3 K_0^2-\Lambda  P_0} \\
 ({\bf U}_0)_{4,16}=\frac{2 \sqrt{P_0} \left(\left((3+\alpha ) \beta ^2-3 \left(1+\beta ^2\right)\right) K_0^2+\Lambda  P_0\right)}{\beta
 \left(3 K_0^2-\Lambda  P_0\right)} & \qquad ({\bf U}_0)_{4,17}=-\frac{2 i k \alpha  K_0 \sqrt{P_0}}{3 K_0^2-\Lambda
 P_0} \\
 ({\bf U}_0)_{5,5}=\frac{2 K_0 \left(1+\frac{\alpha  \beta ^2 K_0^2}{3 K_0^2-\Lambda  P_0}\right)}{\sqrt{P_0}}
& \qquad ({\bf U}_0)_{5,8}=\frac{2 \alpha  \beta ^2 K_0^3}{\sqrt{P_0} \left(-3 K_0^2+\Lambda  P_0\right)} \\
 ({\bf U}_0)_{5,14}=-\frac{2 \alpha  \beta  K_0^2 \sqrt{P_0}}{3 K_0^2-\Lambda  P_0} & \qquad ({\bf U}_0)_{5,16}=\frac{2 i k \alpha  K_0
\sqrt{P_0}}{3 K_0^2-\Lambda  P_0} \\
 ({\bf U}_0)_{5,17}=\frac{2 \sqrt{P_0} \left(\left((3+\alpha ) \beta ^2-3 \left(1+\beta ^2\right)\right) K_0^2+\Lambda  P_0\right)}{\beta
 \left(3 K_0^2-\Lambda  P_0\right)} & \qquad ({\bf U}_0)_{6,1}=-\frac{i k}{\beta  \sqrt{P_0}} \\
 ({\bf U}_0)_{6,2}=-\frac{i k}{\beta  \sqrt{P_0}} & \qquad ({\bf U}_0)_{6,3}=\frac{i k}{\beta  \sqrt{P_0}} \\
 ({\bf U}_0)_{6,6}=\frac{2 K_0 \left(1+\frac{\alpha  \beta ^2 K_0^2}{3 K_0^2-\Lambda  P_0}\right)}{\sqrt{P_0}}
& \qquad ({\bf U}_0)_{6,9}=\frac{2 \alpha  \beta ^2 K_0^3}{\sqrt{P_0} \left(-3 K_0^2+\Lambda  P_0\right)} \\
 ({\bf U}_0)_{6,11}=\frac{2 i k \alpha  K_0 \sqrt{P_0}}{3 K_0^2-\Lambda  P_0} & \qquad ({\bf U}_0)_{6,12}=\frac{2 i k \alpha  K_0
\sqrt{P_0}}{3 K_0^2-\Lambda  P_0} \\
 ({\bf U}_0)_{6,15}=\frac{2 \alpha  \beta  K_0^2 \sqrt{P_0}}{-3 K_0^2+\Lambda  P_0} & \qquad ({\bf U}_0)_{6,18}=\frac{2 \sqrt{P_0}
\left(\left((3+\alpha ) \beta ^2-3 \left(1+\beta ^2\right)\right) K_0^2+\Lambda  P_0\right)}{\beta  \left(3 K_0^2-\Lambda
 P_0\right)} \\
 ({\bf U}_0)_{7,4}=\frac{2 \alpha  \beta ^2 K_0^3}{\sqrt{P_0} \left(-3 K_0^2+\Lambda  P_0\right)} & \qquad ({\bf U}_0)_{7,5}=\frac{2 i
k \left(1+\frac{\alpha  \beta ^2 K_0^2}{3 K_0^2-\Lambda  P_0}\right)}{\beta  \sqrt{P_0}} \\
 ({\bf U}_0)_{7,7}=\frac{2 K_0 \left(1+\frac{\alpha  \beta ^2 K_0^2}{3 K_0^2-\Lambda  P_0}\right)}{\sqrt{P_0}}
& \qquad ({\bf U}_0)_{7,8}=\frac{2 i k \alpha  \beta  K_0^2}{\sqrt{P_0} \left(-3 K_0^2+\Lambda  P_0\right)} \\
 ({\bf U}_0)_{7,13}=\frac{2 \sqrt{P_0} \left(\left((3+\alpha ) \beta ^2-3 \left(1+\beta ^2\right)\right) K_0^2+\Lambda  P_0\right)}{\beta
 \left(3 K_0^2-\Lambda  P_0\right)} & \qquad ({\bf U}_0)_{7,14}=-\frac{2 i k \alpha  K_0 \sqrt{P_0}}{3 K_0^2-\Lambda
 P_0} \\
 ({\bf U}_0)_{7,16}=-\frac{2 \alpha  \left(k^2+\beta ^2 K_0^2\right) \sqrt{P_0}}{\beta  \left(3 K_0^2-\Lambda  P_0\right)}
& \qquad ({\bf U}_0)_{7,17}=\frac{4 i k \alpha  K_0 \sqrt{P_0}}{3 K_0^2-\Lambda  P_0} \\
 ({\bf U}_0)_{8,4}=-\frac{2 i k \left(1+\frac{\alpha  \beta ^2 K_0^2}{3 K_0^2-\Lambda  P_0}\right)}{\beta  \sqrt{P_0}}
& \qquad ({\bf U}_0)_{8,5}=\frac{2 \alpha  \beta ^2 K_0^3}{\sqrt{P_0} \left(-3 K_0^2+\Lambda  P_0\right)} \\
 ({\bf U}_0)_{8,7}=-\frac{2 i k \alpha  \beta  K_0^2}{\sqrt{P_0} \left(-3 K_0^2+\Lambda  P_0\right)} & \qquad ({\bf U}_0)_{8,8}=\frac{2
K_0 \left(1+\frac{\alpha  \beta ^2 K_0^2}{3 K_0^2-\Lambda  P_0}\right)}{\sqrt{P_0}} \\
 ({\bf U}_0)_{8,13}=\frac{2 i k \alpha  K_0 \sqrt{P_0}}{3 K_0^2-\Lambda  P_0} & \qquad ({\bf U}_0)_{8,14}=\frac{2 \sqrt{P_0}
\left(\left((3+\alpha ) \beta ^2-3 \left(1+\beta ^2\right)\right) K_0^2+\Lambda  P_0\right)}{\beta  \left(3 K_0^2-\Lambda
 P_0\right)} \\
 ({\bf U}_0)_{8,16}=-\frac{4 i k \alpha  K_0 \sqrt{P_0}}{3 K_0^2-\Lambda  P_0} & \qquad ({\bf U}_0)_{8,17}=-\frac{2 \alpha  \left(k^2+\beta
^2 K_0^2\right) \sqrt{P_0}}{\beta  \left(3 K_0^2-\Lambda  P_0\right)} \\
 ({\bf U}_0)_{9,1}=\frac{i k}{\beta  \sqrt{P_0}} & \qquad ({\bf U}_0)_{9,2}=-\frac{i k}{\beta  \sqrt{P_0}} \\
 ({\bf U}_0)_{9,3}=\frac{i k}{\beta  \sqrt{P_0}} & \qquad ({\bf U}_0)_{9,6}=\frac{2 \alpha  \beta ^2 K_0^3}{\sqrt{P_0} \left(-3 K_0^2+\Lambda
 P_0\right)} \\
 ({\bf U}_0)_{9,9}=\frac{2 K_0 \left(1+\frac{\alpha  \beta ^2 K_0^2}{3 K_0^2-\Lambda  P_0}\right)}{\sqrt{P_0}}
& \qquad ({\bf U}_0)_{9,11}=-\frac{2 i k \alpha  K_0 \sqrt{P_0}}{3 K_0^2-\Lambda  P_0} \\
 ({\bf U}_0)_{9,12}=-\frac{2 i k \alpha  K_0 \sqrt{P_0}}{3 K_0^2-\Lambda  P_0} & \qquad ({\bf U}_0)_{9,15}=\frac{2 \sqrt{P_0}
\left(\left((3+\alpha ) \beta ^2-3 \left(1+\beta ^2\right)\right) K_0^2+\Lambda  P_0\right)}{\beta  \left(3 K_0^2-\Lambda
 P_0\right)} \\
 ({\bf U}_0)_{9,18}=\frac{2 \alpha  \beta  K_0^2 \sqrt{P_0}}{-3 K_0^2+\Lambda  P_0} & \qquad ({\bf U}_0)_{10,1}=\frac{-2 k^2 \left(1+\beta
^2\right)-\beta ^2 K_0^2-\beta ^2 \Lambda  P_0}{2 \beta  P_0^{3/2}} \\
 ({\bf U}_0)_{10,2}=\frac{\beta  \left(K_0^2+\Lambda  P_0\right)}{2 P_0^{3/2}} & \qquad ({\bf U}_0)_{10,3}=\frac{\beta  \left(K_0^2+\Lambda
 P_0\right)}{2 P_0^{3/2}} \\
 ({\bf U}_0)_{10,10}=-\frac{2 K_0}{\sqrt{P_0}} & \qquad ({\bf U}_0)_{10,15}=-\frac{i k}{\beta  \sqrt{P_0}} \\
 ({\bf U}_0)_{10,18}=\frac{i k}{\beta  \sqrt{P_0}} & \qquad ({\bf U}_0)_{11,1}=\frac{\beta  \left(K_0^2+\Lambda  P_0\right)}{2 P_0^{3/2}}
\\
 ({\bf U}_0)_{11,2}=\frac{2 k^2 \left(1+\beta ^2\right)-\beta ^2 K_0^2-\beta ^2 \Lambda  P_0}{2 \beta  P_0^{3/2}} & \qquad ({\bf U}_0)_{11,3}=\frac{-2
k^2 \left(1+\beta ^2\right)+\beta ^2 K_0^2+\beta ^2 \Lambda  P_0}{2 \beta  P_0^{3/2}} \\
 ({\bf U}_0)_{11,11}=-\frac{2 K_0}{\sqrt{P_0}} & \qquad ({\bf U}_0)_{11,15}=-\frac{i k}{\beta  \sqrt{P_0}} \\
 ({\bf U}_0)_{11,18}=-\frac{i k}{\beta  \sqrt{P_0}} & \qquad ({\bf U}_0)_{12,1}=\frac{\beta  \left(K_0^2+\Lambda  P_0\right)}{2 P_0^{3/2}}
\\
 ({\bf U}_0)_{12,2}=\frac{-2 k^2 \left(1+\beta ^2\right)+\beta ^2 K_0^2+\beta ^2 \Lambda  P_0}{2 \beta  P_0^{3/2}} & \qquad ({\bf U}_0)_{12,3}=\frac{2
k^2 \left(1+\beta ^2\right)-\beta ^2 K_0^2-\beta ^2 \Lambda  P_0}{2 \beta  P_0^{3/2}} \\
 ({\bf U}_0)_{12,12}=-\frac{2 K_0}{\sqrt{P_0}} & \qquad ({\bf U}_0)_{12,15}=\frac{i k}{\beta  \sqrt{P_0}} \\
 ({\bf U}_0)_{12,18}=\frac{i k}{\beta  \sqrt{P_0}} & \qquad ({\bf U}_0)_{13,4}=-\frac{2 \alpha  \beta ^3 K_0^4}{P_0^{3/2} \left(-3 K_0^2+\Lambda
 P_0\right)} \\
 ({\bf U}_0)_{13,7}=\frac{\beta  \left(\left((-3+2 \alpha ) \beta ^2+3 \left(1+\beta ^2\right)\right) K_0^4+2 \Lambda  K_0^2 P_0-\Lambda
^2 P_0^2\right)}{P_0^{3/2} \left(-3 K_0^2+\Lambda  P_0\right)} & \qquad ({\bf U}_0)_{13,13}=\frac{2 \left((-3+\alpha ) \beta
^2+3 \left(1+\beta ^2\right)\right) K_0^3-2 \Lambda  K_0 P_0}{\sqrt{P_0} \left(-3 K_0^2+\Lambda  P_0\right)}
\\
 ({\bf U}_0)_{13,16}=\frac{2 \alpha  \beta ^2 K_0^3}{\sqrt{P_0} \left(3 K_0^2-\Lambda  P_0\right)} & \qquad ({\bf U}_0)_{13,17}=-\frac{2
i k \left(1+\frac{\alpha  \beta ^2 K_0^2}{3 K_0^2-\Lambda  P_0}\right)}{\beta  \sqrt{P_0}} \\
 ({\bf U}_0)_{14,5}=-\frac{2 \alpha  \beta ^3 K_0^4}{P_0^{3/2} \left(-3 K_0^2+\Lambda  P_0\right)} & \qquad ({\bf U}_0)_{14,8}=\frac{\beta
 \left(\left((-3+2 \alpha ) \beta ^2+3 \left(1+\beta ^2\right)\right) K_0^4+2 \Lambda  K_0^2 P_0-\Lambda ^2 P_0^2\right)}{P_0^{3/2}
\left(-3 K_0^2+\Lambda  P_0\right)} \\
 ({\bf U}_0)_{14,14}=\frac{2 \left((-3+\alpha ) \beta ^2+3 \left(1+\beta ^2\right)\right) K_0^3-2 \Lambda  K_0 P_0}{\sqrt{P_0}
\left(-3 K_0^2+\Lambda  P_0\right)} & \qquad ({\bf U}_0)_{14,16}=\frac{2 i k \left(1+\frac{\alpha  \beta ^2 K_0^2}{3 K_0^2-\Lambda
 P_0}\right)}{\beta  \sqrt{P_0}} \\
 ({\bf U}_0)_{14,17}=\frac{2 \alpha  \beta ^2 K_0^3}{\sqrt{P_0} \left(3 K_0^2-\Lambda  P_0\right)} & \qquad ({\bf U}_0)_{15,6}=\frac{k^2
\left(1+\beta ^2\right)-\frac{2 \alpha  \beta ^4 K_0^4}{-3 K_0^2+\Lambda  P_0}}{\beta  P_0^{3/2}} \\
 ({\bf U}_0)_{15,9}=\frac{\beta ^4 K_0^2+\left(1+\beta ^2\right) \left(k^2-\beta ^2 K_0^2\right)-\beta ^2 \Lambda  P_0+\frac{2
\alpha  \beta ^4 K_0^4}{-3 K_0^2+\Lambda  P_0}}{\beta  P_0^{3/2}} & \qquad ({\bf U}_0)_{15,11}=\frac{2 i k \left(1+\frac{\alpha
 \beta ^2 K_0^2}{3 K_0^2-\Lambda  P_0}\right)}{\beta  \sqrt{P_0}} \\
 ({\bf U}_0)_{15,12}=-\frac{2 i k \alpha  \beta  K_0^2}{\sqrt{P_0} \left(-3 K_0^2+\Lambda  P_0\right)} & \qquad ({\bf U}_0)_{15,15}=\frac{2
K_0 \left(-1+\frac{\alpha  \beta ^2 K_0^2}{-3 K_0^2+\Lambda  P_0}\right)}{\sqrt{P_0}} \\
 ({\bf U}_0)_{15,18}=\frac{2 \alpha  \beta ^2 K_0^3}{\sqrt{P_0} \left(3 K_0^2-\Lambda  P_0\right)} & \qquad ({\bf U}_0)_{16,4}=\frac{\beta
 \left(\left((-3+2 \alpha ) \beta ^2+3 \left(1+\beta ^2\right)\right) K_0^4+2 \Lambda  K_0^2 P_0-\Lambda ^2 P_0^2\right)}{P_0^{3/2}
\left(-3 K_0^2+\Lambda  P_0\right)} \\
 ({\bf U}_0)_{16,7}=-\frac{2 \alpha  \beta ^3 K_0^4}{P_0^{3/2} \left(-3 K_0^2+\Lambda  P_0\right)} & \qquad ({\bf U}_0)_{16,13}=\frac{2
\alpha  \beta ^2 K_0^3}{\sqrt{P_0} \left(3 K_0^2-\Lambda  P_0\right)} \\
 ({\bf U}_0)_{16,16}=\frac{2 K_0 \left(-1+\frac{\alpha  \beta ^2 K_0^2}{-3 K_0^2+\Lambda  P_0}\right)}{\sqrt{P_0}}
& \qquad ({\bf U}_0)_{16,17}=-\frac{2 i k \alpha  \beta  K_0^2}{\sqrt{P_0} \left(-3 K_0^2+\Lambda  P_0\right)} \\
 ({\bf U}_0)_{17,5}=\frac{\beta  \left(\left((-3+2 \alpha ) \beta ^2+3 \left(1+\beta ^2\right)\right) K_0^4+2 \Lambda  K_0^2 P_0-\Lambda
^2 P_0^2\right)}{P_0^{3/2} \left(-3 K_0^2+\Lambda  P_0\right)} & \qquad ({\bf U}_0)_{17,8}=-\frac{2 \alpha  \beta ^3 K_0^4}{P_0^{3/2}
\left(-3 K_0^2+\Lambda  P_0\right)} \\
 ({\bf U}_0)_{17,14}=\frac{2 \alpha  \beta ^2 K_0^3}{\sqrt{P_0} \left(3 K_0^2-\Lambda  P_0\right)} & \qquad ({\bf U}_0)_{17,16}=\frac{2
i k \alpha  \beta  K_0^2}{\sqrt{P_0} \left(-3 K_0^2+\Lambda  P_0\right)} \\
 ({\bf U}_0)_{17,17}=\frac{2 K_0 \left(-1+\frac{\alpha  \beta ^2 K_0^2}{-3 K_0^2+\Lambda  P_0}\right)}{\sqrt{P_0}}
& \qquad ({\bf U}_0)_{18,6}=\frac{\beta ^4 K_0^2+\left(1+\beta ^2\right) \left(k^2-\beta ^2 K_0^2\right)-\beta ^2 \Lambda  P_0+\frac{2
\alpha  \beta ^4 K_0^4}{-3 K_0^2+\Lambda  P_0}}{\beta  P_0^{3/2}} \\
 ({\bf U}_0)_{18,9}=\frac{k^2 \left(1+\beta ^2\right)-\frac{2 \alpha  \beta ^4 K_0^4}{-3 K_0^2+\Lambda  P_0}}{\beta  P_0^{3/2}}
& \qquad ({\bf U}_0)_{18,11}=\frac{2 i k \alpha  \beta  K_0^2}{\sqrt{P_0} \left(-3 K_0^2+\Lambda  P_0\right)} \\
 ({\bf U}_0)_{18,12}=-\frac{2 i k \left(1+\frac{\alpha  \beta ^2 K_0^2}{3 K_0^2-\Lambda  P_0}\right)}{\beta  \sqrt{P_0}}
& \qquad ({\bf U}_0)_{18,15}=\frac{2 \alpha  \beta ^2 K_0^3}{\sqrt{P_0} \left(3 K_0^2-\Lambda  P_0\right)} \\
\end{align*}}
All nonzero matrix elements in ${\bf U}_1{}^\rho_{\ \nu}(\t,{k})$ are given by
{\tiny
\begin{align*}
 ({\bf U}_1)_{1,11}=\frac{i k \sqrt{P_0}}{2 \beta } & \qquad ({\bf U}_1)_{1,12}=\frac{i k \sqrt{P_0}}{2 \beta } \\
 ({\bf U}_1)_{2,7}=-\frac{i k K_0 \left(\left(3+\alpha  \beta ^2\right) K_0^2-\Lambda  P_0\right)}{2 \sqrt{P_0} \left(-3
K_0^2+\Lambda  P_0\right)} & \qquad ({\bf U}_1)_{2,8}=\frac{i k \alpha  \beta ^2 K_0^3}{\sqrt{P_0} \left(6 K_0^2-2 \Lambda
 P_0\right)} \\
 ({\bf U}_1)_{2,10}=-\frac{i k \sqrt{P_0}}{2 \beta } & \qquad ({\bf U}_1)_{2,16}=\frac{i k \sqrt{P_0} \left(-\left(3+\alpha  \beta ^2\right) K_0^2+\Lambda
 P_0\right)}{6 \beta  K_0^2-2 \beta  \Lambda  P_0} \\
 ({\bf U}_1)_{2,17}=-\frac{i k \alpha  \beta  K_0^2 \sqrt{P_0}}{6 K_0^2-2 \Lambda  P_0} & \qquad ({\bf U}_1)_{3,7}=\frac{i k \alpha  \beta
^2 K_0^3}{\sqrt{P_0} \left(6 K_0^2-2 \Lambda  P_0\right)} \\
 ({\bf U}_1)_{3,8}=-\frac{i k K_0 \left(\left(3+\alpha  \beta ^2\right) K_0^2-\Lambda  P_0\right)}{2 \sqrt{P_0} \left(-3
K_0^2+\Lambda  P_0\right)} & \qquad ({\bf U}_1)_{3,10}=-\frac{i k \sqrt{P_0}}{2 \beta } \\
 ({\bf U}_1)_{3,16}=-\frac{i k \alpha  \beta  K_0^2 \sqrt{P_0}}{6 K_0^2-2 \Lambda  P_0} & \qquad ({\bf U}_1)_{3,17}=\frac{i k \sqrt{P_0}
\left(-\left(3+\alpha  \beta ^2\right) K_0^2+\Lambda  P_0\right)}{6 \beta  K_0^2-2 \beta  \Lambda  P_0} \\
 ({\bf U}_1)_{4,5}=-\frac{\beta  \left(\left(-3+2 (-3+\alpha ) \beta ^2\right) K_0^4+2 \left(-1+\beta ^2\right) \Lambda  K_0^2 P_0+\Lambda
^2 P_0^2\right)}{4 \sqrt{P_0} \left(-3 K_0^2+\Lambda  P_0\right)} & \qquad ({\bf U}_1)_{4,6}=\frac{\alpha  \beta ^3 K_0^4}{\sqrt{P_0}
\left(6 K_0^2-2 \Lambda  P_0\right)} \\
 ({\bf U}_1)_{4,7}=\frac{i k \alpha  \beta ^2 K_0^3}{2 \sqrt{P_0} \left(-3 K_0^2+\Lambda  P_0\right)} & \qquad ({\bf U}_1)_{4,8}=-\frac{\beta
 \left(K_0^2+\Lambda  P_0\right)}{4 \sqrt{P_0}} \\
 ({\bf U}_1)_{4,14}=\frac{1}{2} K_0 \sqrt{P_0} \left(-2 \left(1+\beta ^2\right)+\frac{\alpha  \beta ^2 K_0^2}{-3 K_0^2+\Lambda
 P_0}\right) & \qquad ({\bf U}_1)_{4,15}=-\frac{\alpha  \beta ^2 K_0^3 \sqrt{P_0}}{6 K_0^2-2 \Lambda  P_0} \\
 ({\bf U}_1)_{4,16}=\frac{i k \alpha  \beta  K_0^2 \sqrt{P_0}}{6 K_0^2-2 \Lambda  P_0} & \qquad ({\bf U}_1)_{4,17}=\frac{K_0
\sqrt{P_0} \left(k^2 \alpha +3 K_0^2-\Lambda  P_0\right)}{6 K_0^2-2 \Lambda  P_0} \\
 ({\bf U}_1)_{5,4}=\frac{\beta  \left(\left(-3+2 (-3+\alpha ) \beta ^2\right) K_0^4+2 \left(-1+\beta ^2\right) \Lambda  K_0^2 P_0+\Lambda
^2 P_0^2\right)}{4 \sqrt{P_0} \left(-3 K_0^2+\Lambda  P_0\right)} & \qquad ({\bf U}_1)_{5,7}=\frac{\beta  \left(K_0^2+\Lambda
 P_0\right)}{4 \sqrt{P_0}} \\
 ({\bf U}_1)_{5,8}=\frac{i k \alpha  \beta ^2 K_0^3}{2 \sqrt{P_0} \left(-3 K_0^2+\Lambda  P_0\right)} & \qquad ({\bf U}_1)_{5,9}=\frac{\alpha
 \beta ^3 K_0^4}{2 \sqrt{P_0} \left(-3 K_0^2+\Lambda  P_0\right)} \\
 ({\bf U}_1)_{5,13}=\frac{1}{2} K_0 \sqrt{P_0} \left(2+2 \beta ^2+\frac{\alpha  \beta ^2 K_0^2}{3 K_0^2-\Lambda  P_0}\right)
& \qquad ({\bf U}_1)_{5,16}=\frac{K_0 \sqrt{P_0} \left(-k^2 \alpha -3 K_0^2+\Lambda  P_0\right)}{6 K_0^2-2 \Lambda 
P_0} \\
 ({\bf U}_1)_{5,17}=\frac{i k \alpha  \beta  K_0^2 \sqrt{P_0}}{6 K_0^2-2 \Lambda  P_0} & \qquad ({\bf U}_1)_{5,18}=\frac{\alpha  \beta
^2 K_0^3 \sqrt{P_0}}{6 K_0^2-2 \Lambda  P_0} \\
 ({\bf U}_1)_{6,1}=-\frac{k^2}{4 \beta  \sqrt{P_0}} & \qquad ({\bf U}_1)_{6,4}=-\frac{\beta  \left(K_0^2+\Lambda  P_0\right)}{4 \sqrt{P_0}}
\\
 ({\bf U}_1)_{6,5}=\frac{i k K_0 \left(\left(3+\alpha  \beta ^2\right) K_0^2-\Lambda  P_0\right)}{2 \sqrt{P_0} \left(-3
K_0^2+\Lambda  P_0\right)} & \qquad ({\bf U}_1)_{6,7}=-\frac{\beta  \left(\left(-3+2 (-3+\alpha ) \beta ^2\right) K_0^4+2 \left(-1+\beta
^2\right) \Lambda  K_0^2 P_0+\Lambda ^2 P_0^2\right)}{4 \sqrt{P_0} \left(-3 K_0^2+\Lambda  P_0\right)}
\\
 ({\bf U}_1)_{6,8}=-\frac{\alpha  \beta ^2 K_0^3 (i k+\beta  K_0)}{2 \sqrt{P_0} \left(-3 K_0^2+\Lambda  P_0\right)}
& \qquad ({\bf U}_1)_{6,13}=\frac{1}{2} K_0 \sqrt{P_0} \\
 ({\bf U}_1)_{6,14}=\frac{i k \alpha  \beta  K_0^2 \sqrt{P_0}}{6 K_0^2-2 \Lambda  P_0} & \qquad ({\bf U}_1)_{6,16}=\frac{1}{2} K_0
\sqrt{P_0} \left(-2 \left(1+\beta ^2\right)+\frac{\alpha  \left(k^2-\beta ^2 K_0^2\right)}{3 K_0^2-\Lambda  P_0}\right)
\\
 ({\bf U}_1)_{6,17}=-\frac{\alpha  \beta  K_0^2 (i k+\beta  K_0) \sqrt{P_0}}{6 K_0^2-2 \Lambda  P_0} & \qquad ({\bf U}_1)_{7,1}=-\frac{i
k K_0}{4 \sqrt{P_0}} \\
 ({\bf U}_1)_{7,2}=-\frac{i k K_0}{4 \sqrt{P_0}} & \qquad ({\bf U}_1)_{7,3}=\frac{i k K_0}{4 \sqrt{P_0}} \\
 ({\bf U}_1)_{7,4}=-\frac{i k K_0}{2 \sqrt{P_0}} & \qquad ({\bf U}_1)_{7,5}=\frac{-k^2 \left(3+\alpha  \beta ^2\right) K_0^2+\alpha  \beta ^4 K_0^4+k^2
\Lambda  P_0}{2 \beta  \sqrt{P_0} \left(-3 K_0^2+\Lambda  P_0\right)} \\
 ({\bf U}_1)_{7,6}=\frac{-k^2 \left(1+\beta ^2\right)+\left(\beta ^2+2 \beta ^4\right) K_0^2+\beta ^2 \Lambda  P_0+\frac{2 \alpha  \beta
^4 K_0^4}{-3 K_0^2+\Lambda  P_0}}{4 \beta  \sqrt{P_0}} & \qquad ({\bf U}_1)_{7,9}=\frac{-k^2 \left(1+\beta ^2\right)+\beta ^2
\left(K_0^2+\Lambda  P_0+\frac{2 i k \alpha  \beta  K_0^3}{-3 K_0^2+\Lambda  P_0}\right)}{4 \beta  \sqrt{P_0}}
\\
 ({\bf U}_1)_{7,11}=-\frac{i k \sqrt{P_0}}{2 \beta } & \qquad ({\bf U}_1)_{7,13}=\frac{i k \sqrt{P_0}}{\beta } \\
 ({\bf U}_1)_{7,14}=\frac{\alpha  K_0 \left(-k^2+\beta ^2 K_0^2\right) \sqrt{P_0}}{6 K_0^2-2 \Lambda  P_0} & \qquad
({\bf U}_1)_{7,15}=\frac{1}{2} K_0 \sqrt{P_0} \left(2+2 \beta ^2+\frac{\alpha  \beta ^2 K_0^2}{3 K_0^2-\Lambda  P_0}\right)
\\
 ({\bf U}_1)_{7,18}=\frac{K_0 \sqrt{P_0} \left(i k \alpha  \beta  K_0-3 K_0^2+\Lambda  P_0\right)}{6 K_0^2-2
\Lambda  P_0} & \qquad ({\bf U}_1)_{8,1}=-\frac{i k K_0}{4 \sqrt{P_0}} \\
 ({\bf U}_1)_{8,2}=\frac{i k K_0}{4 \sqrt{P_0}} & \qquad ({\bf U}_1)_{8,3}=-\frac{i k K_0}{4 \sqrt{P_0}} \\
 ({\bf U}_1)_{8,4}=\frac{k^2 \left(3+\alpha  \beta ^2\right) K_0^2-\alpha  \beta ^4 K_0^4-k^2 \Lambda  P_0}{2 \beta  \sqrt{P_0}
\left(-3 K_0^2+\Lambda  P_0\right)} & \qquad ({\bf U}_1)_{8,5}=-\frac{i k K_0}{2 \sqrt{P_0}} \\
 ({\bf U}_1)_{8,6}=\frac{k^2 \left(1+\beta ^2\right)+\beta ^2 \left(-\Lambda  P_0+K_0^2 \left(-1+\frac{2 i k \alpha  \beta  K_0}{-3
K_0^2+\Lambda  P_0}\right)\right)}{4 \beta  \sqrt{P_0}} & \qquad ({\bf U}_1)_{8,9}=\frac{k^2 \left(1+\beta ^2\right)-\left(\beta ^2+2
\beta ^4\right) K_0^2-\beta ^2 \Lambda  P_0-\frac{2 \alpha  \beta ^4 K_0^4}{-3 K_0^2+\Lambda  P_0}}{4
\beta  \sqrt{P_0}} \\
 ({\bf U}_1)_{8,12}=-\frac{i k \sqrt{P_0}}{2 \beta } & \qquad ({\bf U}_1)_{8,13}=\frac{\alpha  K_0 \left(k^2-\beta ^2 K_0^2\right) \sqrt{P_0}}{6
K_0^2-2 \Lambda  P_0} \\
 ({\bf U}_1)_{8,14}=\frac{i k \sqrt{P_0}}{\beta } & \qquad ({\bf U}_1)_{8,15}=\frac{K_0 \sqrt{P_0} \left(i k \alpha  \beta  K_0+3 K_0^2-\Lambda
 P_0\right)}{6 K_0^2-2 \Lambda  P_0} \\
 ({\bf U}_1)_{8,18}=\frac{1}{2} K_0 \sqrt{P_0} \left(-2 \left(1+\beta ^2\right)+\frac{\alpha  \beta ^2 K_0^2}{-3 K_0^2+\Lambda
 P_0}\right) & \qquad ({\bf U}_1)_{9,1}=\frac{k^2}{4 \beta  \sqrt{P_0}} \\
 ({\bf U}_1)_{9,4}=\frac{i k K_0 \left(\left(3+\alpha  \beta ^2\right) K_0^2-\Lambda  P_0\right)}{2 \sqrt{P_0} \left(-3
K_0^2+\Lambda  P_0\right)} & \qquad ({\bf U}_1)_{9,5}=\frac{\beta  \left(K_0^2+\Lambda  P_0\right)}{4 \sqrt{P_0}} \\
 ({\bf U}_1)_{9,7}=\frac{\alpha  \beta ^2 K_0^3 (-i k+\beta  K_0)}{2 \sqrt{P_0} \left(-3 K_0^2+\Lambda  P_0\right)}
& \qquad ({\bf U}_1)_{9,8}=\frac{\beta  \left(\left(-3+2 (-3+\alpha ) \beta ^2\right) K_0^4+2 \left(-1+\beta ^2\right) \Lambda  K_0^2 P_0+\Lambda
^2 P_0^2\right)}{4 \sqrt{P_0} \left(-3 K_0^2+\Lambda  P_0\right)} \\
 ({\bf U}_1)_{9,13}=\frac{i k \alpha  \beta  K_0^2 \sqrt{P_0}}{6 K_0^2-2 \Lambda  P_0} & \qquad ({\bf U}_1)_{9,14}=-\frac{1}{2} K_0
\sqrt{P_0} \\
 ({\bf U}_1)_{9,16}=\frac{\alpha  \beta  K_0^2 (-i k+\beta  K_0) \sqrt{P_0}}{6 K_0^2-2 \Lambda  P_0} & \qquad ({\bf U}_1)_{9,17}=\frac{1}{2}
K_0 \sqrt{P_0} \left(2 \left(1+\beta ^2\right)+\frac{\alpha  \left(k^2-\beta ^2 K_0^2\right)}{-3 K_0^2+\Lambda
 P_0}\right) \\
 ({\bf U}_1)_{10,2}=\frac{i k \beta  \left(K_0^2+\Lambda  P_0\right)}{8 P_0^{3/2}} & \qquad ({\bf U}_1)_{10,3}=\frac{i k \beta  \left(K_0^2+\Lambda
 P_0\right)}{8 P_0^{3/2}} \\
 ({\bf U}_1)_{10,15}=\frac{k^2}{4 \beta  \sqrt{P_0}} & \qquad ({\bf U}_1)_{10,18}=-\frac{k^2}{4 \beta  \sqrt{P_0}} \\
 ({\bf U}_1)_{11,1}=-\frac{i k \beta  \left(K_0^2+\Lambda  P_0\right)}{8 P_0^{3/2}} & \qquad ({\bf U}_1)_{12,1}=-\frac{i k \beta  \left(K_0^2+\Lambda
 P_0\right)}{8 P_0^{3/2}} \\
 ({\bf U}_1)_{13,5}=\frac{\alpha  \beta ^4 K_0^5}{P_0^{3/2} \left(6 K_0^2-2 \Lambda  P_0\right)} & \qquad ({\bf U}_1)_{13,6}=-\frac{\beta
^2 K_0 \left(\left(3+2 \alpha  \beta ^2\right) K_0^4+2 \Lambda  K_0^2 P_0-\Lambda ^2 P_0^2\right)}{4
P_0^{3/2} \left(-3 K_0^2+\Lambda  P_0\right)} \\
 ({\bf U}_1)_{13,7}=\frac{i k \beta  \left(\left(3+2 \alpha  \beta ^2\right) K_0^4+2 \Lambda  K_0^2 P_0-\Lambda ^2 P_0^2\right)}{4
P_0^{3/2} \left(-3 K_0^2+\Lambda  P_0\right)} & \qquad ({\bf U}_1)_{13,14}=\frac{\beta  K_0^2 \left(-1-\beta ^2+\frac{2 \alpha
 \beta ^2 K_0^2}{-3 K_0^2+\Lambda  P_0}\right)}{4 \sqrt{P_0}} \\
 ({\bf U}_1)_{13,15}=\frac{\beta  \left(\left(3+2 \alpha  \beta ^2\right) K_0^4+2 \Lambda  K_0^2 P_0-\Lambda ^2 P_0^2\right)}{4
\sqrt{P_0} \left(-3 K_0^2+\Lambda  P_0\right)} & \qquad ({\bf U}_1)_{13,16}=\frac{i k \alpha  \beta ^2 K_0^3}{\sqrt{P_0}
\left(6 K_0^2-2 \Lambda  P_0\right)} \\
 ({\bf U}_1)_{13,17}=\frac{k^2 \left(-\left(3+\alpha  \beta ^2\right) K_0^2+\Lambda  P_0\right)}{2 \beta  \sqrt{P_0} \left(-3 K_0^2+\Lambda
 P_0\right)} & \qquad ({\bf U}_1)_{14,4}=\frac{\alpha  \beta ^4 K_0^5}{2 P_0^{3/2} \left(-3 K_0^2+\Lambda  P_0\right)}
\\
 ({\bf U}_1)_{14,8}=\frac{i k \beta  \left(\left(3+2 \alpha  \beta ^2\right) K_0^4+2 \Lambda  K_0^2 P_0-\Lambda ^2 P_0^2\right)}{4
P_0^{3/2} \left(-3 K_0^2+\Lambda  P_0\right)} & \qquad ({\bf U}_1)_{14,9}=\frac{\beta ^2 K_0 \left(\left(3+2 \alpha  \beta ^2\right)
K_0^4+2 \Lambda  K_0^2 P_0-\Lambda ^2 P_0^2\right)}{4 P_0^{3/2} \left(-3 K_0^2+\Lambda  P_0\right)}
\\
 ({\bf U}_1)_{14,13}=\frac{\beta  K_0^2 \left(1+\beta ^2-\frac{2 \alpha  \beta ^2 K_0^2}{-3 K_0^2+\Lambda  P_0}\right)}{4
\sqrt{P_0}} & \qquad ({\bf U}_1)_{14,16}=\frac{k^2 \left(\left(3+\alpha  \beta ^2\right) K_0^2-\Lambda  P_0\right)}{2 \beta  \sqrt{P_0}
\left(-3 K_0^2+\Lambda  P_0\right)} \\
 ({\bf U}_1)_{14,17}=\frac{i k \alpha  \beta ^2 K_0^3}{\sqrt{P_0} \left(6 K_0^2-2 \Lambda  P_0\right)} & \qquad ({\bf U}_1)_{14,18}=-\frac{\beta
 \left(\left(3+2 \alpha  \beta ^2\right) K_0^4+2 \Lambda  K_0^2 P_0-\Lambda ^2 P_0^2\right)}{4 \sqrt{P_0}
\left(-3 K_0^2+\Lambda  P_0\right)} \\
 ({\bf U}_1)_{15,7}=\frac{K_0 \left(k^2 \left(1+\beta ^2\right)-\frac{2 \alpha  \beta ^4 K_0^4}{-3 K_0^2+\Lambda  P_0}\right)}{4
P_0^{3/2}} & \qquad ({\bf U}_1)_{15,8}=\frac{K_0 \left(-k^2 \left(1+\beta ^2\right)+\beta ^2 \left(K_0^2+\Lambda  P_0-\frac{2
\alpha  \beta ^2 K_0^4}{-3 K_0^2+\Lambda  P_0}\right)\right)}{4 P_0^{3/2}} \\
 ({\bf U}_1)_{15,16}=\frac{-\left(1+\beta ^2\right) \left(k^2+\beta ^2 K_0^2\right)+\frac{2 \alpha  \beta ^4 K_0^4}{-3 K_0^2+\Lambda
 P_0}}{4 \beta  \sqrt{P_0}} & \qquad ({\bf U}_1)_{15,17}=\frac{k^2 \left(1+\beta ^2\right)+\beta ^2 \left(-K_0^2-\Lambda  P_0+\frac{2
\alpha  \beta ^2 K_0^4}{-3 K_0^2+\Lambda  P_0}\right)}{4 \beta  \sqrt{P_0}} \\
 ({\bf U}_1)_{16,4}=-\frac{i k \beta  \left(\left(3+2 \alpha  \beta ^2\right) K_0^4+2 \Lambda  K_0^2 P_0-\Lambda ^2 P_0^2\right)}{4
P_0^{3/2} \left(-3 K_0^2+\Lambda  P_0\right)} & \qquad ({\bf U}_1)_{16,5}=\frac{\beta ^2 K_0 \left(\left(3+2 \alpha  \beta ^2\right)
K_0^4+2 \Lambda  K_0^2 P_0-\Lambda ^2 P_0^2\right)}{4 P_0^{3/2} \left(-3 K_0^2+\Lambda  P_0\right)}
\\
 ({\bf U}_1)_{16,6}=\frac{\alpha  \beta ^4 K_0^5}{2 P_0^{3/2} \left(-3 K_0^2+\Lambda  P_0\right)} & \qquad ({\bf U}_1)_{16,13}=\frac{i
k \alpha  \beta ^2 K_0^3}{2 \sqrt{P_0} \left(-3 K_0^2+\Lambda  P_0\right)} \\
 ({\bf U}_1)_{16,14}=-\frac{\beta  \left(\left(3+2 \alpha  \beta ^2\right) K_0^4+2 \Lambda  K_0^2 P_0-\Lambda ^2 P_0^2\right)}{4
\sqrt{P_0} \left(-3 K_0^2+\Lambda  P_0\right)} & \qquad ({\bf U}_1)_{16,15}=\frac{\beta  K_0^2 \left(1+\beta ^2-\frac{2 \alpha
 \beta ^2 K_0^2}{-3 K_0^2+\Lambda  P_0}\right)}{4 \sqrt{P_0}} \\
 ({\bf U}_1)_{17,4}=-\frac{\beta ^2 K_0 \left(\left(3+2 \alpha  \beta ^2\right) K_0^4+2 \Lambda  K_0^2 P_0-\Lambda ^2
P_0^2\right)}{4 P_0^{3/2} \left(-3 K_0^2+\Lambda  P_0\right)} & \qquad ({\bf U}_1)_{17,5}=-\frac{i k \beta  \left(\left(3+2 \alpha
 \beta ^2\right) K_0^4+2 \Lambda  K_0^2 P_0-\Lambda ^2 P_0^2\right)}{4 P_0^{3/2} \left(-3 K_0^2+\Lambda
 P_0\right)} \\
 ({\bf U}_1)_{17,9}=\frac{\alpha  \beta ^4 K_0^5}{P_0^{3/2} \left(6 K_0^2-2 \Lambda  P_0\right)} & \qquad ({\bf U}_1)_{17,13}=\frac{\beta
 \left(\left(3+2 \alpha  \beta ^2\right) K_0^4+2 \Lambda  K_0^2 P_0-\Lambda ^2 P_0^2\right)}{4 \sqrt{P_0}
\left(-3 K_0^2+\Lambda  P_0\right)} \\
 ({\bf U}_1)_{17,14}=\frac{i k \alpha  \beta ^2 K_0^3}{2 \sqrt{P_0} \left(-3 K_0^2+\Lambda  P_0\right)} & \qquad ({\bf U}_1)_{17,18}=\frac{\beta
 K_0^2 \left(-1-\beta ^2+\frac{2 \alpha  \beta ^2 K_0^2}{-3 K_0^2+\Lambda  P_0}\right)}{4 \sqrt{P_0}}
\\
 ({\bf U}_1)_{18,7}=\frac{K_0 \left(k^2 \left(1+\beta ^2\right)+\beta ^2 \left(-K_0^2-\Lambda  P_0+\frac{2 \alpha  \beta ^2 K_0^4}{-3
K_0^2+\Lambda  P_0}\right)\right)}{4 P_0^{3/2}} & \qquad ({\bf U}_1)_{18,8}=\frac{K_0 \left(-k^2 \left(1+\beta ^2\right)+\frac{2
\alpha  \beta ^4 K_0^4}{-3 K_0^2+\Lambda  P_0}\right)}{4 P_0^{3/2}} \\
 ({\bf U}_1)_{18,16}=\frac{-k^2 \left(1+\beta ^2\right)+\beta ^2 \left(K_0^2+\Lambda  P_0-\frac{2 \alpha  \beta ^2 K_0^4}{-3 K_0^2+\Lambda
 P_0}\right)}{4 \beta  \sqrt{P_0}} & \qquad ({\bf U}_1)_{18,17}=\frac{\left(1+\beta ^2\right) \left(k^2+\beta ^2 K_0^2\right)-\frac{2
\alpha  \beta ^4 K_0^4}{-3 K_0^2+\Lambda  P_0}}{4 \beta  \sqrt{P_0}} \\
\end{align*}}

\bibliographystyle{jhep}

\bibliography{muxin}

\end{document}